\documentclass[journal]{IEEEtran}

\usepackage{cite}
\usepackage{graphicx}
\usepackage{amsmath,amssymb}
\usepackage{array}
\usepackage{booktabs,tabularx}
\usepackage{booktabs}
\usepackage{multirow}
\usepackage{color}
\usepackage{comment}
\PassOptionsToPackage{hyphens}{url}
\usepackage{xurl}
\usepackage{hyperref}
\usepackage{booktabs}
\usepackage{tabularx}
\usepackage{array}
\usepackage{makecell}
\usepackage{amssymb}

\usepackage{longtable}
\usepackage{array}
\usepackage{ragged2e}
\usepackage{hyperref}

\usepackage[table]{xcolor}
\usepackage{array}

\definecolor{primary}{HTML}{f7c593}  
\definecolor{shared}{HTML}{2dc9a2}   
\definecolor{support}{HTML}{DDDDDD}  

\usepackage{graphicx}    
\usepackage{xcolor}      
\usepackage{graphicx}    
\usepackage{xcolor}      

\usepackage{tikz}        
\usetikzlibrary{positioning, shapes, arrows.meta} 

\usepackage{graphicx} 
\usepackage{tabularx,booktabs,ragged2e,array}
\newcolumntype{Y}{>{\RaggedRight\arraybackslash}X}
\title{AI Data Centers Meet Electrical Grids: A Review of Power Challenges and Coordinated Solutions}

\author{Mariam Mughees, Yuzhuo Li,~\IEEEmembership{Member,~IEEE}, Yize Chen,~\IEEEmembership{Member,~IEEE}, \\and Yunwei Ryan Li,~\IEEEmembership{Fellow,~IEEE} 
\thanks{ 
The authors are with the Department of Electrical and Computer Engineering, University of Alberta, email: \{mughees, yuzhuo, yize.chen, yunwei.li\}@ualberta.ca.}}

\date{September 2026}

\begin{document}

\maketitle

\begin{abstract}
     Artificial intelligence (AI) workloads are creating large, power-electronic loads whose effects on power systems depend on location, temporal variability, controllability, and
  interactions with other loads and resources. Annual electricity consumption alone cannot capture these effects. This review examines how AI data centers affect power systems stability, renewable-energy integration, and low-carbon planning by tracing the power-delivery chain from accelerators and workloads through rack converters, uninterruptible power supplies (UPS), storage, and microgrids to the transmission point of interconnection. Reported oscillatory events and
  multi-gigawatt load transfers highlight how facility responses can amplify disturbances, while programmable workloads and controllable power converters offer opportunities for demand flexibility. Challenges and mitigation strategies are organized across device, rack, facility, and power systems levels, spanning millisecond-to-year timescales. A bandwidth-matching framework relates disturbances to the response capabilities of mitigation resources. The review distinguishes computational energy efficiency from power systems and sustainability outcomes: lower energy consumption per token does not necessarily reduce peak demand, electrical disturbances, or carbon emissions. It examines how coordinated workload scheduling, grid-interactive UPS systems, storage, high-voltage DC distribution, and flexible interconnection can support reliability, efficiency, and decarbonization. Key research needs include dynamic load models, high-bandwidth telemetry, carbon- and grid-aware scheduling, and mechanisms for valuing flexible computational demand.


\end{abstract}

\section{Introduction}

\label{sec:intro}

Artificial intelligence (AI) workloads are reshaping both the scale and dynamic characteristics of data center electricity demand at the same time that power systems are undergoing rapid decarbonization through increasing deployment of renewable and inverter-based resources (IBRs). Conventional data centers are already significant electricity consumers, but in a new generation of AI data centers, AI training and inference concentrate substantially greater computing capacity within individual racks and clusters, producing higher power densities and more rapidly varying demand at facility and campus scales. Individual developments are approaching gigawatt scale, while regional project pipelines can exceed the available capacity of multiple substations and transmission corridors. Large computational loads must connect without undermining grid reliability, increasing dependence on carbon-intensive marginal generation, or constraining renewable utilization. Thus,  electricity supply is not the sole part of this challenge. 
Evaluating new AI demand requires location, firmness, load characteristics, reliability and controllability to be assessed together~\cite{chen2025electricity, NERC2025LargeLoads}. Table~\ref{tab:large-load-grid-events}  summarizes recent grid disturbances involving data centers and other large computational loads.

\begin{table*}[t]
\centering
\caption{Recent grid events involving AI and large computational loads (2022--2026).}
\label{tab:large-load-grid-events}
\renewcommand{\arraystretch}{1.15}
\scriptsize
\begin{tabularx}{\textwidth}{
    >{\raggedright\arraybackslash}p{0.07\textwidth}
    >{\raggedright\arraybackslash}p{0.17\textwidth}
    >{\raggedright\arraybackslash}p{0.11\textwidth}
    >{\raggedright\arraybackslash}X
    >{\raggedright\arraybackslash}p{0.07\textwidth}}
\toprule
\textbf{Year} &
\textbf{Event type} &
\textbf{Load impact} &
\textbf{Event description} &
\textbf{Source} \\
\midrule

2022 &
Impulsive load loss &
1,500--1,700 MW &
On 7 December 2022, a delayed-clearing fault in West Texas caused a
load reduction in excess of 1,500 MW across roughly 400 facilities,
including substantial losses from oil and gas sites. &
\cite{ercot_lfltf_2025} \\

2022--2023 &
Sustained oscillation &
Not reported &
Dominion data center load exhibited sustained oscillations at
approximately 14.7--14.8 ~Hz. Synchrophasor measurements indicated that
the behavior persisted over multiple months. &
\cite{mishra2025understanding, Dominion2024} \\

2023--2025 &
Repeated ride-through loss &
$>$100 MW each &
ERCOT identified 26 large electronic load ride-through events with
indirect load loss exceeding 100 MW from crypto facilities between
January 2023 and September 2025, with reductions ranging from roughly
17\% to 95\% of pre-disturbance consumption. &
\cite{NERC_CryptoIncident_2026} \\

2024 &
Impulsive load loss &
1,500 MW &
A 230 kV transmission fault in July 2024 caused approximately 1,500 MW
of data center load loss in the Eastern Interconnection, distributed
across roughly 25--30 substations and 60 load points. &
\cite{NERC_Incident_2025, NERC2025LargeLoads} \\

2025--2026 &
Impulsive load loss &
$>$1,000 MW &
Repeated customer-initiated load reductions in which more than 1,000 MW
of computational load disconnected within seconds, motivating NERC's
subsequent Level~3 alert. &
\cite{NERC_Alert_L3_2026} \\

2026 &
Impulsive load loss &
3,000 MW &
A July 2026 transmission line fault in the PJM/Northern Virginia area
disconnected more than 3 GW of data center load, the largest such
event recorded to date, causing regional voltage to rise before
operators restored normal conditions. &
\cite{PJM2026DominionLoadTransfer} \\

\bottomrule
\end{tabularx}
\end{table*}

Correlated customer-side protection responses have twice caused gigawatt-scale reductions in grid-supplied demand within the same Northern Virginia corridor. On 10 July 2024, a fault on a 230-kV transmission line was followed by an approximately \(1.5~\mathrm{GW}\) reduction in data center load as multiple facilities transferred their demand to behind-the-meter backup systems. In 2026, a comparable, normally cleared 230-kV fault in PJM triggered a substantially larger two-stage transfer, causing an unexpected disconnection and transfer of approximately \(3.8~\mathrm{GW}\) load~\cite{NERC2025LargeLoads, PJM2026DominionLoadTransfer}. The abrupt loss of demand created a significant generation--load imbalance and a widespread voltage disturbance. ERCOT identified at least 26 events between 2023–2025 in which large computational loads tripped during routine voltage disturbances~\cite{mclaughlin2026texasgrid}. Although transmission protection and operator actions prevented the event from escalating into a cascading outage, it exposed important gaps in large-load ride-through capability, dynamic modeling, utility--customer protection coordination, and controlled load-reconnection requirements.

These observations are primarily reliability events, but they also expose a sustainability constraint. A large computational load that is poorly coordinated with the host network may require greater operating reserves, network reinforcement, storage headroom, or fast-ramping generation even when its annual electricity consumption is supplied contractually from renewable sources. A controllable data center that rides through disturbances, limits ramps, shifts flexible computation, and coordinates internal storage with grid conditions can instead support variable renewable integration, with no change to the computational service delivered.

Beyond scale, the temporal behavior of AI workloads introduces a qualitatively new class of electrical load. Large-scale training jobs and coordinated inference pipelines generate rapid, correlated power fluctuations spanning multiple timescales—from sub-second device-level switching to minute-level workload orchestration. These dynamics differ fundamentally from those of conventional industrial or commercial loads. They challenge assumptions underlying traditional power systems planning, stability analysis, and operational control.

Siting and interconnection compound the problem, as expanding
computational demand must be coordinated with the development of
generation, storage, and transmission infrastructure. The scale of
this expansion is illustrated by U.S. data center electricity
consumption, which reached approximately 176~TWh in 2023; Berkeley
Lab's 2024 assessment projected annual consumption of
325--580~TWh by 2028~\cite{shehabi20242024}. Meanwhile, by the
end of 2025, approximately 1,312~GW of generation and 749~GW of
storage were actively seeking transmission interconnection in the
U.S. alone. 
These supply-side queues exclude data center load connection
requests, but illustrate the lead times and uncertainty associated
with bringing new resources online to serve growing demand.
Proposed capacity therefore cannot be treated as immediately
available supply, motivating coordinated planning of computational
loads, electricity resources, and network expansion.

For an individual data center, state-level generation surplus,
electricity cost, or installed capacity does not establish whether
a particular bus can securely accommodate a large campus.
Deliverable capacity also depends on transmission limits,
short-circuit strength, reactive power capability, transformer
availability, and the dynamic behavior of nearby resources.
Consequently, an energy-rich region may still offer little firm,
dynamically secure headroom at the proposed connection point.
This distinction motivates evaluating workload flexibility,
on-site storage, and coordinated facility controls alongside
network reinforcement, including whether conditional or phased
connections can accommodate new demand while respecting local
operating limits.

 \begin{figure*}[h!]
    \centering
    \includegraphics[width=0.93\linewidth]{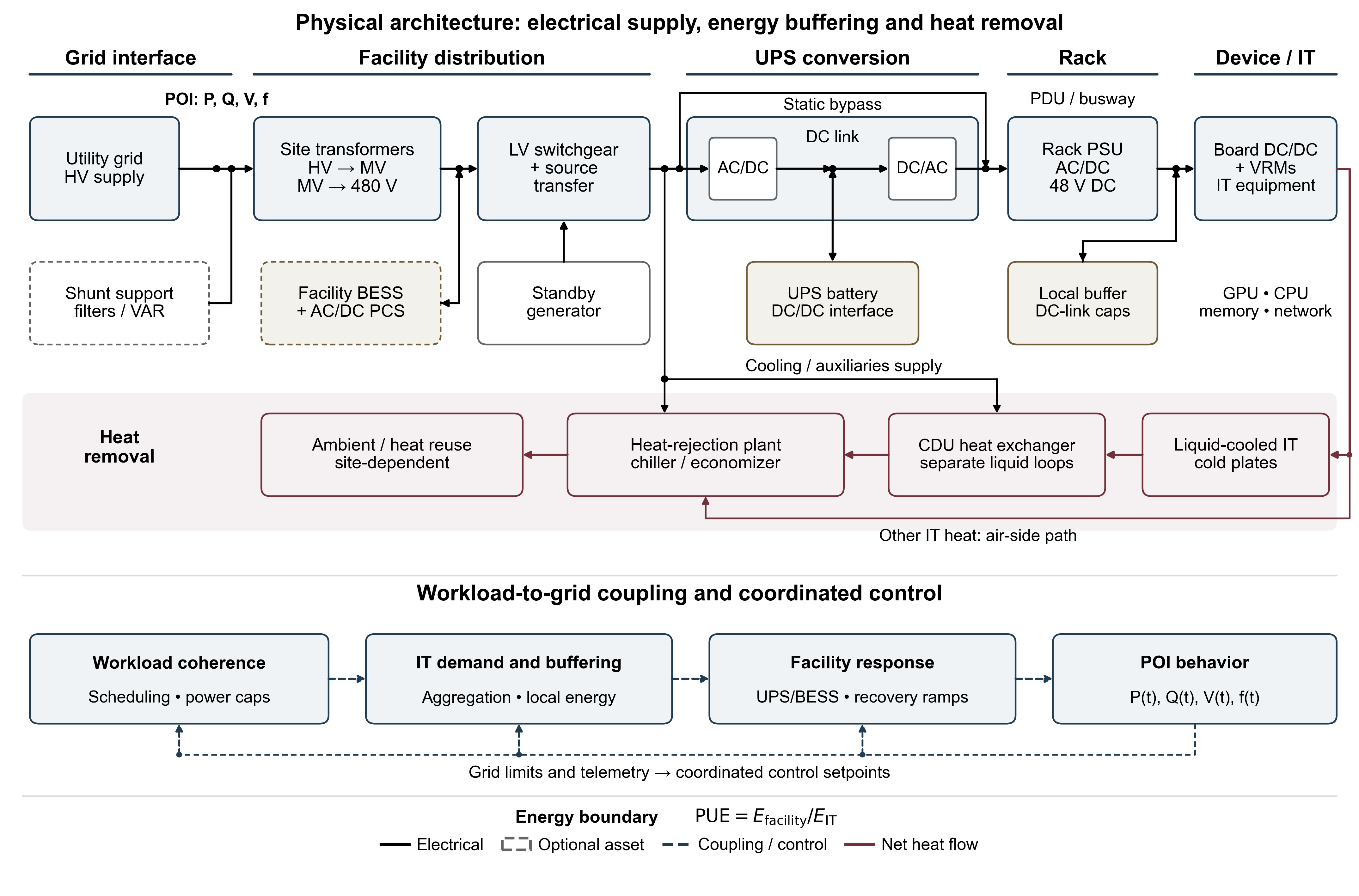}
\caption{AI data center power-delivery architecture, showing the electrical path from the utility grid through facility distribution, UPS conversion, rack power supplies, and device-level voltage regulation, together with backup generation, energy storage, and cooling. The lower panel shows workload-to-grid coupling and coordinated control; the energy boundary defines power usage effectiveness~\cite{schneider2026rd111}.}
    \label{fig:power_arch}
\end{figure*}

These challenges arise from interactions across technical domains that are usually studied separately. Power system studies commonly represent large data centers as static or slowly varying active- and reactive-power injections and focus on adequacy, interconnection, voltage, frequency, and stability. Computer systems research instead emphasizes throughput, latency, utilization, cooling, power capping, and energy efficiency, often without explicitly representing the dynamic limits of the transmission network supplying the facility. A third body of research on sustainable data centers considers renewable procurement, carbon-aware computing, cooling efficiency, energy management, and demand response, but typically evaluates these objectives at minute-to-hour or annual timescales~\cite{tabbakh2024towards, chen2025electricity}. The resulting gap is not an absence of research on data center energy use, but a lack of an end-to-end framework connecting software-generated electrical dynamics to both power systems reliability and sustainable energy integration. Additional transmission does not by itself stabilize an adverse converter interaction, and reducing accelerator energy consumption leaves a coherent ramp intact. A complete assessment must therefore follow the electrical pathway from workload timing through device and rack conversion, facility controls and storage, and finally the point of interconnection with the host grid.

This review identifies and addresses that gap by examining the electrical pathway from workload timing and device behavior through rack and facility power
conversion to the host grid. In particular:
\begin{itemize}
  \item At the \emph{power grid level}, AI loads amplify instability in low-inertia systems.
  \item At the \emph{facility and rack levels}, power-electronic architectures determine whether workload-driven fluctuations are absorbed internally or transmitted as destabilizing disturbances.
  \item At the \emph{device and workload levels}, scheduling policies and dynamic power management directly modulate real-power transients across timescales from milliseconds to minutes.
\end{itemize}

These cross-layer interactions cannot be mitigated solely by single-layer solutions. A range of software- and hardware-based power-modulation schemes is emerging across the power and IT hierarchies within data centers. These schemes include the use of power assets such as generators and energy storage; hardware-based controls such as server throttling and shutdown; and application-based approaches such as model routing and workload migration. The framework organizes challenges and solutions across four
layers: 
\begin{itemize}
    \item Grid layer: power grid stability, interconnection, system strength
mechanisms, and the siting of demand relative to renewable resources and
transmission headroom;
\item Facility layer: microgrids, uninterruptible power supplies (UPS), energy storage, and on-site generation;
\item Rack layer: power architectures, high-voltage DC (HVDC) systems, and conversion technologies;
\item Device and workload layer: graphics processing unit (GPU) behavior, scheduling policies, and power-aware optimization.

\end{itemize}
\label{sec:power_arch}

Throughout, sustainable AI data center operation denotes advancing
four objectives jointly---energy efficiency, carbon reduction, renewable
accommodation, and grid-facing load shaping---without compromising
host-grid reliability. Within this framework, the paper makes the
following contributions:

\begin{itemize}
    \item It characterizes data centers as evolving, dynamic, and power-electronics-rich large loads that may create stability, reliability, and interconnection challenges across multiple timescales.

    \item It develops a cross-layer taxonomy of mitigation strategies spanning grid, facility, rack, device, and workload controls.

    \item It proposes an implementation framework connecting technical requirements at the
point of interconnection to the responsibilities of system operators, utilities, data
center operators, vendors, and AI developers.
\end{itemize}
\vspace{-2pt}
The remainder of the paper follows the electrical pathway from workload to
grid. Section~\ref{sec:profile} characterizes AI load behavior and the data
center power-delivery architecture. Section~\ref{sec:challenges} analyzes the
resulting grid-facing challenges. Section~\ref{sec:solutions} presents the
multi-layer solution stack, mapping each mitigation to the challenge it
addresses and to the regulatory requirements. Section~\ref{sec:stakeholders} provides recommendations on stakeholder responsibilities and concludes with key insights and future research directions, including opportunities for
tighter integration between AI systems and power systems operation.

\section{AI Load Characteristics and Data Center Power Architecture}
\label{sec:profile}
\subsection{AI Load Profiling}

\begin{figure*}
    \centering
    \includegraphics[width=0.89\linewidth]{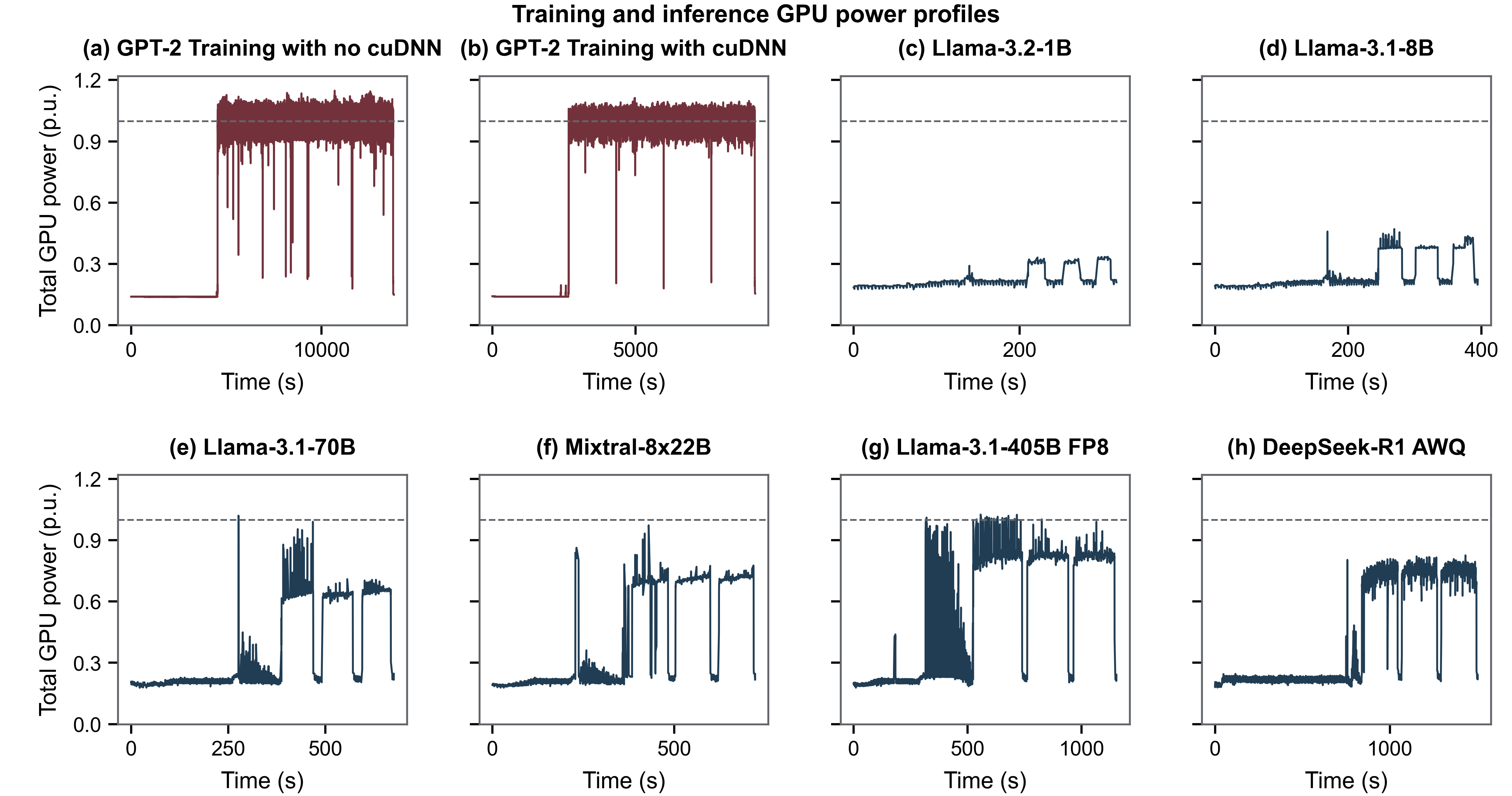}
    \caption{Per-unit GPU power measurements for large language model (LLM) training and inference workloads. The training traces show repeated load reductions, whereas the inference traces show workload-dependent, irregular bursts.}
    \label{fig:AI_profile}
\end{figure*}

AI data center workloads fall into two operational regimes, training and inference, whose absolute power requirements are comparable but whose dispatchability differs sharply. Figure~\ref{fig:AI_profile} shows power signatures on a per-unit basis. Training workloads typically engage thousands of accelerators at or near rated power for periods ranging from several days to multiple months. For example, LLaMA-3~405B was trained using 16{,}000 H100 GPUs~\cite{meta2024llama31}. With accelerator utilization often approaching $80\%$, training therefore resembles a large, relatively firm industrial load at the hourly temporal resolution relevant to power systems planning. Its principal variability occurs over sub-second to second timescales and is determined primarily by software execution rather than by exogenous demand. Collective communication barriers, gradient-synchronization operations, and checkpoint writes can periodically idle substantial portions of the accelerator fleet, causing facility power to fall from near-full utilization toward the $20\%$--$25
\%$ idle-power floor within seconds before recovering just as quickly~\cite{sajadi2026power}. These scheduler-generated transitions have temporal structure that the facility operator can anticipate. The recent NVIDIA Vera Rubin NVL72 system consumes between 190 kW and 230 kW per rack, illustrating the high power density.

Inference exhibits a substantially different operational profile. Although aggregate inference demand may be lower at an individual site, its load envelope is more closely coupled to user activity and therefore exhibits pronounced diurnal and weekly patterns, together with event-driven excursions. Within this broader envelope, however, synchronous inference is effectively non-dispatchable. Individual requests cannot generally be deferred or interrupted without violating latency or other quality-of-service requirements~\cite{sajadi2026power}. Inference variability is stochastic rather than scheduler-induced. It reflects the statistics of request arrivals and the uncertain computational effort of individual requests, whose energy consumption is unknown until generation terminates.

\subsection{Power Architecture of Data Centers}

GPU power demands aggregate at server and rack buses before passing through facility conversion and distribution to the POI. Synchronized workload transitions can reinforce one another across devices, whereas rack buffers, UPS systems, and facility storage can attenuate upstream fluctuations. The POI therefore sees the combined effects of workload correlation, conversion losses, cooling demand, and facility control responses.

Figure~\ref{fig:power_arch} traces the power path from generation to the
processor core. Power is drawn from a mix of conventional and
inverter-based generation and enters the campus at a point of
interconnection (POI) whose short-circuit strength---itself reduced as
synchronous generation is displaced by inverter-based resources---
conditions the stability of every converter downstream of it. From the POI, medium- and low-voltage distribution losses account for 2--3\% of imported energy, with paralleling switchgear
tying the sources and dedicated power-quality compensation countering the
nonlinear, fast-varying load. Standby generation and an online uninterruptible power supply (UPS, with 2--4\% conversion losses) provide outage ride-through,
while battery or hybrid storage supplies backup reserve and increasingly
shapes the upstream ramp; because workload ramps can outrun upstream
generation and network response, that storage delivers most of its
benefit when pre-positioned ahead of a load transition rather than
dispatched after one. Low-voltage distribution then carries power through
rack-level conversion---the rack power supply unit (PSU, with 4--6\% conversion losses) and board-level voltage regulator modules (VRMs, with 8--12\% conversion losses)---to the sub-volt, high-current rails
the accelerators require, buffered locally against the fastest transients. For data centers, power usage effectiveness (PUE) is defined as
\begin{equation}
\mathrm{PUE}=\frac{E_{\mathrm{facility}}}{E_{\mathrm{IT}}}.
\label{eq:pue}
\end{equation}
Here, both energy quantities are evaluated over the same reporting period and at explicitly defined measurement boundaries. Facility overhead, including cooling and electrical losses upstream of the IT-equipment measurement point, therefore accounts for a fraction of $(1-\frac{1}{\mathrm{PUE}})$ of total facility energy. For an assumed PUE range of \(1.1\)--\(1.2\), this fraction is approximately \(9.1\)--\(16.7\%\). These upstream losses must not be counted again when estimating energy delivery to the compute dies. Conversion losses downstream of the IT measurement point must be evaluated separately, with each component's efficiency defined relative to its own input energy.

The fraction of facility energy delivered to the target compute dies is consequently
\begin{equation}
\eta_{\mathrm{facility}\rightarrow\mathrm{die}}=
\frac{1}{\mathrm{PUE}}
\eta_{\mathrm{IT}\rightarrow\mathrm{die}},
\qquad
\eta_{\mathrm{IT}\rightarrow\mathrm{die}}=
\frac{E_{\mathrm{die}}}{E_{\mathrm{IT}}}.
\label{eq:facility_to_die}
\end{equation}
Here, \(\eta_{\mathrm{IT}\rightarrow\mathrm{die}}\) accounts for downstream conversion losses and the allocation of IT energy among the target compute dies and other IT components~\cite{schneider2026rd111}. Note that PUE and conversion efficiency do not capture carbon intensity, water use, renewable availability, or load flexibility; component efficiency and system-level sustainability are distinct.

\section{Challenges of Powering AI Data Centers}\label{sec:challenges}
The dynamic load characteristics and grid responses of AI data centers depend on converter controls, energy storage, renewable integration, AI workload scheduling and synchronization, protection settings, and UPS and peripheral control loops. This section examines the resulting oscillatory, voltage, frequency, reliability, planning, and interconnection challenges spanning timescales from microseconds to multiple years.

\subsection{Sub-Synchronous Challenges in Converter-Dominated Systems}
\label{subsec:ssr}

Sub-synchronous resonance and oscillations (SSR/SSO), historically associated with torsional--electrical interactions between synchronous generators and series-compensated transmission, now arise through a fundamentally different mechanism in converter-dominated grids. Grid-following and grid-forming
inverters introduce frequency-dependent impedance and finite control-loop delays, so network eigenmodes in the 10--40~Hz band can be poorly damped or destabilized under weak-grid conditions~\cite{cheng2022real,NESO2024}. A modern AI campus aggregates thousands of active power-factor-correction (PFC) rectifiers, double-conversion UPS units, and rack-level converters. Tight DC-bus regulation makes the aggregate resemble a constant-power load: when terminal voltage falls, current increases to maintain power. The corresponding negative incremental
conductance can reduce damping at frequencies where upstream controls
cannot respond~\cite{Chakraborty2024Impedance,sun2022dynamic}. The aggregate
admittance also varies with utilization, current limits, enabled control
modes, vendor mix, and firmware.

Workload timing can also supply a forcing function. Communication periods of roughly 25–50 ms place the fundamental frequency in the 20–40~Hz range, with higher harmonics occurring above the fundamental. Training with synchronous stochastic gradient descent (SGD) enforces tight
barrier synchronization across the fleet. Concentrated AI demand weakens grid damping while creating software-controlled disturbances at the frequencies where damping is already weakest.

\begin{figure}[!h]
  \centering
  \includegraphics[width=0.45\textwidth]{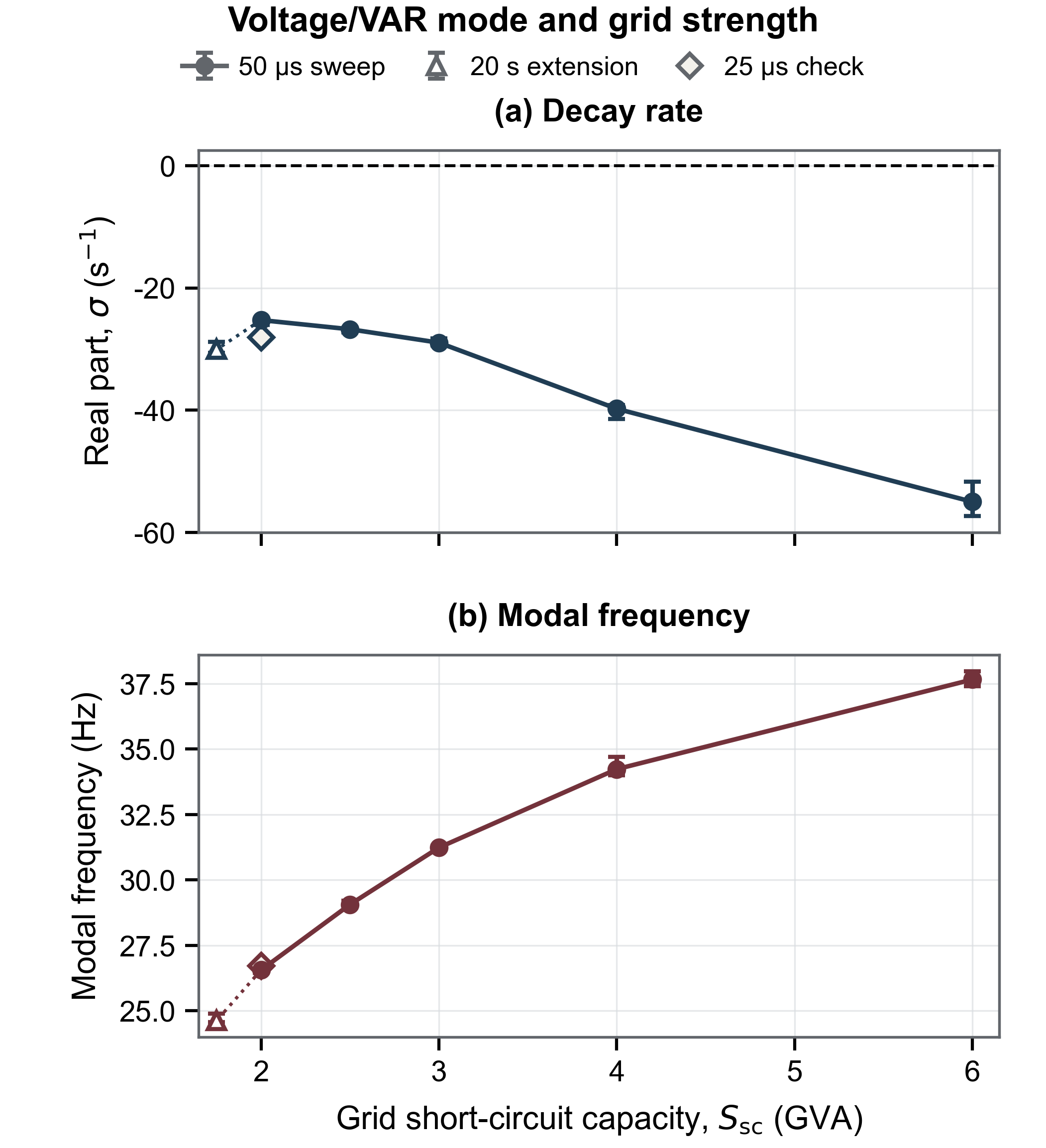}
  \caption{Dominant voltage/VAR mode identified from native PSCAD ringdowns of the 1~GW nominal PNNL central-UPS model. The 50~$\mu$s EMT sweep uses $S_{\mathrm{sc}}=2$, 2.5, 3, 4, and 6~GVA. The triangle is the separately settled 20~s case at 1.75~GVA, and the diamond is the independent 25~$\mu$s check at 2~GVA. Error bars span the fit-window and model-order sensitivity. Every displayed point passed the operating and equipment screens, and all identified real parts are negative~\cite{PNNL38817}.}
  \label{fig:native-strength}
\end{figure}

The effect of grid strength is evaluated using a detailed data center
  model in PSCAD at a fixed nominal facility rating of 1~GW. A small grid
  disturbance is applied at each operating point, and the dominant
  voltage/VAR mode is identified from the resulting POI responses. For
  the modal pole $\lambda=\sigma\pm j2\pi f_m$, a negative $\sigma$
  denotes a stable, decaying response.

  Figure~\ref{fig:native-strength} shows the identified decay rate and modal
  frequency for short-circuit capacities between 2 and 6~GVA. As
  $S_{\mathrm{sc}}$ decreases from 6 to 2~GVA, the real part of the pole moves
  from $-54.99$ to $-25.27~\mathrm{s}^{-1}$, while the modal frequency
  decreases from 37.65 to 26.55~Hz. The corresponding damping ratio
  decreases from 0.227 to 0.150. These results show that weakening the
  grid reduces the damping of the identified voltage/VAR mode, although
  all accepted operating points remain stable.

  The extended 1.75~GVA case also remains stable, while lower-strength
  cases violate operating constraints and are excluded. A
  $25~\mu\mathrm{s}$ time-step check confirms the modal frequency and
  the negative sign of $\sigma$, although the decay-rate magnitude
  remains moderately time-step sensitive. No zero-real-part crossing is
  observed within the feasible range; therefore, the results do not
  establish a universal SCR threshold or maximum stable facility rating.

\begin{figure}[h!]
\centering
\includegraphics[width=0.5\textwidth]{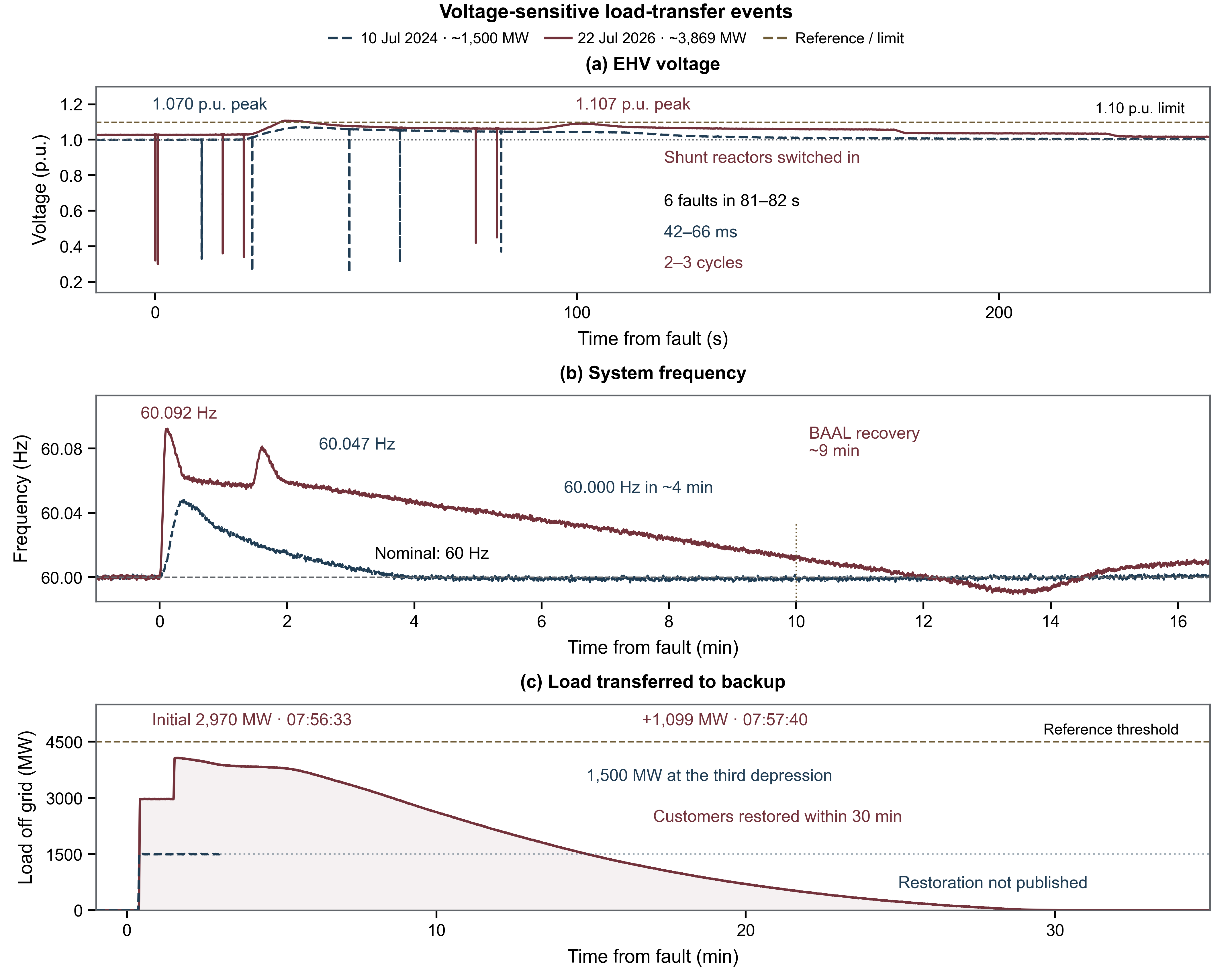}
\caption{Comparison of two voltage-sensitive data center load-transfer events in Northern Virginia: the 10 July 2024 event and the 22 July 2026 event. (a) Reconstructed load-pocket transmission-voltage response following the initial 230-kV faults and subsequent reclosing operations. (b) Reconstructed Eastern Interconnection frequency response, showing peak frequencies of approximately \(60.047~\mathrm{Hz}\) in 2024 and \(60.092~\mathrm{Hz}\) in 2026. (c) Estimated data center load transferred from the utility grid to behind-the-meter backup systems. The 2024 event removed approximately \(1.5~\mathrm{GW}\), primarily following the third voltage depression, whereas the 2026 event occurred in two stages: an initial \(2{,}970~\mathrm{MW}\) transfer followed by an additional \(1{,}099~\mathrm{MW}\), with an estimated net reduction of approximately \(3{,}869~\mathrm{MW}\)~\cite{NERC2025LargeLoads,PJM2026DominionLoadTransfer}.}

\label{fig:multiscale-event}
\end{figure}

A 2025 Dominion Energy investigation documented a sustained $\sim$14.7~Hz oscillation in its 115~kV Northern Virginia data center corridor, a region hosting one of the world's largest hyperscale clusters, traced to a single campus among several served by the affected substations~\cite{mishra2025understanding,Dominion2024}. The event
began with visible voltage flicker reported at three nearby substations
during an overnight period, shortly after the routine ramp-down of four
hydro units at a nearby station that supplies fault current to the corridor. It persisted for nearly
two hours across the wide-area synchrophasor footprint.
The frequency is too low for any plausible mechanical or network LC resonance and is consistent in magnitude with the voltage-loop bandwidth of a UPS or active rectifier.
The campus appears to be creating a new closed-loop mode through its impedance interaction with a weak corridor, not exciting a pre-existing network mode. Such modes propagate over wide geographic areas and are superimposed on existing IBR-driven dynamics, complicating wide-area mode estimation and source attribution~\cite{Dominion2024}.

At present, frequency-domain screening and EMT studies require control-faithful aggregate models of UPS and rectifier behavior, including loop gains, current limits, and ride-through logic~\cite{PNNL38817}. Operators can measure the POI waveform but cannot in general identify which of several co-located facilities is the dominant contributor, so sub-synchronous resonance must now also be studied as a load-side problem.

\subsection{Inter-Area Oscillations}
\label{subsec:interarea}

Inter-area oscillations are coherent electromechanical swings between
geographically separated synchronous generator groups connected by weak
tie-lines. They appear in the 0.1--1~Hz band on
seconds-to-minutes timescales~\cite{klein1991fundamental,NERC2019IOA}. Because the underlying modes involve rotor-angle dynamics and
synchronizing torque, load effects have traditionally been represented
through aggregate load models and statistical damping rather than as coherent forcing sources.

AI hyperscalers concentrate hundreds of megawatts to several gigawatts at a single point of interconnection and vary that demand coherently under software-imposed schedules rather than through independent end-use decisions. Three load-side influences on inter-area dynamics act at the same site: (i) a structural shift in the modes through generation
displacement, (ii) narrow-band forcing through workload spectral
content, and (iii) impulsive excitation through coordinated protection
action.

As synchronous generation is displaced by inverter-based resources, rotor-angle dynamics and synchronizing torque change, altering the characteristics of the inter-area modes and generally reducing their damping~\cite{NERC2025LargeLoads,NERC2020Guideline}. Modes historically
near 0.5--0.8~Hz are migrating toward 0.1--0.3~Hz, into the band where
slower AI workload phenomena have the most spectral energy. 
 Millisecond-scale compute--communication phases sit above the
inter-area band (Section~\ref{subsec:ssr}). Several slower workload structures, however, generate real-power variation within 0.1--1 Hz. These may come from large-batch iteration boundaries, correlated chiller and pump cycling, and dynamic voltage and frequency scaling (DVFS) or power-cap loops with seconds-scale time constants.
Training with synchronous stochastic gradient descent (SGD) enforces near-simultaneous phase transitions across the fleet, so these variations do not average out, and the resulting forcing is narrowband and software-tunable. This concern is also reflected in regulatory practice. For instance, NERC's May 2026 Level~3 Alert
identifies AI training workloads alongside electric arc furnaces as
the load classes most likely to introduce forced oscillations into the
bulk power system~\cite{NERC_Alert_L3_2026}.

Spatial concentration ties these mechanisms together. Mode shapes have geographic structure, as some buses sit at antinodes, others near nodes, then a given megawatt of load couples to wide-area dynamics in proportion to its participation factor.

Figure~\ref{fig:multiscale-event} illustrates the coupling between facility protection and bulk-system power balance. Transmission-voltage disturbances prompted customer-side transfers to backup supplies, reducing grid demand while protected loads could remain supplied locally. The resulting generation surplus contributed to the observed frequency increases, alongside changes in post-disturbance voltage. The reported reductions were approximately 1.5 GW in 2024 and 3.9 GW in 2026, with frequency peaks of 60.047 Hz and 60.092 Hz, respectively. These events demonstrate why ride-through, transfer, and restoration behavior must be represented in large-load studies: the grid-facing contingency depends not only on the initiating fault but also on the collective response of connected facilities. The reconstructed curves illustrate this sequence; they are not suitable for estimating event RoCoF or identifying oscillatory modes~\cite{Dominion2025FIR}.

Conventional screening represents large loads as static or slowly varying P–Q injections, which is insufficient for hyperscale AI sites. Analysis requires spectral characterization of the workload-driven forcing function alongside modal analysis of the host network. 

\subsection{Voltage and Frequency Stability}
\label{subsec:voltage}

AI data centers combine distinct voltage-stability mechanisms into a
single facility: a steady-state reactive-margin problem from constant-power-load behavior, a transient problem created by protection and ride-through logic, and a short-term problem originating in motor-dominated cooling.

The steady-state mechanism follows directly from the impedance properties
discussed in Section~\ref{subsec:ssr}. A campus regulating DC-bus power tightly behaves approximately as a constant-power load, increasing its current draw as terminal voltage falls. Although modern active-PFC front ends operate close to unity power factor, the higher current increases reactive consumption in the upstream transmission corridor through its inductive reactance, reducing voltage margins following disturbances. Weak-grid conditions amplify voltage excursions caused by sudden load ramps. 
The same property that produces load-side SSR at frequencies above the inter-area band also produces a risk of voltage collapse at zero frequency.

The transient mechanism arises from facility protection. During modest voltage or frequency excursions, UPS and control schemes may trip or transfer load to on-site generation, imposing abrupt MW/MVAr steps that complicate post-fault recovery and reduce short-term voltage-stability headroom~\cite{psercDynamicLoads2024}. The protective action mitigates the local disturbance and converts it into a system-level contingency. A sag the campus could have ridden through becomes a step disconnection of hundreds of megawatts. Recent work proposes enhanced
low-voltage ride-through controls for data centers~\cite{xie2025enhancing} and
dynamic composite load models that capture UPS limits and pulsed AI
workloads~\cite{milanoDCmodel2025}, but neither is yet standard in
interconnection studies.

The third voltage stability mechanism originates in the cooling system. Chilled-water pumps, condenser fans, and other induction-motor loads increase reactive draw
during voltage sags; a sustained voltage depression can drive motors toward stalling, deepening the local reactive deficit and precipitating a stall cascade.
 
In an AI campus, the cooling load tracks compute load with a lag, so a workload-driven ramp that elevates motor demand precisely when reactive margin is already
degraded creates a coincident worst case that conventional motor-load
screening, calibrated against weather-driven air-conditioning demand, does
not capture.

Similarly, frequency risk arises from the interaction between declining synchronous
inertia and a contingency whose size depends on facility controls. For an
imbalance $\Delta P$, the initial rate of change of frequency is
approximately

\begin{equation}
\dot{f} = -\frac{\Delta P\, f_0}{2 H S_{\mathrm{base}}},
\label{eq:rocof}
\end{equation}

where $\Delta P$ is the incremental net power deficit (positive for
an increase in demand or a loss of generation), $H$ is the aggregate
inertia constant on the common base $S_{\mathrm{base}}$, and $f_0$ is the
nominal frequency~\cite{entsoe_rocof2020}. A reduction in grid-supplied
load therefore corresponds to $\Delta P<0$ and a positive initial RoCoF.
This aggregate swing-equation approximation does not represent spatial
variations in locally measured frequency.

\begin{figure*}[h!]
\centering
\includegraphics[width=0.9\textwidth]{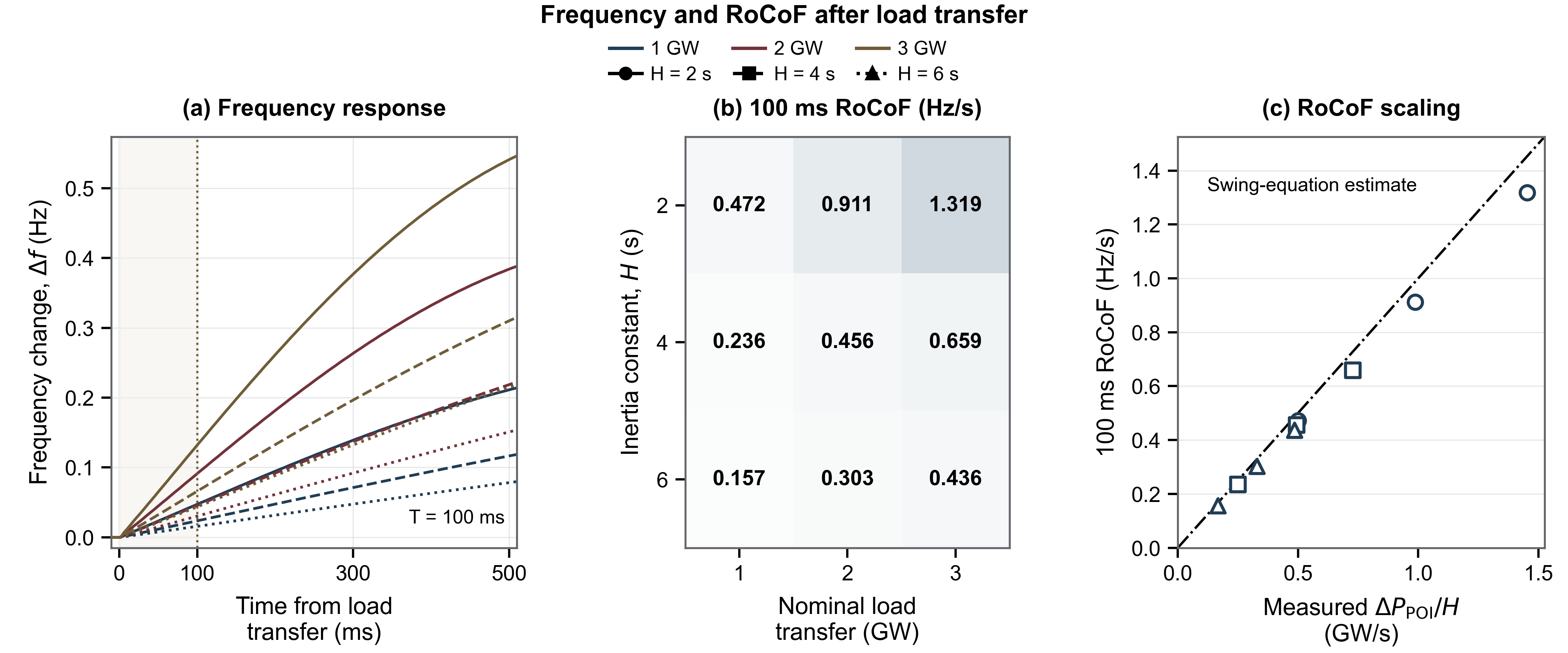}
\caption{Frequency and rate-of-change-of-frequency (RoCoF) responses following load transfer. (a) Event-referenced frequency responses for nominal load
  transfers of $1$, $2$, and $3~\mathrm{GW}$ and inertia constants of $H=2$, $4$, and $6~\mathrm{s}$; the shaded region denotes the $T=100~\mathrm{ms}$
  measurement window. (b) Corresponding $100~\mathrm{ms}$ RoCoF values as functions of load-transfer magnitude and inertia. (c) Measured RoCoF plotted
  against $\Delta P_{\mathrm{POI}}/H$, with the dash-dotted line showing the initial swing-equation estimate~\cite{PNNL38817}.}

\label{fig:freq-rocof}
\end{figure*}

Figure \ref{fig:freq-rocof} illustrates the grid-frequency response when a data center’s UPS transfers its protected IT load from utility supply to battery supply. Before
  the transfer, power flows from the grid through the rectifier, DC link, and inverter to the IT equipment. At transfer, the rectifier is blocked, and the
  battery supplies the DC link through the DC/DC converter while the inverter continues powering the IT load. Thus, IT service is maintained, but grid-supplied demand decreases abruptly. Mechanical input to the synchronous generators cannot decrease immediately, creating a temporary generation surplus.

To examine the initial frequency impact of the transfer mechanism
illustrated in Figure~\ref{fig:freq-rocof}, we conducted a generic EMT sensitivity study
using an adapted PNNL DML v3.2 centralized-UPS model~\cite{PNNL38817}. The
synchronous-generation equivalent is held at $30\,\mathrm{GVA}$ and
$60\,\mathrm{Hz}$, while $H = 2,\,4,\,6\,\mathrm{s}$ is examined for
three campus configurations with nominal protected-IT loads of 1, 2,
and $3\,\mathrm{GW}$. At $t_0 = 5\,\mathrm{s}$, a commanded rectifier
block transfers the protected load to battery-supported inverter
operation; cooling and other campus loads remain connected. This
isolates the frequency consequence of the transfer from the preceding
transmission-fault dynamics. The simulations use a $50\,\mu\mathrm{s}$
integration step and a $250\,\mu\mathrm{s}$ recording interval.

Figure~\ref{fig:freq-rocof}(a) shows the event-referenced frequency increase calculated
from synchronous-machine speed. The signed $100\,\mathrm{ms}$ RoCoF
in Figure~\ref{fig:freq-rocof}(b) increases with transferred demand and decreases with
inertia. For the nominal $2\,\mathrm{GW}$ campus at
$H = 4\,\mathrm{s}$, grid import falls by approximately
$1.976\,\mathrm{GW}$ while mean protected IT power remains
approximately $1.947\,\mathrm{GW}$. The corresponding $R_{100}$
is $0.456\,\mathrm{Hz/s}$, compared with an ideal-step inertial
reference of $0.494\,\mathrm{Hz/s}$.

Figure~\ref{fig:freq-rocof}(c) organizes the response by $\Delta P_{\mathrm{POI}}/H$.
The finite-window results lie below the ideal-step reference because
the frequency change reflects the realized accelerating-power
trajectory rather than an instantaneous, constant imbalance equal
to the later POI-power reduction. The results illustrate the
sensitivity of the initial frequency response to transfer magnitude
and stored synchronous kinetic energy. They represent separately
scaled generic campuses and are not calibrated reproductions of
the Northern Virginia events.

\subsection{Device-Level Reliability}
\label{subsec:device_reliability}

AI workloads stress GPU hardware on a faster timescale than the enterprise computing workloads for which today's device, package, and PSU specifications were calibrated. All-reduce phases, checkpoint boundaries, and large-batch synchronization produce sudden power and thermal transients with sub-millisecond rise times and power peaks that approach nameplate ratings. These excursions begin at the GPU die and its power delivery network (PDN) and propagate
outward through server PSUs and rack distribution~\cite{infineon_scaling_2025,vicor_genai_2024}. These transients can develop faster than the regulators, packaging, and telemetry can respond to or capture them. 

Electrically, fast $di/dt$ events excite PDN resonances and expose voltage-regulator loop latency, producing overshoot and undershoot on the low-voltage core rails. Recent Open Compute Project specifications for AI rack power delivery limit transient output-voltage deviation to approximately $\pm 3\%$ during prescribed high-slew-rate load-step tests
\cite{OCPDiablo400}. Maintaining this tolerance becomes increasingly difficult as
workload-induced load steps approach the GPU's full thermal design power (TDP). At the same time, high instantaneous currents concentrate $I^2R$ dissipation at resistive discontinuities such as connector contacts, VRM solder joints, and busbar terminations. 

The observability layer compounds these effects rather than catching them.
Interfaces such as \texttt{nvidia-smi} sample far more slowly than the
transients they are meant to capture, under-reporting peaks and hiding the link
between workload phase and device stress~\cite{yang_ptsmi_2023}. This has two consequences with respect to runtime control and qualification. For the former, thermal throttling, power capping, and frequency governors act on filtered statistics and miss the worst-case events they exist to prevent, so stress accumulates below the detection threshold. Qualification and burn-in procedures calibrated against this telemetry may also underestimate field stress, allowing devices to pass acceptance testing despite potentially damaging operating conditions.

 \begin{figure*}[h!]
     \centering
     \includegraphics[width=1\linewidth]{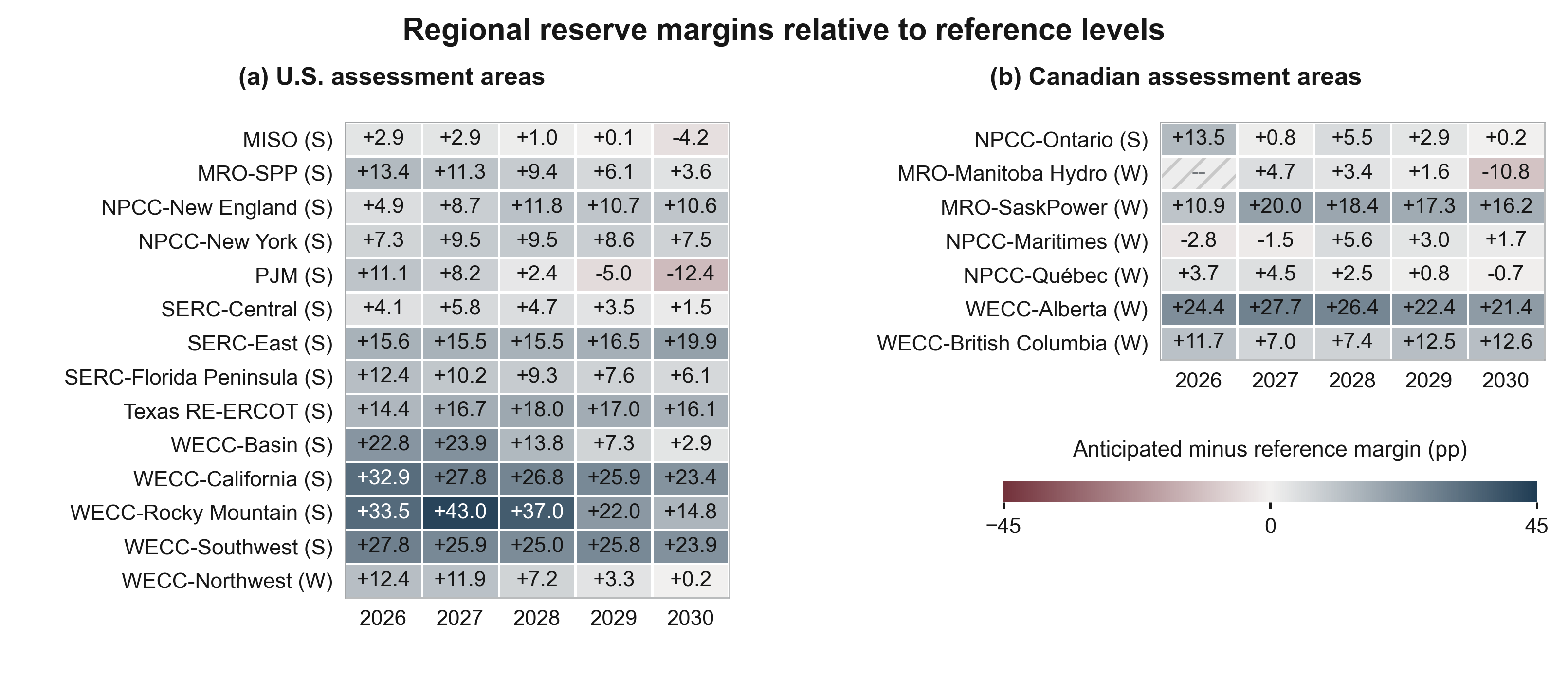}
     \caption{Illustration of regional capacity adequacy under AI load growth. Values show anticipated reserve margin minus the area- and year-specific reference margin, in percentage points, for retained (a) U.S. and (b) Canadian assessment areas. Inputs are from Table~2 of NERC's 2025 Long-Term Reliability Assessment, published January 2026~\cite{NERC2025LTRA2026}. The color scale is symmetric about zero. S and W indicate summer and winter; winter columns use the starting year. SERC-Southeast is excluded and Manitoba winter 2026--27 is unplotted because summary-table and regional-dashboard values conflict. ERCOT follows the summary table's rounded reference of 13.8\%. Geographic labels retain NERC assessment boundaries; Maritimes includes northern Maine. Positive values do not establish connection headroom, dynamic stability, or absence of annual energy shortfalls.}
 \vspace{-1em}
     \label{fig:Map}
 \end{figure*}

\subsection{Planning, Siting, and Interconnection}
\label{subsec:planning}

As AI load growth is significantly faster than grid infrastructure such as transmission upgrades, substations, and large transformers, the central interconnection question is whether very large loads receive firm service, connect under curtailable or non-firm arrangements, or pair with dedicated generation and storage. System operators such as PJM, ERCOT, and AESO have adopted different approaches. While approaches vary based on their market designs, adequacy criteria, and inter-tie capabilities \cite{PJM2025,TexasSB6-2025,AESO2025}. The regional capacity-adequacy indicator in Figure ~\ref{fig:Map}
is the anticipated reserve margin relative to the applicable reference margin:
\begin{equation}
\Delta M_{r,t}=M^{\mathrm{ant}}_{r,t}-M^{\mathrm{ref}}_{r,t}.
\label{eq:reserve_margin_difference}
\end{equation}
Here, $M^{\mathrm{ant}}_{r,t}$ and $M^{\mathrm{ref}}_{r,t}$ are the anticipated
and reference reserve margins, respectively, reported in percent for assessment
area $r$ and the indicated seasonal period $t$. Consequently, $\Delta M_{r,t}$
is expressed in percentage points (pp), rather than relative percentage change.
For example, margins of 25\% and 15\% yield $\Delta M_{r,t}=+10$~pp.
Positive and negative values indicate margins above and below the reference,
respectively; zero indicates equality. Inputs are taken from NERC's
2025 Long-Term Reliability Assessment~\cite[p.~14]{NERC2025LTRA2026}.

Recent FERC, DOE, and NERC actions reflect the absence of a settled
framework for large controllable and co-located loads
\cite{FERC2025ShowCause,DOE2025,NERC2025LargeLoads}. Planning studies must therefore represent more than peak demand. Relevant characteristics include location, service firmness, ramp and ride-through limits, self-supply and reactive capability, and dispatchable flexibility. Study
validity must also be maintained after commissioning when firmware, workload composition, or operating policies change.

\subsection{Social and Community Challenges}

Rapid expansions of AI data centers is creating growing social and community challenges worldwide. Their high demand for electricity, water, land, and supporting infrastructure can intensify competition with local households, agriculture, and other essential services, while nearby communities may also experience noise, traffic, pollution, and changes to local landscapes. In some regions, concerns have emerged over Indigenous and land rights, limited public consultation, unequal distribution of environmental burdens, and the relatively small number of permanent jobs compared with the scale of public resources and incentives provided. These impacts raise broader questions of environmental justice and community consent. Table~\ref{tab:social_community_ai_dc} lists the main challenges and provides examples.

\begin{table*}[h!]
\centering
\caption{Social and community challenges associated with AI-driven data center expansion.}
\label{tab:social_community_ai_dc}

\scriptsize
\setlength{\tabcolsep}{3.5pt}
\renewcommand{\arraystretch}{1.08}

\begin{tabularx}{\textwidth}{
    >{\raggedright\arraybackslash}p{0.20\textwidth}
    >{\raggedright\arraybackslash}p{0.31\textwidth}
    >{\raggedright\arraybackslash}X}
\toprule

\textbf{Community challenge} &
\textbf{How AI data centers produce it} &
\textbf{Global examples} \\

\midrule

Electricity cost shifting &
Large AI loads require new generation, transmission, and grid upgrades whose costs may ultimately affect consumers. &
\textbf{Ireland/EU:} Rising data center demand has been associated with higher system costs and concerns that grid-expansion costs may be transferred to households~\cite{Ireland2026,EUGrid2026}. \\

Ghost demand and grid planning &
Speculative connection requests can reserve scarce grid capacity and trigger infrastructure planning for projects that may never be built. &
\textbf{EU:} Grid operators increasingly report speculative or immature connection applications, while rapid AI/data center growth is creating exceptionally large connection queues~\cite{EUGrid2026}. \\

Water justice &
Cooling high-density computing can increase competition for water between data centers, households, agriculture, and ecosystems. &
\textbf{South Africa, Kenya, India:} Communities have raised concerns over allocating scarce water to large data centers in Johannesburg/Cape regions, Naivasha, and Visakhapatnam~\cite{APSouthAfrica2026,Kenya2026,India2026}. \\

Local pollution from power generation &
Grid constraints can encourage dedicated gas generation or extensive diesel backup systems, concentrating emissions near host communities. &
\textbf{Australia:} Proposed Northern Territory AI facilities are linked to gas generation. \textbf{New Zealand:} Makarewa plans include up to 84 diesel backup generators~\cite{ReutersAustralia2026,NZ2026}. \\

Land use and rural industrialization &
Large campuses, substations, transmission infrastructure, and cooling systems can transform agricultural or residential landscapes. &
\textbf{New Zealand:} Makarewa occupies roughly 49 ha. \textbf{Japan:} Inzai residents have opposed a 52-meter data center building near housing~\cite{NZ2026,Japan2026}. \\

Few permanent jobs relative to footprint &
Data centers create major construction activity but comparatively limited long-term employment despite substantial land, energy, and water requirements. &
\textbf{New Zealand:} Makarewa is projected to create about 1,200 construction jobs but only around 50 permanent positions while requiring approximately 280 MW~\cite{NZ2026}. \\

Noise and light disturbance &
Cooling systems, transformers, generators, and 24/7 operation can introduce persistent industrial noise and lighting. &
\textbf{Uruguay, New Zealand, Japan:} Residents near data center developments have reported or anticipated noise, lighting, and generator disturbances~\cite{Uruguay2026,NZ2026,Japan2026}. \\

Off-site infrastructure burdens &
AI facilities can require additional transmission lines, power generation, pipelines, roads, and water infrastructure beyond the data center site. &
\textbf{Australia:} Northern Territory AI projects are becoming linked with new gas and power infrastructure. Similar grid-expansion pressures are emerging across Europe~\cite{ReutersAustralia2026,EUGrid2026}. \\

Unequal distribution of impacts &
Environmental and resource burdens may fall disproportionately on Indigenous, rural, or socioeconomically disadvantaged communities. &
\textbf{Brazil:} Anac\'e communities have contested impacts on traditional territory. \textbf{South Africa:} Civil-rights groups question resource allocation amid existing inequalities~\cite{Brazil2026,APSouthAfrica2026}. \\

\bottomrule
\end{tabularx}

\end{table*}

\section{Opportunities and Solution Stacks}
\label{sec:solutions}
\begin{figure*}[h!]
    \centering
    \includegraphics[width=0.9\linewidth]{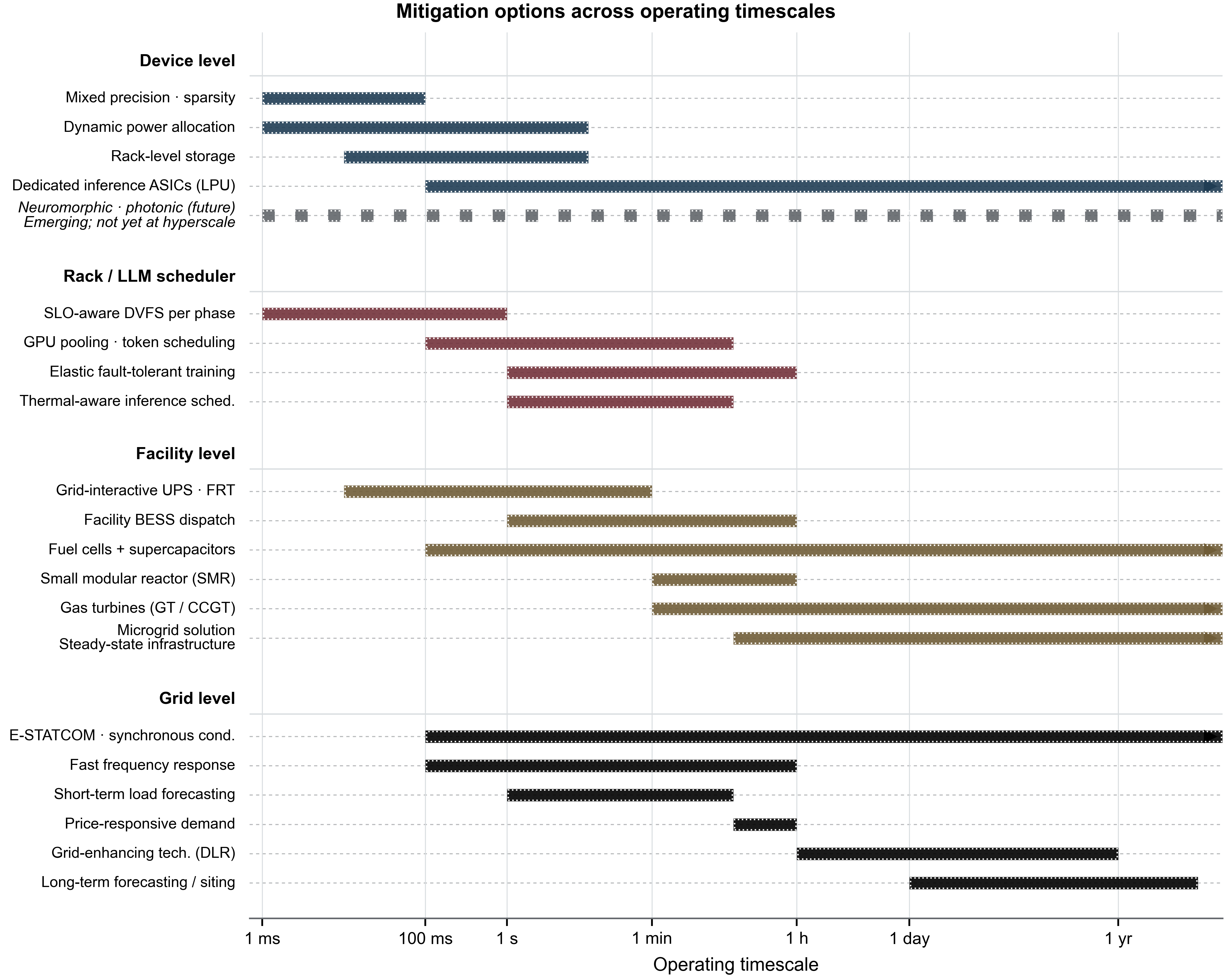}
      \caption{Grid-, facility-, scheduler-, and device-level solutions mapped to
    their operating timescales. Each bar spans the range over which a technique
    acts, from sub-second device and protection response to multi-year siting.}
    \label{fig:solution_dimensions}
\end{figure*}

Beyond reliably supplying adequate power, the solution stack aims to improve controllability and reshape data centers' electrical behavior so that grid-facing demand becomes observable, regulated, and stability-supportive. Figure~\ref{fig:solution_dimensions} organizes the full stack along two axes: the \emph{layer} at which a mitigation acts---device, rack/scheduler, facility, or grid---and the \emph{timescale}
over which it can adjust power. These timescales range from milliseconds for device-level control to multiple years for siting decisions.

No single layer spans the full disturbance spectrum. The fast device- and rack-level transients identified in Section~\ref{sec:challenges} cannot be contained by grid-level
resources alone. Effective mitigation instead matches each disturbance to a resource of comparable bandwidth---the cross-layer coordination this paper advocates. The densest
overlap falls in the seconds-to-minutes band, where device, facility, and
grid resources all operate; there, coordination rather than redundancy
determines whether they reinforce one another or interfere with one another. A microgrid coordinates resources across the facility-to-grid boundary. Its power-converter interface allows the data center to behave as a single coherent system, mediating between internal load dynamics and grid requirements.

\subsection{Grid-Level Solutions}
\label{subsec:grid_sol}
 
Grid-level measures address stability, transfer capability, and planning
constraints outside the facility boundary. Fast-response resources regulate
voltage and frequency; forecasting improves visibility; grid-enhancing
technologies (GETs) relieve thermal congestion; flexible connection models
expose controllability; and spatial planning reduces long-term network stress.
 
\subsubsection{Fast Frequency and Voltage Response Solutions}
\label{sec:grid_fast_response}
 
As Section~\ref{subsec:voltage} establishes, large power-electronic loads exacerbate frequency excursions, complicate voltage control, and intensify system-strength constraints when synchronous inertia and fault-current
contribution are declining~\cite{NERC_Incident_2025,NERC_SOR_2025,
NERC_GFM_2021,ESIG_GFM_Report_2022}. Effective mitigation matches each disturbance to a resource of comparable bandwidth. This is the cross-layer coordination the paper advocates.

\noindent \emph{STATCOMs and E-STATCOMs: dynamic voltage support and active-power injection}: 
Static synchronous compensators (STATCOMs) are a mature element of this
toolkit. As voltage-sourced converters, they provide sub-cycle reactive injection and
absorption, improving voltage regulation~\cite{Siemens_SVCPLUS}.
Energy-supported STATCOMs (E-STATCOMs) extend this capability by adding
short-duration storage, e.g., supercapacitors or high-power batteries, to the
DC link. The same converter can then provide both fast reactive support
and brief active-power injections. 
Energy duration is limited to seconds, so E-STATCOMs arrest the initial transient without contributing to sustained adequacy.
 
\noindent \emph{Synchronous condensers: inertia and short-circuit strength.}
Synchronous condensers add physical inertia, short-circuit current, and continuous reactive capability to the system. These improve voltage stability, protection performance, and fault ride-through
in IBR-dominated regions~\cite{ESIG_GFM_Report_2022}. Modular packages, sometimes paired with flywheels,
are increasingly deployed at EHV substations feeding large load clusters,
including AI data centers, where they give downstream power-electronic
equipment a stronger electrical reference.
Synchronous condensers stiffen the grid; the data center load profile is unchanged.
 
\noindent \emph{Transmission-connected BESS: fast active-power balancing.}
Transmission-connected battery energy storage systems (BESS) deliver
sub-second frequency response, ramping support, and peak shaving, with
field experience confirming their suitability for fast frequency response
and inertia-like services. Sited at substations supplying AI clusters, a BESS can
reduce the net frequency impact of fast-changing demand and reshape the
apparent POI profile. Reserve headroom is critical: a BESS optimized for arbitrage may be unavailable during reliability events, so AI-serving dispatch rules
must reserve fast-response capacity for contingencies and post-fault
recovery.

\subsubsection{Short-Term Load Forecasting}
\label{sec:grid_stlf}

The nonstationary, behind-the-meter dynamics of
Section~\ref{subsec:voltage} leave the grid seeing only the
aggregate POI injection; internal scheduling,
auto scaling, and thermal control remain opaque to operators.

Forecasting at a resolution of seconds is an observability
mechanism, not an energy-management tool. 
It supplies operators with net-power trajectories, ramp-rate distributions, and uncertainty bounds for reserve activation, AGC, congestion management, and FFR procurement~\cite{mughees2025short}. These forecasts should capture event-triggered changes, cooling-related delays, and post-disturbance load restoration. It becomes useful when linked to pre-agreed actions such as arming the BESS, applying power caps, deferring flexible training, or changing UPS operating modes.

\subsubsection{Grid-Enhancing Technologies}
\label{sec:grid_gets}

GETs offer a near-term pathway for
accommodating AI-driven demand growth on existing transmission
corridors. Because large AI facilities can materialize with
hundreds of megawatts of connected capacity within a few years,
they outpace the 7--10-year timelines associated with conventional
transmission planning, permitting, and construction~\cite{sun2022dynamic}.
Dynamic line rating, advanced power-flow control, and topology
optimization can increase utilization of existing assets, alleviate
congestion, and defer the need for new
rights-of-way~\cite{ESIG2025GETs}.

GETs expand transfer capability without changing the dynamic
behavior of the connected load: dynamic line rating raises thermal
limits and topology optimization redistributes flows, but neither
mitigates sub-synchronous oscillation, voltage instability, or
protection-induced load steps. GETs help most where thermal congestion is the binding constraint. They do little under weak-grid, low-inertia, or low-observability conditions.

\subsubsection{Emerging Connection Types for AI Data Centers}
\label{subsubsec:connections}

\noindent \emph{Shiftable and non-capacity-backed connections}: These are emerging as a new class of interconnection agreement for very large AI loads~\cite{google_flexible_2025}. Under this model, a data center connects without securing full firm capacity rights, provided that the system operator receives enforceable authority to curtail load during system stress, enabling faster connection while protecting reliability for existing customers.

The reliability value of this model depends on the data center
being technically controllable. Curtailment must be automated,
verifiable, and fast enough to be useful during scarcity,
congestion, or emergency conditions. This requires telemetry,
defined curtailment blocks, coordination with on-site generation
and storage, and the ability to shed or defer non-critical compute
without compromising essential services. Once configured, the AI data center behaves as a large controllable demand resource that the system can reduce when it needs relief. As an operational approach relevant to such connections, a hierarchical receding-horizon controller is proposed~\cite{sreekumar2026multi}, which coordinates fast worker throttling with slower server shutdown and restart decisions to accommodate time-varying power caps while minimizing service impact.

\noindent \emph{Price-responsive connections}: These expose AI data centers to granular system price signals rather than fixed, fully hedged tariffs~\cite{jiang2021optimal}. Operators adjust load in response to wholesale prices, scarcity prices, or explicit demand-response signals. The central advantage is alignment with system conditions: AI
workloads can absorb surplus renewable generation during low-price periods and
back off during scarcity. Price response is too slow for sub-second reliability events. It operates on medium timescales and cannot substitute for FFR, ride-through, or automated emergency control. A further concern is rebound behavior. If many data centers respond to low prices by simultaneously increasing training or
inference workloads, they may create new synchronized ramps;
similarly, deferred workloads may return together after a high-price
interval ends.

\subsubsection{Long-Term Load Forecasting and Spatial Siting}
\label{sec:grid_ltlf}

Long-term forecasting for AI data centers characterizes multi-year
demand trajectories by accounting for AI adoption, compute
intensity, hardware and cooling efficiency, siting patterns, and
the balance between grid-supplied and self-supplied energy. As hyperscale operators procure capacity and power-purchase agreements across multiple regions, spatially resolved forecasts become necessary to align contractual portfolios with physically deliverable capacity, informing generation expansion, transmission reinforcement, resource-adequacy planning, and the design of connection arrangements that distinguish firm from flexible demand~\cite{WRI2025_ForecastingTricky}.

Forecasting practice is beginning to formalize this approach. Under PJM's large load
adjustment process, projects of 50~MW or greater are submitted annually in both
capacity and demand terms and de-rated according to contractual firmness: loads
energizing within three years require an electric service obligation or
construction commitment, loads three to eight years out are treated as non-firm
and discounted against milestone evidence, and, in the absence of supporting data, PJM applies
default values for capacity-to-demand conversion and for the multi-year
energization ramp~\cite{pjm_lar_impl}. Applying these rules lowered PJM's
near-term peak forecast through 2032 relative to the previous cycle
\cite{pjm_lf2026}.

  Project-based forecasting translates requested connection capacity into
  expected demand by accounting for project maturity, operating utilization,
  and energization schedules. The California Energy Commission applies
  utilization factors, project-confidence levels, and ramping schedules to
  utility-reported data center requests. It then uses hourly load profiles
  to estimate their contribution to system peak
  demand~\cite{CEC2026DataCenterMethodology}.
  Similarly, the Northern Virginia Electric Cooperative (NOVEC) develops
  building-level demand forecasts using utilization estimates and
  energization timelines, aggregates these forecasts geographically,
  and excludes or derates projects with substantial development
  risks~\cite{NOVEC2025DataCenterForecast}.
  These practices distinguish announced capacity from the demand likely
  to materialize within the planning horizon.

  Probabilistic approaches can additionally represent uncertainty in
  project completion, demand magnitude, and energization timing.
  Georgia Power's load-realization model, as described in regulatory
  testimony, uses Monte Carlo sampling to represent project realization,
  served demand relative to announced capacity, and delays in commercial
  operation~\cite{Roumpani2025GPCIRPTestimony}.
  For planning applications, such scenarios should also consider shared
  constraints, including transmission upgrades and equipment availability,
  that may delay several projects simultaneously. This would allow the
  forecast to represent correlated development risks alongside
  project-specific uncertainty.

  The planning requirement extends beyond aggregate demand to its
  location, firmness, and flexibility. The system implications of an
  AI campus depend on local transmission capacity, network strength,
  and available supply. Operational flexibility should therefore be
  characterized alongside the demand forecast, including the magnitude,
  duration, and availability of load modulation. Flexible computing
  workloads, grid-interactive UPS systems, and on-site storage can
  support grid operation when their response is coordinated with
  system needs~\cite{Colangelo2026GridInteractiveAI}. Long-term forecasting should therefore separate firm base load demand, flexible compute demand, interruptible demand, and behind-the-meter generation dependency. Such forecasts also shape spot-market evolution and renewable
integration, identifying where AI loads can anchor new wind, solar,
storage, or clean firm generation~\cite{PJM2025_LTLoad}.

\subsection{Facility-Level Solutions}
\label{subsec:facility}
Facility-level assets are built and dispatched within the data center
operator's control envelope. Coordinated operation at the POI matters more than the number of assets installed. Each resource should be matched to the disturbance timescale it can serve.

\subsubsection{Microgrid-Enabled Behind-the-Meter Operation}
\label{subsubsec:microgrid}

A growing number of AI data center operators are deploying behind-the-meter (BTM) microgrids as a response to the seven-to-ten-year timelines associated with utility interconnection, permitting, and transmission expansion~\cite{su2025grid}. A microgrid turns the data center from a passive single-point load into an actively managed energy system. A typical configuration integrates one or more of the following: fast-start gas turbines or reciprocating engines, facility-scale battery energy storage systems (BESS), and, increasingly, fuel cells and renewable generation. These resources are coordinated by a supervisory controller behind a single point of interconnection (POI)~\cite{choukse_power_stabilization}. 
A particularly advantageous configuration is the hybrid AC/DC topology, in which an AC subgrid fed by the utility and on-site generation coexists with one or more DC microgrids connected through bidirectional interfacing converters~\cite{li2023smart}.
Because modern AI racks are inherently DC loads at 48--800 V, and because battery storage, photovoltaics, and fuel cells all operate locally at DC, a shared DC link can reduce conversion stages and provide a common point for energy buffering, while bidirectional converters regulate the AC interface~\cite{LiHe2014_HarmonicComp}. 
Timescale separation between control loops is essential. Recent real-time simulation work shows that forecast-aided hybrid microgrids, in which one-second-ahead AI load
predictions pre-position storage setpoints, substantially reduce frequency deviations under abrupt transients compared with reactive-only control~\cite{LinMughees2025_ForecastMG, li2025ai}.

 Data center microgrids may combine synchronous generation with converter-interfaced resources using grid-following or grid-forming controls. Grid-forming capability can be provided by appropriately controlled resources within the microgrid; its location and contribution depend on the electrical architecture and operating strategy~\cite{NERC_GFM_2021,ESIG_GFM_Report_2022}.
 
The grid-supportive
character of the microgrid  depends on the explicit
specification of the interface requirements: ramp-rate limits on
active-power change during islanding and reconnection. A fault ride through (FRT) 
envelope under which the microgrid remains synchronized. Islanding without these constraints produces an impulsive POI step comparable to the July 2024 Northern Virginia event. Grid-support performance depends on coordinated generation, storage, converter controls, and protection across the microgrid~\cite{ESIG_GFM_Report_2022}.

The BTM microgrid model also offers regulatory and commercial
flexibility. Operators can negotiate non-capacity-backed or
curtailable connection agreements
(Section~\ref{subsubsec:connections}) because the on-site portfolio
provides an alternative supply path under curtailment, and the
microgrid offers a natural integration point for future small modular reactors (SMRs), hydrogen fuel cells, or on-site renewable generation without grid-side
upgrades~\cite{IAEA2024_SMR,FCHEADataCenters}. The regulatory
category under which co-located load--generation operates is the
subject of active policymaking: FERC's 2025 show-cause
order~\cite{FERC2025ShowCause} and the Department of Energy's
Large Loads ANOPR~\cite{NERC2025FERCComments} indicate that interconnection,
telemetry, and protection obligations remain under review.

\subsubsection{Grid-Interactive UPS}
\label{sec:grid_interactive_ups}

Performance requirements for large converter-based loads are
shifting from local protection to grid-stability contribution.
Within this framework, two concepts are central: fault ride-through (FRT) and post-fault active power recovery (PFAPR)~\cite{FingridKJV2026,ENTSOEDCC}. FRT requires converter-based
demand to remain connected and behave in a controlled manner during
voltage depressions or short-duration faults; PFAPR specifies a
ramp-bounded restoration of active power following fault clearance,
preventing large, rapid demand changes that destabilize frequency and voltage.

\begin{figure}[h!]
    \centering
    \includegraphics[width=0.89\linewidth]{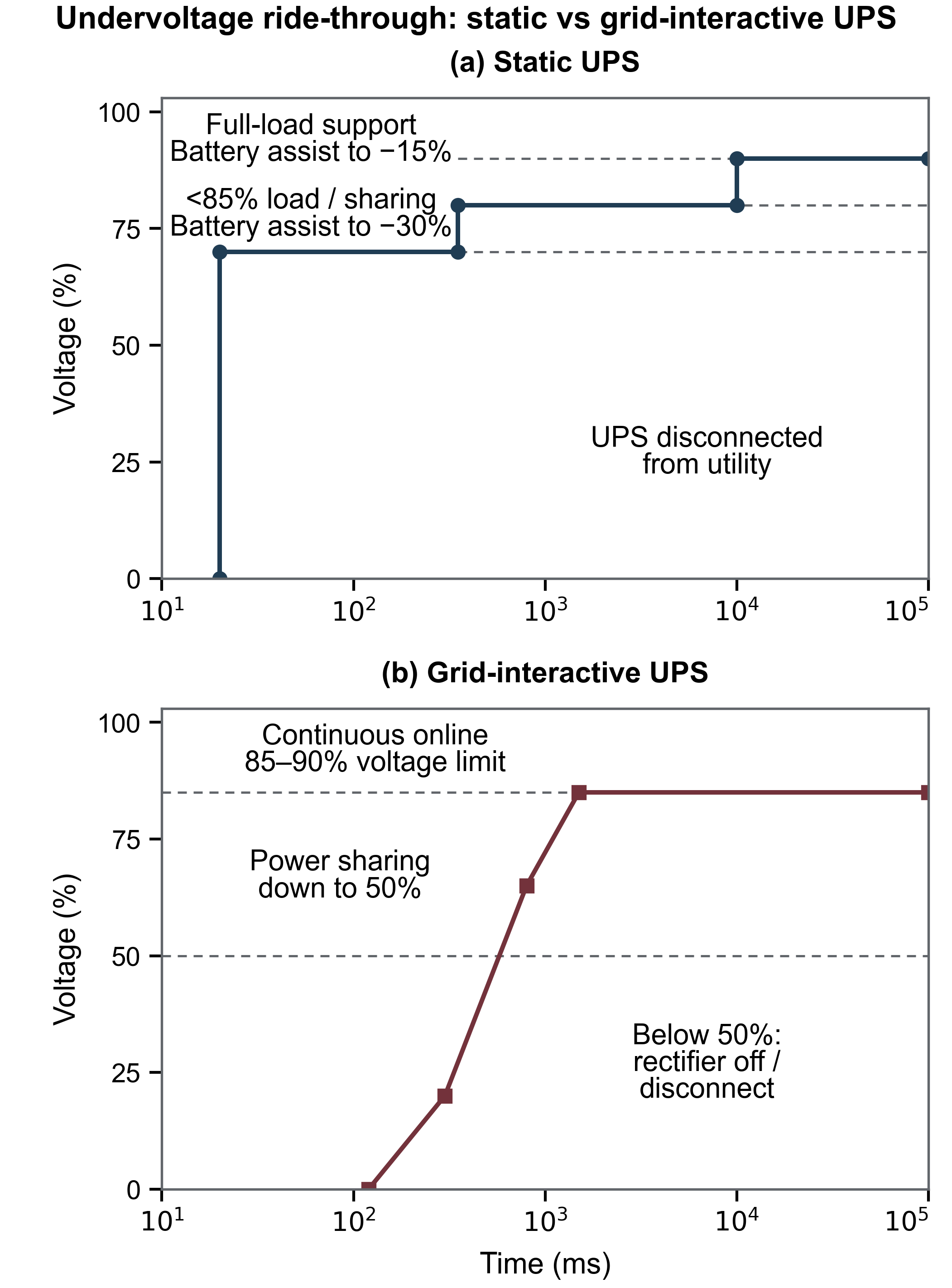}
    \caption{Static and grid-interactive UPS FRT behavior. (a) A static UPS disconnects at moderate sag depths, producing an abrupt POI
    step. (b) A grid-interactive UPS rides through under a defined FRT
    envelope and restores active power along a controlled PFAPR
    ramp~\cite{takci2025,ERCOTNOGRR282,Eaton2025}.}
    \label{fig:frt_static_gi}
\end{figure}

Grid-interactive uninterruptible power supplies (UPS) provide a practical mechanism for satisfying these obligations. During a fault, the UPS limits grid current draw, maintaining continuity of supply to the IT load and preserving the grid connection within the FRT envelope; after clearance, the rectifier ramps grid draw along the prescribed PFAPR profile. Grid-interactive architectures let large converter-based facilities behave as stable, predictable demand resources~\cite{takci2025,ERCOTNOGRR282,Eaton2025}.
Figure~\ref{fig:frt_static_gi} contrasts this behavior with a static UPS that
transfers the load when voltage crosses a fixed threshold. The grid-interactive UPS (bottom) instead remains synchronized across a defined FRT envelope, maintaining continuous online operation down to approximately 50\% sag depth and sharing power from
internal storage rather than tripping; after fault clearance, it restores active
power along a controlled PFAPR ramp rather than reconnecting as a step.

The hardware capability that satisfies these obligations also
positions grid-interactive UPS systems as candidates for emerging
sub-second flexibility markets. The UPS battery is dedicated energy
storage with millisecond response, and its participation in FFR or
dynamic reactive support can be reconciled with ride-through
reservation through an explicit state-of-charge
policy~\cite{paananen2023grid,AESO_FFR_Pilot_2023}. Institutional and market-design constraints currently limit this
capability: most jurisdictions do not yet offer a market product priced
at the sub-second timescale at which grid-interactive UPS naturally
operates.

\subsubsection{Facility-Level BESS Dispatch}
\label{subsubsec:facility_bess}

A central medium-voltage BESS and the distributed UPS fleet serve different
roles. The central BESS supports fast ramp control, peak shaving, generator
start-up, and microgrid energy management; the UPS protects critical IT loads
and supplies ride-through over the initial milliseconds to minutes~\cite{sakalkar2020data}. Coordinated control should assign the fastest disturbances to the UPS or other high-power storage on  milliseconds level, then transfer the sustained energy requirement to
the central BESS and the on-site generation portfolio described in
Section~\ref{sec:onsite_generation}.

BESS can provide multiple services, including AI-load ramp smoothing, backup power, peak shaving, and frequency and voltage support.

The selection of battery chemistry embeds related trade-offs. For instance, lithium iron phosphate (LFP) dominates new deployments because of its safety and cycle life, but at the cost of energy density; high-energy nickel chemistries offer greater energy density, while thermal-runaway risk has limited hyperscale adoption; lithium titanate (LTO) offers long cycle life at prohibitive cost per kWh; flow batteries suit long-duration profiles unlike those of AI facilities; and solid-state chemistries remain pre-deployment.

\begin{table}[h!]
\centering
\caption{On-site generation for AI microgrids, by ramp bandwidth,
deployment scale, and build-cycle position.}
\label{tab:onsite_gen}
\renewcommand{\arraystretch}{1.3}
\setlength{\tabcolsep}{4pt}
\scriptsize
\begin{tabular}{@{}p{1.4cm}p{1.5cm}p{1.3cm}p{1.9cm}p{0.7cm}@{}}
\toprule
\textbf{Technology} & \textbf{Ramp rate / response} & \textbf{Build time / deployment} &
\textbf{Role / status} & \textbf{Ref.} \\
\midrule
Simple-cycle gas / recip. &
20--60 MW/min &
1--2 yr &
Fast load-following anchor; mature at $\sim$GW scale &
\cite{ge_dc_microgrid} \\
Combined-cycle gas &
8--20 MW/min &
1--2 yr &
Efficient baseload in a hybrid microgrid; needs a fast-ramp layer &
\cite{siemens_onsite} \\
SMR (LWR) &
0.5--1 \%/min &
2028--2032 &
Low-carbon baseload; \$90--170/MWh; needs complementary fast assets &
\cite{IAEA2024_SMR} \\
SMR (adv. HT) &
$\sim$5 \%/min &
2028--2032 &
Faster-following baseload; earlier-stage licensing &
\cite{Xenergy_ERA_2025} \\
PEM fuel cell &
Seconds to rated output &
Near-term &
Low-emission diesel-backup replacement; tens-of-MW scale &
\cite{FCHEADataCenters} \\
SOFC (+ supercapacitor) &
40--100\% step &
Near-term &
Variable prime power; CCGT-class CO$_2$ on gas, clean on H$_2$ &
\cite{BloomLoadFollowing2024} \\
\bottomrule
\end{tabular}
\end{table}

\subsubsection{On-Site Generation: Gas, Nuclear, and Fuel Cells}
\label{sec:onsite_generation}

On-site generation for AI microgrids is a bandwidth-matching problem more than a fuel choice. Each technology occupies a distinct ramp-rate band relative to the workload variability it must follow, and a distinct position in the capacity build-out timeline. Table~\ref{tab:onsite_gen} summarizes the relevant parameters; the discussion below develops the role each
technology plays in the stack.
The reliability ranking of these technologies does not coincide with their
sustainability ranking. Gas supplies rapid firm capacity at the ramp rates AI
demand requires but retains operational emissions for the life of the asset, along
with water and air-permitting constraints and exposure to fuel infrastructure;
SMRs and low-carbon hydrogen offer emissions reductions but face licensing, cost,
water, and fuel-infrastructure constraints, and neither follows the workload
directly.

Table~\ref{tab:onsite_gen} reveals a bandwidth–sustainability trade-off. Fast-start gas and reciprocating units provide the most mature near-term load-following capability, whereas SMRs and fuel cells are better suited to firm or baseload roles and require storage or flexible computation to absorb faster variations~\cite{IAEA2024_SMR,schlegel2024small,Xenergy_ERA_2025, FCHEADataCenters,BloomLoadFollowing2024}. Gigawatt-scale microgrids will therefore require portfolios of complementary resources rather than a single generation technology.

\subsection{Workload-Scheduling and LLM-Serving Solutions}
\label{subsec:workload}

The scheduler acts at the source of demand variation. Micro-batch boundaries,
collective-communication barriers, optimizer steps, checkpointing, batching,
auto scaling, and admission decisions determine when accelerators enter
high- and low-power phases and how strongly those phases align across devices.
Device-, rack-, and facility-level controls can attenuate the resulting
variation, but they cannot alter its source timing and can mask its grid-facing
effect only within finite power, energy, and bandwidth limits. This distinction
links scheduler design to the converter-driven and inter-area mechanisms in
Sections~\ref{subsec:ssr} and~\ref{subsec:interarea}.

The literature is most clearly organized along two axes. The first is control
timescale: cluster-level decisions act over minutes or longer, instance-level
decisions act over seconds, and iteration-level decisions can act within a
second. The second is objective: an \emph{energy-reduction} mechanism lowers
average joules per token, whereas a \emph{load-shaping} mechanism changes the
ramp, peak-to-mean ratio, or temporal coherence of demand. A mechanism may
contribute to both objectives, but the two effects should not be treated as
equivalent~\cite{stojkovic2025tapas,liu2025greenllm,basit2026dualscale,hankendi2026pals}.

\begin{figure*}[t]
  \centering
  
  \includegraphics[width=0.88\linewidth]{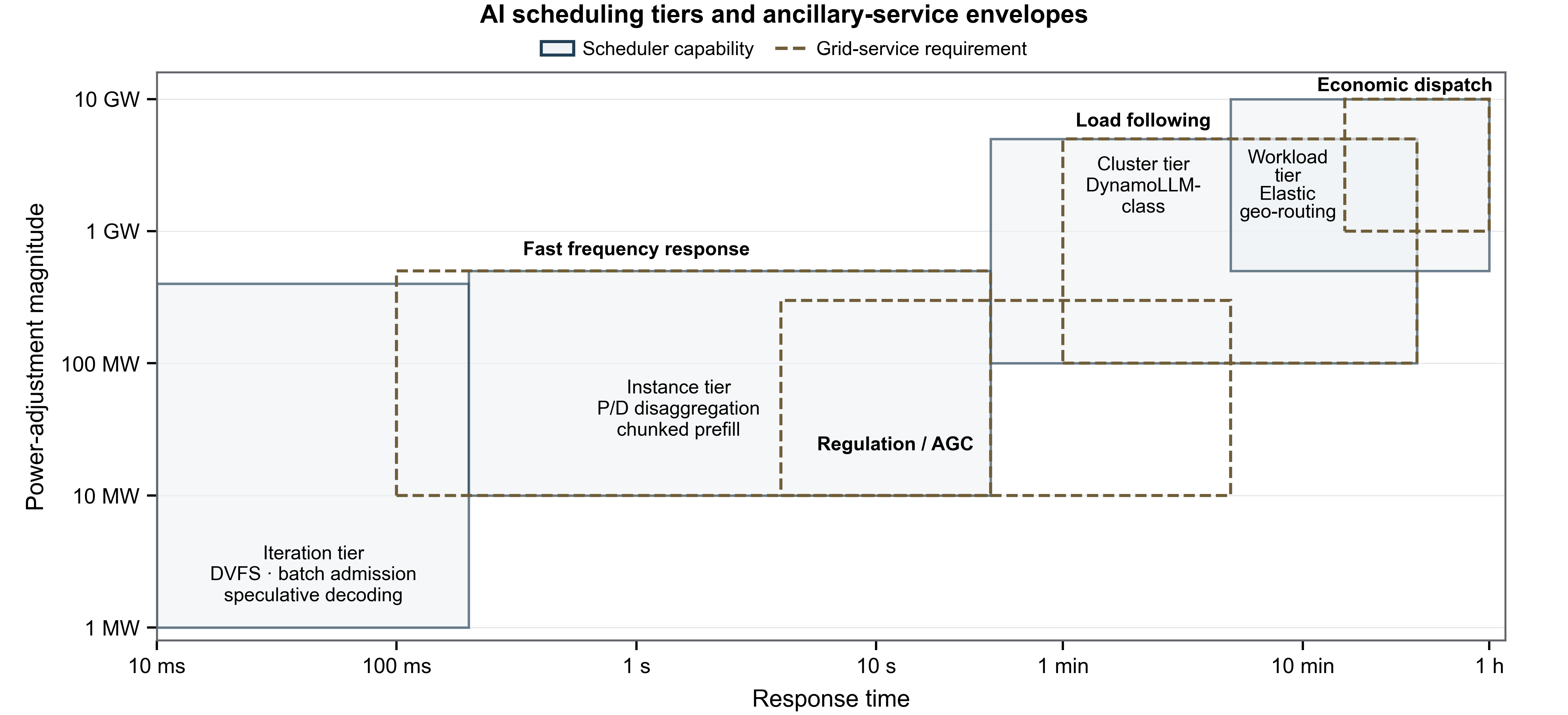}
  \caption{Conceptual mapping of scheduler controls by response time and
  adjustable power. Iteration-level controls provide the fastest adjustment, while instance-level controls shape demand over seconds.}
  \label{fig:scheduler_atlas}
\end{figure*}

\subsubsection{Training: Reducing Coherent Forcing}
\label{subsubsec:training_scheduler}

Synchronous data-parallel training can align compute, collective
communication, optimizer, and checkpoint phases across large accelerator
fleets. The resulting electrical excursions are correlated and therefore need
not average out at the point of interconnection (POI). Communication-period
harmonics may enter the 10--40~Hz converter-driven band, while slower optimizer
and checkpoint cycles may overlap the 0.1--1~Hz inter-area band
\cite{kozhu2025,mishra2025understanding}. These mechanisms should be evaluated for their effects on ramp rate, peak demand, and load coherence, alongside throughput and energy efficiency.

Existing training systems provide useful control levers. Elastic
resizing changes the number of devices contributing coherently; fault-tolerant
recovery can shorten or stage restart-related transients; and topology-aware
placement redistributes variation across racks, feeds, or sites
\cite{wu2024lazarus,fftrainer2025,narayanan2020heterogeneity,
jiang2024megascale}. These mechanisms are currently optimized mainly for
throughput and availability. A grid-aware extension could penalize checkpoint
or collective-communication schedules that repeatedly align with weakly damped
system modes, subject to explicit performance and reliability constraints.

\subsubsection{Inference: Timescale-Aware Control Hierarchy}
\label{subsubsec:inference_scheduler}

Inference demand is more stochastic than training demand~\cite{li2024unseen}. Bursty arrivals, phase-asymmetric computation, batching, and coordinated auto scaling can still create substantial seconds-to-minutes variability. The available controls should therefore be separated by response time.

\emph{Cluster tier (minutes to hours): provisioning and routing}.  Cluster-level systems resize resource pools, adjust parallelism and baseline frequencies, and route requests according to energy, thermal, carbon, or price
signals~\cite{stojkovic2025tapas,jaiswal2025sageserve,li2024sprout,sheshanarayana2026gar, moore2026marlin}. This tier provides the operational mechanism for the flexible connection arrangements in
Section~\ref{subsubsec:connections}, where workload deferral or geographic routing can reduce power imports during high-price, congested, or curtailed intervals. Its response is appropriate for load following and economic dispatch, and may support slower regulation if validated, but not sub-second ride-through.

\emph{Instance tier (seconds): phase separation, batching, and caching}. Prefill/decode disaggregation places the two phases on separate accelerator pools and enables phase-specific control~\cite{patel2025n,zhong2024distserve,basit2026dualscale,
hankendi2026pals}. Chunked prefill interleaves long prefill stages with ongoing decode, converting a bustier compute phase into a more uniform mixed-phase load~\cite{agrawal2024taming}. For long-context workloads, tiering the key–value (KV) cache across high-bandwidth memory (HBM), CPU memory, and storage can reduce demand on HBM~\cite{zheng2026adaptive}. These mechanisms could complement one another: disaggregation exposes phase-specific control, while chunking and cache management reduce or redistribute the resulting variation. 

\emph{Iteration tier (sub-second): DVFS and decoding strategy}.
Stage-aware dynamic voltage and frequency scaling (DVFS) adjusts accelerator
frequency to queue state and workload phase, reducing energy without imposing
a uniform performance penalty~\cite{liu2025greenllm,basit2026dualscale}. Speculative decoding
is more regime-dependent: it can reduce energy when memory-bound decoding
dominates, but verification overhead can increase energy at larger batch sizes
\cite{leviathan2023fast,dutta2026benchmarking,qin2024optimized}. The
higher-level controller should therefore select iteration-level policies using
batch size, latency target, and available power headroom rather than treating
one method as universally optimal.

The following two workload structures cut across all three tiers.

\emph{Mixture-of-Experts and Routing-Induced Coherence}. Mixture-of-experts (MoE) models reduce average executed computation through
sparse activation, but routing may also create correlated hardware-activation
patterns when similar requests arrive together
\cite{liu2026survey,hankendi2026pals}. Whether this behavior produces
narrow-band power content relevant to the oscillatory mechanisms in
Section~\ref{subsec:ssr} has not been established. Measurements at the
accelerator, rack, and POI are needed before MoE routing is classified as
either a reliability risk or a load-shaping opportunity. Candidate controls,
if problematic coherence is observed, include admission staggering,
routing-aware batching, and diversified expert placement.

\emph{Shaping the Agentic Power Signature}. A growing share of deployed workloads involves agentic execution rather than monolithic inference. Capacity itself does not resemble an execution pattern. In \emph{agentic} execution, the model sits inside a control loop,
so a request expands into a data-dependent graph of inferences and tool
calls, and execution alternates between accelerator- and host-bound phases~\cite{yang2026architectural}. This burst structure is a property of the harness, not the model. Fan-out width, tool-call batching, stage boundaries, and retry/backoff policy set the interval between load transitions and hence the frequency band in which power fluctuations are concentrated. These parameters are software-settable and already
visible to the runtime~\cite{yang2026architectural, pham2026agentstop}, so software, not the workload itself, sets the spectral content of agentic load.

Each tier acts on a different attribute of the POI power profile; Figure~\ref{fig:scheduler_atlas} summarizes their overlapping timescales. The
scheduler acts on the timing that generates these features, whereas
storage, converters, and grid resources act on the spectrum once
produced. Because scheduler operation is constrained by service-level obligations and depends on access to grid-event information, scheduling complements rather than replaces storage, converters, and grid resources.

\subsection{Rack-Level Power Architecture}
\label{subsec:rack}

Here, the rack layer encompasses the power-conversion chain
  between the rack bus and the upstream medium-voltage interface.
  Its electrical behavior depends on the bus voltage, the design and location of the medium-voltage interface, and the power-semiconductor devices used. Beyond
  efficiency and power density, these choices influence input
  impedance, response to disturbances, and power quality.
  Their effects at the point of interconnection (POI) depend
  on interactions with facility transformers, UPS systems,
  storage, and supervisory controls.

\subsubsection{High-Power PSUs and 48--54~V Rack Buses}
\label{subsubsec:psu}

Industry reports describe high-density AI racks approaching or
  exceeding 100~kW~\cite{Subzero2025}. Open Rack V3 specifies a
  48~V rack bus~\cite{OCP_ORv3_2022}. At the same delivered power,
  increasing the distribution voltage from 12~V to 48~V reduces
  current to one-quarter and, for unchanged conductor resistance,
  reduces the corresponding $I^2R$ conduction loss to one-sixteenth.
  These ratios describe distribution conduction losses, not overall
  PSU conversion efficiency.

The reduction in current relative to a 12~V architecture does not eliminate the high-current challenge. A 100~kW rack supplied at 48~V draws approximately 2.1~kA before accounting for conversion losses. Consequently, reliable operation requires accurate current sharing among parallel PSUs, low-resistance busbars and connectors, selective fault isolation, and coordinated protection.
Although larger shared PSUs may improve conversion efficiency, concentrating capacity into fewer units can increase the amount of equipment affected by a single failure. Modular redundancy and electrical segmentation are therefore needed to prevent single-point fault from disabling the entire rack.

\begin{figure*}[h!]
    \centering
    \includegraphics[width=0.95\linewidth]{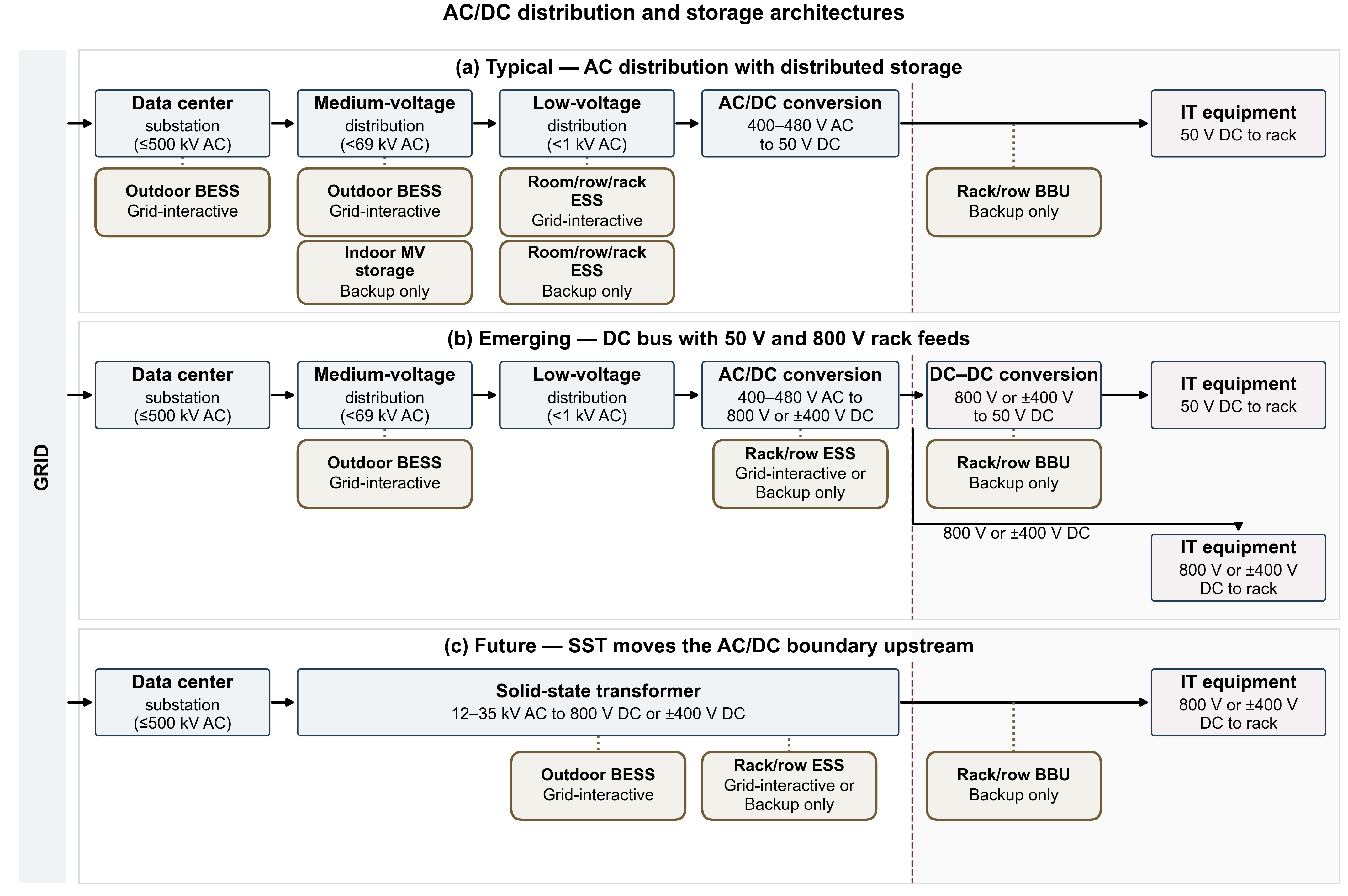}
   \caption{Comparison of data center power-distribution
  architectures: (a) AC distribution with downstream conversion
  to 50~V DC; (b) low-voltage AC/DC conversion supplying an 800~V or $\pm400$~V DC bus, with optional conversion to
  50~V; and (c) SST-based medium-voltage-to-DC conversion.
  Storage locations and operating roles are illustrative.}
    \label{fig:power_chain}
\end{figure*}

\subsubsection{Solid-State Transformers for Medium-Voltage-to-DC Conversion}
  \label{subsubsec:hvdc}

Supplying high-voltage DC racks requires an interface between
  the facility's AC supply and the DC distribution bus. For
  800~V DC distribution, this interface can use a conventional
  line-frequency transformer and rectifier or a solid-state
  transformer (SST). NVIDIA's 800~V DC initiative considers both approaches~\cite{nvidia_800v_hvdc}, while PNNL discusses
  SST-based conversion as a prospective architecture for supplying
  DC racks~\cite{PNNL38817}. Figure~\ref{fig:power_chain} compares
  AC distribution with downstream 50~V conversion, conventional
  AC/DC conversion supplying a high-voltage DC bus, and SST-based
  medium-voltage-to-DC conversion. It also shows alternative
  storage locations for backup and grid-interactive operation.
   An SST combines power-electronic conversion with high-frequency magnetic isolation to provide voltage transformation and a controlled output. Its
   topology and number of conversion stages depend on the application. For example, Xu et al. investigate a 10 kV AC-to-800 VDC architecture
   comprising an H-bridge AC/DC stage and an isolated dual-active-bridge DC/DC stage~\cite{Zhang_SST_2026}. Their implementation uses coordinated rectifier voltage/current
   control and DC/DC phase-shift control to regulate the DC bus. Closed-loop response depends on the converter dynamics, controller tuning, filters, and
   operating limits; switching frequency constrains the achievable response.
  
   The AC-interface control mode should be specified separately from the conversion topology. Grid-following controls regulate current relative to the
   existing grid voltage, whereas grid-forming controls establish an internal voltage reference and support voltage and frequency dynamics. The services
   available from either implementation depend on its control design and current, power, and energy limits ~\cite{NERC_GFM_2021,ESIG_GFM_Report_2022}. Regulation of the downstream DC bus
   is a separate control objective.
  
   DC links provide connection points for batteries or supercapacitors through suitable interfaces. When integrated with appropriately rated storage and
   coordinated controls, an SST-based supply can buffer load fluctuations and support the DC bus during upstream disturbances. The support duration is
   bounded by usable stored energy and the power deficit, while converter ratings constrain the instantaneous contribution. For power systems assessment,
   the relevant configuration is therefore the complete converter–storage system, including its control mode, protection, and energy capability. These
   attributes should be specified when evaluating upstream power fluctuations and disturbance response.

\subsubsection{Wide-Bandgap Devices}
\label{subsubsec:wbg}

The compact, high-bandwidth converters that make the architectures 
above feasible at AI rack power densities are themselves enabled by 
the transition from silicon to wide-band gap (SiC and GaN) power 
devices. At kilovolt-class blocking voltages, silicon-based MV front 
ends (diode and thyristor rectifiers, IGBT-based converters) are 
limited by their device figures of merit, with elevated conduction and 
switching losses, restricted switching frequency, and constrained 
thermal headroom. SiC MOSFETs at the MV interface combine lower 
specific on-resistance, negligible reverse-recovery behavior, and 
reliable operation at elevated junction 
temperatures~\cite{NavitasWhitepaper2025, She2013, Lesnicar2003}; GaN 
devices in the 100--650~V range enable switching at hundreds of kHz 
to MHz, supporting compact multi-kilowatt DC--DC stages at $\sim$98\% 
efficiency at 10~kW~\cite{Infineon800V, MistriGaNReview, MouserGaN}. Together, these device advances enable higher switching frequencies, reduce passive-component volume, improve power density, and lower total conversion losses across the MV-to-load chain.

SiC and GaN devices enable higher switching frequencies and
potentially wider control bandwidths, but the resulting small-signal input
impedance is governed by the full converter design rather than the
semiconductor alone: current and DC-link controllers, synchronization
method, feedforward paths, filter, and operating point. Under weak-grid
conditions, non-passive converter behavior can interact with network
impedance or lightly damped sub-synchronous modes. The relevant frequency
range must therefore be established by impedance measurement or validated
small-signal and electromagnetic-transient models
\cite{Chakraborty2024Impedance,sun2022dynamic,PNNL38817}.

\subsection{Device-Level Solutions}
\label{subsec:device_sol}

The device layer is the point in the cross-layer stack at which the
spectral content of the AI campus load is first determined. Every
mitigation discussed in this subsection operates simultaneously as
an energy-efficiency mechanism and as a load-shaping operator: it
reduces average power, and it alters the ramp rate, transient
amplitude, and spectral distribution of the power drawn by the
compute substrate.

\subsubsection{Profiling and Tuning Numerical Precision}
\label{sec:device_precision}

Numerical precision and sparsity are the primary device-level tuning
knobs for LLM training and inference.

Reduced-precision formats can reduce memory traffic and increase arithmetic throughput when supported by the target hardware. For example, NVIDIA H100 Tensor Cores support FP8 arithmetic~\cite{nvidia_h100_arch}, while NVIDIA A100 Tensor Cores support INT8 and INT4 arithmetic. In AI model serving domain, control knobs such as precision, batch size, concurrency, model sparsity, and parallelism should be selected jointly to improve throughput while satisfying service level objectives (SLOs). Indeed, precision and sparsity are load-shaping levers as well as average-power levers. Decisions about bit width and sparsity patterns set the magnitude of the workload-driven forcing function that Sections~\ref{subsec:ssr} and \ref{subsec:interarea} identify as the root coupling between AI workloads and grid oscillation modes. It is noteworthy few current design practice exploits this coupling. The machine learning literature selects precision and sparsity against a joules-per-token objective; the power systems literature treats the workload spectrum as exogenous, and absorbs it in thedownstream through energy storage and utilization-aware scheduling.

Data movement is another device-level lever. Modern accelerators and CPUs provide dedicated (de)compression engines supporting formats such as Snappy and LZ4. Selecting suitable formats, chunk sizes, and prefetching strategies while offloading compression and decompression from general-purpose cores eases bandwidth pressure and improves system-level performance per watt for data-intensive inference~\cite{nvidia_blackwell_nvcomp}. Similar to precision and sparsity, compression offload reduces both average power and the amplitude of the memory-bound power excursions that alternate with compute-bound phases.

\subsubsection{Dedicated AI Inference Accelerators}
\label{sec:device_asic}

Dedicated inference accelerators tailor arithmetic, memory, and control to neural-network execution. Dense
   multiply–accumulate arrays and local data reuse reduce the overhead of executing matrix operations, while simplified control reduces the resources
   devoted to general-purpose processing. Google’s first-generation tensor processing unit illustrates this approach through low-precision matrix
   arithmetic and software-managed on-chip memory, demonstrating improved performance per watt on production inference workloads. These gains depend
   on matching the model’s computational and memory requirements to the hardware~\cite{GoogleTPU_Jouppi2017}.  Compiler-managed execution provides another approach to specialization. Groq’s tensor streaming processor, for example, schedules computation and data movement explicitly, using distributed on-chip SRAM and closely coupled matrix and vector units. 
   
 For LLM inference, prefill benefits from parallel computation, while decoding at small batch sizes is often limited by memory bandwidth. Longer contexts
  increase key–value cache demands. Accelerator selection and model placement should therefore balance compute capacity, memory bandwidth, and communication
  against batch size, context length, and latency requirements~\cite{pope2023efficiently}.
  
   For power systems planning, these architectural trade-offs suggest evaluating the complete serving configuration, including the number of accelerators
   and their memory and interconnect resources. At a fixed inference demand and service quality, lower energy per token can reduce average power. At a
   power-constrained site, the same efficiency gain may instead support greater throughput within the existing power budget. The resulting power
   fluctuations depend on how computation and communication are scheduled across devices. A useful evaluation should therefore report energy per token
   alongside peak rack demand and power variability under matched service requirements.

\subsubsection{Device-Level Load Smoothing}
\label{sec:device_smoothing}

The mitigation strategies in Sections~\ref{sec:device_precision} and~\ref{sec:device_asic} reshape the workload that drives the power
transient. This subsection examines a complementary path:
absorbing the transient at the rack's electrical boundary. 

In large AI training jobs, thousands of GPUs switch synchronously
between idle and high-power states. The resulting power swings
outpace what conventional generation can follow and appear to
neighboring customers as voltage spikes or sags. Rack-level energy
storage with intelligent power control addresses this mechanism
directly. Capacitor banks in power shelves absorb energy during brief
power dips---checkpointing or communication phases---and release it
when GPUs ramp back up. The power curve is thereby flattened at the
rack input rather than passed through to the facility and grid.

NVIDIA reports a direct comparison under identical AI workloads.
Legacy racks pass GPU power spikes through to the AC side. GB300~NVL72
racks with energy-storage shelves reduce grid-facing peak power by
approximately 30\%, with the power profile seen by the GPUs
unchanged~\cite{choukse_power_stabilization}.

Rack-level storage is the most developed of the device-level options. Its usable smoothing band is bounded from above by the closed-loop
bandwidth of the bidirectional DC/DC stage interfacing the buffer to the rack
bus, and from below by usable energy: a buffer absorbing an excursion of
magnitude $\Delta P$ with stored energy $E$ sustains it for only
$\tau \approx E/\Delta P$, so content below $1/\tau$ passes through
unattenuated. The upper bound lies well above the spectral content of the transients in Section~\ref{subsec:ssr} and the impulsive
steps in Section~\ref{subsec:voltage}. Rack-level storage filters the excursion before it propagates upstream of the rack boundary rather than reducing it at the source.

Three limitations nonetheless bound its role. First, the energy
capacity of capacitor-based shelves supports smoothing on
millisecond-to-seconds timescales but does not extend to the minutes-long horizon served by facility-level battery energy storage systems (BESS)
(Section~\ref{subsubsec:facility_bess}). Device-level storage
therefore complements rather than replaces facility storage. Second,
adding active electronics and storage inside the rack introduces new
rack-internal failure modes and protection-coordination requirements.
Third, capital cost per rack scales with deployment. Its
value---an avoided contribution to facility- and grid-level
disturbances---is difficult to attribute under current market
structures.  Rack-level storage is technically mature and effective. Reliability integration and cost allocation could be the barriers to adoption.

\subsubsection{Dynamic Power Allocation and Power-Aware Profiles}
\label{sec:device_dynamic_power}

Static power allocation fixes the CPU--GPU power split irrespective
of workload behavior. For AI workloads that alternate between
CPU-bound and GPU-bound phases, a fixed partition rarely matches the
current bottleneck: when the CPU is on the critical path, it may lack sufficient power while the GPU has unused headroom; the reverse occurs when the GPU is the bottleneck.
Because end-to-end performance is governed by
$\max(T_{\text{CPU}}, T_{\text{GPU}})$, fixed splits systematically
under-provision the bottleneck and over-provision non-critical
components. 

HPC-scale studies show that moderate power caps reduce energy consumption and temperature with limited performance impact, though the optimal cap
depends on model architecture, batch size, and parallelism
strategy~\cite{zhao2024sustainable}. Power-allocation strategies are
accordingly moving from static caps toward dynamic caps,
\emph{workload-aware power profiles}, and ML-driven closed-loop
controllers that adjust power based on utilization, bandwidth, and thermal signals, achieving double-digit energy savings with modest
slowdowns~\cite{Narayanaswamy2025,Desai2025}.
 Standardized power and energy telemetry enables power- and
carbon-aware scheduling~\cite{RedHatKepler2023}.

   GPU power capping can reduce power draw and operating temperature with limited performance loss~\cite{zhao2024sustainable}. Processor reliability studies further show that
  dynamic voltage and frequency changes can induce thermal cycling, motivating control policies that explicitly account for temperature variations~\cite{haghbayan2020thermal}. At the data center level, synchronized AI workloads require control of power-fluctuation magnitude and ramp rates. GPU
   power-smoothing mechanisms address these requirements through programmable ramp limits, minimum power floors, and stop delays. Together, these
   findings motivate coordinated power-management policies that consider both device thermal cycling and aggregate load variability.

The device-level mitigation strategies surveyed above address the mechanisms discussed in Section~\ref{subsec:device_reliability} to varying degrees. Rack-level energy storage complements workload shaping by absorbing
fast transients at the rack's electrical boundary. Workload-aware power profiles and precision/sparsity tuning mitigate thermomechanical fatigue by reducing the amplitude of thermal cycling, though neither eliminates cumulative damage. Heterogeneous inference fabrics act on the sub-synchronous and inter-area mechanisms by reducing workload coherence at the source. No device-level mitigation, however, closes the
telemetry-bandwidth gap: standard interfaces sample device power well below the bandwidth of the transients that determine device stress, so runtime control and qualification continue to operate on filtered statistics. Closing this gap requires package-level current and voltage instrumentation capable of resolving sub-millisecond transients and a vendor-neutral telemetry standard. 

\begin{table*}[h!]
\centering
\caption{Regulatory instruments, reliability standards, and grid-connection
frameworks governing large computational loads.}
\label{tab:policy}

\scriptsize
\begingroup
\setlength{\tabcolsep}{3.4pt}
\renewcommand{\arraystretch}{1.12}

\begin{tabularx}{\textwidth}{
    @{}
    >{\raggedright\arraybackslash}p{0.150\textwidth}
    >{\raggedright\arraybackslash}p{0.110\textwidth}
    Y
    >{\raggedright\arraybackslash}p{0.070\textwidth}
    >{\raggedright\arraybackslash}p{0.140\textwidth}
    @{}
}
\toprule
\textbf{Instrument}
&
\textbf{Body (region)}
&
\textbf{Requirement or contribution}
&
\textbf{Regulatory force}
&
\textbf{Status / date}
\\
\midrule

\multicolumn{5}{@{}c}{\textit{Reliability standards and ride-through requirements}}
\\
\midrule

Large Loads Task Force outputs and Reliability Guideline
&
NERC (North America)
&
Characterization of emerging large-load behavior, gap assessment against
existing reliability standards, and recommended practices for load-composition
data sharing, voltage ride-through design, protection coordination, and model
validation, addressed to load owners and equipment vendors as well as utilities.
&
Voluntary
&
Task force formed Aug.~2024; white papers 2025; guideline published May~1, 2026
\cite{NERC_LLWG_GapPaper2026}
\\

Level~2 and Level~3 Alerts
&
NERC
&
Escalating essential actions on computational-load modeling, planning and
operational studies, high-speed instrumentation, commissioning, communications,
and protection and control.
&
Mandatory response
&
Level~2 Sep.~2025; Level~3 May~2026, responses due Aug.~3, 2026
\cite{NERC_Alert_L2_2025,NERC_Alert_L3_2026}
\\

Project~2026-02 (CLO-001-1)
&
NERC
&
Development of enforceable modeling, ride-through, verification, and
coordination requirements for qualifying computational loads.
&
Proposed mandatory
&
Draft standards under formal comment
\cite{NERCProject2026_02}
\\

Directive on computational-load standards
&
FERC
&
Converts NERC's accelerated standards schedule into an enforceable federal
requirement and opens the question of registering large loads as entities
directly accountable under FPA Section~215.
&
Mandatory
&
Order issued July~2026
\cite{FERC_NERC_Directive2026}
\\

NOGRR282 and NPRR1308
&
ERCOT (Texas)
&
Frequency and voltage ride-through requirements and associated protocol
provisions for Large Computational Loads.
&
Mandatory
&
Approved July~2026; effective Aug.~1, 2026
\cite{ERCOTNOGRR282,ERCOTNPRR1308}
\\

Grid-readiness standards needs analysis
&
IEEE SA
&
Review of existing industry efforts and standards for data center grid
integration, with identification of gaps warranting new standardization work.
&
Advisory
&
Published Jan.~2026
\cite{IEEE_DCStandardsGap2026}
\\

\midrule
\multicolumn{5}{@{}c}{\textit{Interconnection and connection frameworks}}
\\
\midrule

ANOPR on large load interconnection (RM26-4)
&
FERC / U.S.~DOE
&
Consideration of federal rules for interconnecting loads above 20~MW to the
interstate transmission system, following a direction issued under Section 403 of the DOE Organization Act
&
Proposed
&
DOE direction Oct.~2025; FERC action June~2026
\cite{FERC_RM26_4}
\\

Federal Power Act Section~206 large load proceedings
&
FERC
&
Six RTOs/ISOs were directed to justify or reform tariff provisions governing
the timely and reliable connection and service of data centers and other
large loads.
&
Mandatory
&
Orders issued June~18, 2026; compliance filings due within 60~days
\cite{ferc2026largeload}
\\

Co-located load and behind-the-meter generation rules (EL25-49)
&
FERC / PJM
&
Requires PJM to define how co-located load may be served, to establish three
new transmission services with energy-withdrawal caps, and to adjust capacity
interconnection rights to reflect reduced net injection when on-site generation
serves local demand.
&
Mandatory
&
Order Dec.~18, 2025; PJM compliance filing Feb.~2026
\cite{FERC_EL25_49}
\\

High Impact Large Load process
&
SPP
&
Dedicated study process for large loads together with electrically proximate
generation, treating load and its supply as a single interconnection question.
&
Mandatory
&
Approved by FERC
\cite{SPP_HILL2026}
\\

Large Load Integration Program
&
AESO (Alberta)
&
Interim connection allocation and development of a long-term framework covering
interconnection, planning, operations, markets, tariff, reliability, and
bring-your-own-generation pathways.
&
Mixed
&
Phase~I complete; Phase~II ongoing
\cite{AESOLargeLoad2026}
\\

Demand connections reform
&
Ofgem, NESO, DESNZ (Great Britain)
&
Response to a demand connection queue that grew from 41~GW to 125~GW against a
45~GW system peak, of which roughly 140 data centers represent about 50~GW;
under consideration are phased, ramped, and non-firm connections and
prioritization of projects able to demonstrate flexibility.
&
Proposed
&
Call for input Q1~2026; further consultation autumn~2026
\cite{Ofgem_DemandConnections2026}
\\

Large Energy User connection policy
&
CRU and EirGrid (Ireland)
&
Conditions connection offers on minimum performance and availability of on-site
generation or storage, initiates Grid Code work on fault ride-through for large
users, and treats islanded fossil-fueled developments as contrary to national
policy.
&
Mandatory
&
Policy Dec.~2025; system operator proposals due Mar.~2026
\cite{CRU_LEU_Policy2025}

\\

\midrule
\multicolumn{5}{@{}c}{\textit{Flexibility classification frameworks}}
\\
\midrule

Flex MOSAIC
&
EPRI with more than 65 utilities, system operators, and AI facility operators
&
Voluntary classification of large-load flexibility by the magnitude, timing,
duration, and frequency of the response a facility can commit to, intended as a
common vocabulary for interconnection studies and planning in place of bespoke
bilateral terms.
&
Voluntary
&
Announced Mar.~2026
\cite{EPRI_FlexMOSAIC2026}
\\

\bottomrule
\end{tabularx}
\endgroup
\end{table*}

\subsection{Regulations and Policymaking}

Regulatory and grid-code requirements exhibit the same timescale
separation as the technical solution stack. At planning timescales, rules should define service firmness, upgrade responsibilities, cost allocation, and eligibility for curtailable or non-firm connections. At operating timescales, they should specify POI telemetry, curtailment response, ramp limits, and staged load restoration. At disturbance timescales, converter-interfaced loads should meet verifiable FRT and PFAPR envelopes. Table~\ref{tab:policy} summarizes representative instruments. Across jurisdictions, the common direction is toward treating large computational loads as modeled, controllable, and auditable grid participants rather than static demand.

\subsection{Recent Data, Models, and Demonstrations}
\label{subsec:recent_developments}

 Recent activity has moved from policy and modeling toward measurement and field validation. Table~\ref{tab:resources} groups representative efforts into public datasets and benchmarks, workload-to-grid models, and demonstrations or commercial flexibility agreements. Current evidence has progressed from traces and component models to field demonstrations, although gigawatt-scale flexibility remains more commonly contracted or being commissioned than dynamically validated. Together, these efforts show the transition from estimating AI power demand to testing dispatch, observability, and quality-of-service-preserving response in operational settings. The examples are illustrative, and status is reported as of August 2026.

\begin{table*}[h!]
\centering
\caption{Open datasets, simulation models, and field demonstrations relevant to the grid integration of AI data centers.}
\label{tab:resources}

\scriptsize
\begingroup
\setlength{\tabcolsep}{3.4pt}
\renewcommand{\arraystretch}{1.12}

\begin{tabularx}{\textwidth}{
    @{}
    >{\raggedright\arraybackslash}p{0.150\textwidth}
    >{\raggedright\arraybackslash}p{0.105\textwidth}
    Y
    >{\raggedright\arraybackslash}p{0.150\textwidth}
    >{\raggedright\arraybackslash}p{0.100\textwidth}
    @{}
}
\toprule
\textbf{Resource or activity}
&
\textbf{Lead organizations}
&
\textbf{Contribution to data center--grid integration}
&
\textbf{Scale, resolution, or measured effect}
&
\textbf{Status}
\\
\midrule

\multicolumn{5}{@{}c}{\textit{Measured datasets and benchmarks}}
\\
\midrule

GenAI workload power-profile dataset
&
National Lab of the Rockies
&
Measured power profiles for training and inference workloads, and
facility-level profiles generated using a bottom-up,
event-driven model.
&
5 and 10~Hz sampling; facility-level synthesis
&
Public; 2026
\cite{NLRGenAIPower2026}
\\

MIT SuperCloud Dataset
&
MIT Lincoln Laboratory
&
Anonymized scheduler records and CPU/GPU time-series measurements from a
production HPC environment.
&
100~ms GPU telemetry
&
Public; 2021
\cite{Samsi2021MITSupercloud}
\\

ML.ENERGY Benchmark
&
ML.ENERGY Initiative
&
Standardized LLM and diffusion-model inference measurements, including energy,
throughput, latency, output length, and per-request power timelines.
&
Per-request timelines across model and GPU configurations
&
Open dataset and toolkit
\cite{Chung2025MLEnergy}
\\

BurstGPT
&
Wang et al.
&
Multi-month traces from production LLM services characterizing request arrivals,
burstiness, response lengths, and failures for workload replay.
&
Request-level; no facility power
&
Public
\cite{Wang2025BurstGPT}
\\

Azure LLM inference traces
&
Microsoft
&
Production inference request-arrival traces with input and output token counts, widely used to replay serving workloads and provide arrival processes as inputs to power models.
&
Request-level; no power channel
&
Public; 2023, 2024
\cite{AzureLLMTrace}
\\

AcmeTrace
&
Shanghai AI Laboratory
&
Multi-month traces from production LLM clusters spanning
pretraining, fine-tuning, and evaluation jobs, with utilization and failure
records.
&
Two production clusters
&
Public; 2024
\cite{Hu2024AcmeTrace}
\\

TokenPowerBench
&
Niu et al.
&
Inference power benchmark that sweeps the configuration space operators
actually tune: batch size, tensor and pipeline parallelism, context length,
and quantization.
&
16 H100 GPUs; models with 180B--480B parameters
&
Public; 2025
\cite{Niu2025TokenPowerBench}
\\

Watt Counts
&
Fadel Argerich et al.
&
Open energy-aware inference dataset spanning heterogeneous accelerators in
both batch and online serving regimes.
&
50 models $\times$ 10 GPU types
&
Public; 2026
\cite{WattCounts2026}
\\

\midrule
\multicolumn{5}{@{}c}{\textit{Models, predictors, and co-simulation}}
\\
\midrule

OpenG2G
&
Chung et al.
&
Open-source co-simulation linking AI workload and data center models,
distribution-grid simulation, and feedback controllers to assess trade-offs
between computing performance and grid conditions.
&
Workload-to-feeder co-simulation
&
Open source; 2026
\cite{Chung2026OpenG2G}
\\

Logistic power--concurrency characterization
&
Chung et al.; Liang et al.
&
Compact closed-form map from in-flight sequence count to accelerator power,
suitable for embedding in dynamic load models rather than treating GPU power
as a fixed fraction of TDP.
&
H100-SXM5: 300~W idle to 600~W nominal, $<$3\% fit error
&
Published; 2026
\cite{chung2026mlenergy}
\\

WattGPU
&
Fadel Argerich et al.
&
Predicts mean inference power and inter-token latency for GPU--model pairs not
seen during training, removing the need to profile every deployment
combination.
&
42 LLMs, 8 GPUs; $\leq$3.4\% (offline) and $\leq$13.5\% (server) median absolute percentage error (APE)
&
Public; 2026
\cite{WattGPU2026}
\\

\midrule
\multicolumn{5}{@{}c}{\textit{Test beds, demonstrations, and flexibility agreements}}
\\
\midrule

Agora large load test bed
&
NLR; U.S.~DOE; industry partners
&
Dedicated controlled environment replicating the technical complexity of a
large data center interconnection, operated alongside the laboratory's
grid-emulation platform and DC-architecture test capability.
&
Hardware-in-the-loop at interconnection scale
&
Launched May~2026
\cite{NLRAgora2026}
\\

DCFlex initiative
&
EPRI with hyperscalers, utilities, and vendors
&
Coordinated portfolio of field demonstrations at operating data centers,
generating reference architectures and shared results across markets rather
than one-off pilots.
&
Nine sites across the U.S. and Europe
&
Launched Oct.~2024; ongoing
\cite{EPRI_DCFlex2026}
\\

Grid-interactive UPS deployment in Ireland
&
Microsoft; Eaton; Enel~X; EirGrid
&
Data center UPS batteries provide grid services while retaining their primary
power-conditioning and backup-power functions.
&
Market participation of installed UPS fleet
&
Deployed 2022
\cite{MicrosoftGridInteractiveUPS2022}
\\

Phoenix AI-flexibility demonstration
&
Emerald AI; EPRI; Oracle; NVIDIA; SRP
&
Software-based workload orchestration during a peak-demand event while
maintaining computing quality-of-service requirements.
&
256 GPUs; 25\% power reduction sustained for 3~h
&
Field demonstration; peer-reviewed
\cite{Colangelo2026GridInteractiveAI}
\\

London power-flexibility trial
&
Emerald AI; National Grid; EPRI; Nebius; NVIDIA
&
Cluster response to a large number of simulated grid events without disrupting
critical workloads.
&
96 GPUs; $>$200 events over 5~days; $>$1/3 reduction in $<$1~min
&
Trial Dec.~2025
\cite{NationalGridEmerald2026}
\\

Utility-side observability pilot
&
Silicon Valley Power; NVIDIA; Emerald AI
&
Bidirectional platform giving the utility real-time visibility of data center
load and a dispatch path back to it, addressing the observability side of
flexibility rather than the actuation side alone.
&
Municipal utility serving 58 data centers
&
Pilot; 2026
\cite{SVP_EmeraldPilot2026}
\\

Aurora AI Factory
&
EPRI; NVIDIA; Emerald AI; partners
&
Facility designed for flexibility from the outset rather than retrofitted,
intended to validate workload flexibility at full campus scale.
&
96~MW, Manassas, Virginia
&
Commissioning late 2026 
\cite{AuroraAIFactory2026}
\\

Contracted demand response portfolio
&
Google with five U.S.~utilities
&
Machine-learning workload curtailment embedded in long-term supply contracts,
establishing flexibility as a commercial term of interconnection rather than a
demonstration.
&
1~GW aggregate contracted capacity
&
Milestone announced Mar.~2026
\cite{Google1GWDR2026}
\\

\bottomrule
\end{tabularx}
\endgroup
\end{table*}

\section{Recommendations and Conclusions}
\label{sec:stakeholders}
\subsection{Recommendations}

Reliable integration of AI data centers requires responsibilities to be distributed across the grid, facility, and workload layers. \textbf{Grid operators and utilities} should treat AI data centers as dynamic large loads rather than static P--Q injections and incorporate converter behaviors, UPS and protection logic, AI-workload-driven ramps, system strength, and ride-through characteristics into interconnection studies. Grid operators should define measurable requirements for fault ride-through, post-fault power recovery, ramp rates, reactive capability, and telemetry, while utilities should provide the required transmission capacity, compensation, and system-strength infrastructure.  

\textbf{Data center operators} should ensure that facility assets are operated in grid-supportive modes rather than solely for backup or economic optimization. Fast-timescale disturbances could be absorbed locally through rack-level buffering and grid-interactive UPS systems, while facility BESS, microgrids, and on-site generation should address progressively slower and longer-duration variations. Adequate power and state-of-charge headroom should be reserved for reliability events, and islanding, transfer, and reconnection should follow controlled ramp limits to avoid creating large grid-facing power steps.

\textbf{AI developers and cloud tenants} should contribute flexibility at the source of demand. Where service-level requirements permit, training and inference workloads can reduce synchronized power excursions through elastic scheduling, staggered checkpointing, batching, DVFS, power capping, and geographic workload routing. These mechanisms should increasingly be evaluated not only for throughput and energy efficiency but also for their effects on ramp rate, peak demand, and load coherence.

Finally, \textbf{policymakers and regulators} should establish standards and market arrangements that align these technical capabilities with system needs. Flexible and curtailable connections, compensation for fast demand response, clear cost-allocation rules, and periodic verification of declared facility behavior can provide incentives for controllability. Across all stakeholders, mitigation should follow the bandwidth-matching principle: assign each disturbance to a resource that responds on a comparable timescale.

\subsection{Conclusion}

AI data centers could not be well represented as passive, static loads any longer. Their gigawatt-scale concentration, power-electronic interfaces, and software-coordinated workloads allow internal device, converter, protection, and scheduling dynamics to propagate to the electricity transmission and distribution systems. Recent oscillatory events and coordinated gigawatt-scale load transfers demonstrate that facility-side behavior can become a bulk-system disturbance. Consequently, reliable AI integration cannot be achieved through generation and transmission expansion alone; planning and operation must also account for how these facilities respond dynamically to both workload changes and grid disturbances.

As AI workloads are programmable, and modern data centers contain fast converters, UPS systems, energy storage, microgrids, and workload controls capable of reshaping grid-facing demand, opportunities emerge to turn AI-driven demand from a perceived burden into a proactive building block for reliable and sustainable energy systems. Effective integration therefore requires a cross-layer, \textit{bandwidth-matched} architecture in which device and rack controls suppress the fastest transients, facility resources manage second-to-minute variations and ride-through, workload scheduling provides controllable flexibility, and grid-level measures address system strength, congestion, and long-term adequacy. As securing more electricity capacity is not the sole objective for future AI infrastructure, large computational demand must also become observable, controllable, and dynamically predictable at the point of interconnection. AI data centers can become grid-supportive demand resources that follow variable generation instead of requiring firm capacity to follow them. 

\bibliographystyle{IEEEtran}
\bibliography{ref1_deduplicated}

@article{chen2025electricity,
  title={Electricity demand and grid impacts of AI data centers: Challenges and prospects},
  author={Chen, Xin and Wang, Xiaoyang and Colacelli, Ana and Lee, Matt and Xie, Le},
  journal={arXiv preprint arXiv:2509.07218},
  year={2025}
}

@misc{ercot_lfltf_2025,
    author       = {Gravois, Patrick},
    title        = {{ERCOT} Large Load Loss/Reduction Events 2020--2024},
    howpublished = {Electric Reliability Council of Texas (ERCOT), PDCWG meeting presentation},
    year         = {2024},
    month        = Nov.,
    note         = {Nov 19, 2024; slide 5 documents the Dec. 7, 2022 West Texas event},
    url          = {https://www.ercot.com/files/docs/2024/11/18/ERCOT%20Large%20Load%20Events_PDCWG_19Nov2024.pdf}
  }

@techreport{NERC2025LargeLoads,
  author      = {{North American Electric Reliability Corporation (NERC)}},
  title       = {Characteristics and Risks of Emerging Large Loads},
  institution = {North American Electric Reliability Corporation},
  type        = {Large Loads Task Force White Paper},
  month       = Jul.,
  year        = {2025},
  url         = {https://www.nerc.com/globalassets/who-we-are/standing-committees/rstc/whitepaper-characteristics-and-risks-of-emerging-large-loads.pdf},
  note        = {[Online]}
}

@techreport{NERC_CryptoIncident_2026,
  author      = {{North American Electric Reliability Corporation (NERC)}},
  title       = {Incident Review: Considering Voltage-Sensitive Crypto Load Reductions},
  institution = {North American Electric Reliability Corporation},
  type        = {Event Analysis Report},
  year        = {2026},
  note        = {Reviews 26 ERCOT large electronic load ride-through events exceeding 100~MW, Jan. 2023--Sep. 2025},
  url         = {https://www.nerc.com/globalassets/our-work/reports/event-reports/incident_review_considering_voltage-sensitive_crypto_load_reductions.pdf}
}

@techreport{NERC_Alert_L3_2026,
  author      = {{North American Electric Reliability Corporation (NERC)}},
  title       = {Essential Action to Industry --- Computational Load Modeling, Studies, Instrumentation, Commissioning, Operations, Protection, and Control},
  institution = {North American Electric Reliability Corporation},
  type        = {Level 3 Essential Action Alert},
  month       = May,
  year        = {2026},
  note        = {Issued May 4, 2026; seven Essential Actions, responses due Aug. 3, 2026},
  url         = {https://www.nerc.com/globalassets/programs/bpsa/alerts/level-3-computational-load-alert.pdf}
}

@techreport{NERC_Alert_L2_2025,
  author      = {{North American Electric Reliability Corporation (NERC)}},
  title       = {Industry Recommendation: Large Load Interconnection, Study, Commissioning, and Operations},
  institution = {North American Electric Reliability Corporation},
  type        = {Level 2 Alert},
  year        = {2025},
  url         = {https://www.nerc.com/globalassets/programs/bpsa/alerts/2025/nerc-alert-level-2--large-loads.pdf}
}

@misc{FERC_RM26_4,
  author       = {{Federal Energy Regulatory Commission (FERC)}},
  title        = {Interconnection of Large Loads to the Interstate Transmission System},
  howpublished = {Docket No. RM26-4-000, Advance Notice of Proposed Rulemaking proceeding},
  month        = Oct.,
  year         = {2025},
  note         = {ANOPR proposed by the U.S. Secretary of Energy on Oct. 23, 2025; FERC notice inviting comments issued Oct. 27, 2025},
  url          = {https://www.ferc.gov/rm26-4}
}

@article{mishra2025understanding,
  title={Understanding the inception of 14.7 Hz oscillations emerging from a data center},
  author={Mishra, Chetan and Vanfretti, Luigi and Delaree Jr, Jaime and Purcell, TJ and Jones, Kevin D},
  journal={Sustainable Energy, Grids and Networks},
  volume={43},
  pages={101735},
  year={2025},
  publisher={Elsevier}
}

@misc{PJM2026DominionLoadTransfer,
  author       = {{PJM Interconnection}},
  title        = {{Jul. 22, 2026 Dominion Load Transfer Event}},
  year         = {2026},
  month        = Aug.,
  day          = {6},
  howpublished = {Presentation to the PJM Operating Committee},
  note         = {Presented by Matthew Wharton; includes preliminary findings from PJM Interconnection and Dominion Energy regarding the Jul. 22, 2026 Northern Virginia load-transfer event},
  url          = {https://www.pjm.com/-/media/DotCom/committees-groups/committees/oc/2026/20260806/20260806-item-05---dominion-Jul.y-loss-of-load-event---presentation.pdf},
  urldate      = {2026-08-27}
}

@electronic{meta2024llama31,
  author       = {{Meta AI}},
  title        = {Introducing {Llama} 3.1: Our most capable models to date},
  howpublished = {Meta AI Blog},
  month        = Jul.,
  year         = {2024},
  note         = {Published Jul. 23, 2024; accessed Aug. 29, 2026},
  url          = {https://ai.meta.com/blog/meta-llama-3-1/}
}

@article{tabbakh2024towards,
  author  = {A. Tabbakh and L. {Al Amin} and M. Islam and G. M. I. Mahmud and
             I. K. Chowdhury and M. S. H. Mukta},
  title   = {Towards sustainable {AI}: a comprehensive framework for
             {Green AI}},
  journal = {Discover Sustainability},
  volume  = {5},
  number  = {1},
  pages   = {408},
  year    = {2024}
 
}

@misc{kozhu2025,
  author        = {M.-S. Ko and H. Zhu},
  title         = {Wide-Area Power System Oscillations from Large-Scale {AI}
                   Workloads},
  eprint        = {2508.16457},
  archiveprefix = {arXiv},
  primaryclass  = {eess.SY},
  month         = Aug.,
  year          = {2025},
  note          = {Version 2 posted Apr. 20, 2026},
  doi           = {10.48550/arXiv.2508.16457},
  url           = {https://arxiv.org/abs/2508.16457}
}

@techreport{NESO2024,
  author      = {{National Grid Electricity System Operator (ESO)}},
  title       = {Sub-synchronous oscillations in {GB}: Current state and plans for future management},
  institution = {National Grid Electricity System Operator},
  month       = May,
  year        = {2024},
  note        = {Accessed Aug. 29, 2026},
  url         = {https://www.neso.energy/document/319056/download}
}

@inproceedings{Chakraborty2024Impedance,
  title = {Impedance scan of inverter-based resources and diesel generator for stability analysis},
  author = {Chakraborty, Soham and Wang, Jing and Ganguly, Subhankar and Kroposki, Benjamin},
  booktitle = {2024 IEEE Power \& Energy Society General Meeting (PESGM)},
  pages = {1--5},
  year = {2024},
  organization = {IEEE}
}

@article{sun2022dynamic,
  author  = {J. Sun and S. Wang and J. Wang and L. M. Tolbert},
  title   = {Dynamic Model and Converter-Based Emulator of a Data Center Power
             Distribution System},
  journal = {IEEE Transactions on Power Electronics},
  volume  = {37},
  number  = {7},
  pages   = {8420--8432},
  month   = Jul.,
  year    = {2022}
  
}

@inproceedings{Dominion2024,
  author    = {T. J. Purcell IV and R. E. Tuck and C. Mishra and M. J. Till},
  title     = {Analysis of Real-World Sub-Synchronous Oscillation Events in a
               Power System with High Penetration of Inverter-Based Resources},
  booktitle = {Proc. Transient Recorders Users Council (TRUC) Conference},
  organization = {Dominion Energy},
  address   = {Richmond, VA, USA},
  year      = {2024},
  url       = {https://truc.org/wp-content/uploads/Analysis-of-Real-World-Sub-Synchronous-Oscillation-Events-in-a-Power-System-with-High-Penetration-of-Inverter-Based-Resources.pdf}
}

@techreport{schneider2026rd111,
  author      = {{Schneider Electric}},
  title       = {Data Center Reference Design 111: 7,536~{kW}, Tier {III}, {ANSI}, Chilled Water, Liquid-Cooled {AI} Clusters ({NVIDIA} {GB300})},
  institution = {Schneider Electric},
  type        = {Reference Design},
  number      = {RD111DSR0},
  note        = {Version 2.0},
  month       = {Mar.},
  year        = {2026},
  url         = {https://www.se.com/us/en/download/document/RD111DSR0/}
}

@article{klein1991fundamental,
  author = {M. Klein and G. J. Rogers and P. Kundur},
  title = {A fundamental study of inter-area oscillations in power systems},
  journal = {IEEE Transactions on Power Systems},
  volume = {6},
  number = {3},
  pages = {914--921},
  year = {1991}
  
}

@techreport{NERC2019IOA,
  author      = {{North American Electric Reliability Corporation (NERC)}},
  title       = {Interconnection Oscillation Analysis},
  institution = {North American Electric Reliability Corporation},
  month       = Jul.,
  year        = {2019},
  url         = {https://www.nerc.com/comm/RSTC/SMWG/Interconnection_Oscillation_Analysis.pdf}
}

@techreport{NERC2020Guideline,
  author      = {{North American Electric Reliability Corporation (NERC)}},
  title       = {Oscillation Analysis for Monitoring and Mitigation},
  institution = {North American Electric Reliability Corporation},
  month       = May,
  year        = {2020},
  url         = {https://www.nerc.com/comm/RSTC_Reliability_Guidelines/Oscillation_Analysis_for_Monitoring_And_Mitigation_TRD.pdf}
}

@techreport{Dominion2025FIR,
  author      = {{Dominion Energy}},
  title       = {Facility Interconnection Requirements},
  institution = {Dominion Energy},
  month       = Sep.,
  year        = {2025},
  note        = {Sep. 25, 2025},
  url         = {https://www.dominionenergy.com/-/media/content/large-business-services/pdfs/virginia/facility-interconnection-requirements.pdf}
}

@article{xie2025enhancing,
  title={Enhancing Data Center Low-Voltage Ride-Through},
  author={Xie, Yiheng and Cui, Wenqi and Wierman, Adam},
  journal={arXiv preprint arXiv:2510.03867},
  year={2025}
}

@misc{milanoDCmodel2025,
  author        = {A. Jim{\'e}nez-Ruiz and F. Milano},
  title         = {Data Center Model for Transient Stability Analysis of Power
                   Systems},
  eprint        = {2505.16575},
  archiveprefix = {arXiv},
  primaryclass  = {eess.SY},
  month         = May,
  year          = {2025},
  doi           = {10.48550/arXiv.2505.16575},
  url           = {https://arxiv.org/abs/2505.16575}
}

@techreport{entsoe_rocof2020,
  author      = {{ENTSO-E}},
  title       = {Inertia and Rate of Change of Frequency ({RoCoF})},
  institution = {ENTSO-E},
  address     = {Brussels, Belgium},
  year        = {2020}
}

@techreport{OCPDiablo400,
  author      = {{Microsoft} and {Meta} and {Google}},
  title       = {Diablo 400 Project: Rack and Power, Version 0.5.2},
  institution = {Open Compute Project (OCP)},
  month       = May,
  year        = {2025},
  url         = {https://www.opencompute.org/documents/ocp-specification-diablo-400-v0p5p2-2025-05-30-pdf}
}

@techreport{infineon_scaling_2025,
  author      = {{Infineon Technologies AG}},
  title       = {Scaling {AI} Data Center Power Delivery with {Si}, {SiC}, and {GaN}},
  institution = {Infineon Technologies AG},
  type        = {White Paper},
  year        = {2025},
  note        = {Accessed: Nov. 2, 2025},
  url         = {https://www.infineon.com/assets/row/public/documents/24/59/whitepaper-scaling-ai-datacenter-power-delivery-with-si-sic-and-gan.pdf}
}

@article{sajadi2026power,
  title={Power Grid Infrastructure for AI Data Centers},
  author={Sajadi, Amir and Za’ter, Muhy Eddin and Vabson, Maria and Baker, Kyri and Hodge, Bri-Mathias},
  journal={Computer},
  volume={59},
  number={8},
  pages={50--59},
  year={2026},
  publisher={IEEE}
}

@misc{yang_ptsmi_2023,
  author       = {Z. Yang and K. Adamek and W. Armour},
  title        = {Part-time Power Measurements: \texttt{nvidia-smi}'s Lack of Attention},
  howpublished = {arXiv preprint arXiv:2312.02741},
  month        = Dec.,
  year         = {2023},
  note         = {Accessed: Nov. 2, 2025},
  url          = {https://arxiv.org/abs/2312.02741}
}

@techreport{PJM2025,
  author      = {{PJM Interconnection}},
  title       = {Large Load Additions -- Critical Issue Fast Path: Executive Summary},
  institution = {PJM Interconnection},
  type        = {Members Committee Filing},
  month       = Nov,
  year        = {2025},
  url         = {https://www.pjm.com/-/media/DotCom/committees-groups/committees/mc/2025/20251119/20251119-item-01b---imm-package---executive-summary.pdf}
}

@misc{TexasSB6-2025,
  author       = {{Texas Legislature}},
  title        = {Senate Bill 6 --- Relating to the planning, interconnection and operation of certain large electrical loads},
  howpublished = {89th Legislature},
  year         = {2025},
  url          = {https://capitol.texas.gov/tlodocs/89R/billtext/html/SB00006S.htm}
}

@techreport{AESO2025,
  author      = {{Alberta Electric System Operator (AESO)}},
  title       = {Connection Requirements for Transmission-Connected Data Centres},
  institution = {Alberta Electric System Operator},
  type        = {Draft Technical Requirements},
  year        = {2025},
  url         = {https://aesoengage.aeso.ca/connection-requirements-for-transmission-connected-data-centres}
}

@misc{FERC2025ShowCause,
  author       = {{Federal Energy Regulatory Commission (FERC)}},
  title        = {Notice of Action Taken --- Co-Located Large Loads at Generating Facilities (Docket No. {EL25-49}, {AD24-11}, {EL25-20}, {PJM} Interconnection, {L.L.C.})},
  month        = Feb.,
  year         = {2025},
  url          = {https://www.ferc.gov/news-events/news/ferc-orders-action-co-location-issues-related-data-centers-running-ai}
}

@misc{DOE2025,
  author = {{U.S. Department of Energy}},
  title = {Secretary of Energy's Direction that the Federal Energy Regulatory Commission Initiate Rulemaking Procedures and Proposal Regarding the Interconnection of Large Loads Pursuant to the Secretary's Authority Under Section 403 of the Department of Energy Organization Act},
  howpublished = {DOE filing to FERC},
  month = {Oct.},
  year = {2025},
  url = {https://www.energy.gov/sites/default/files/2025-10/403%20Large%20Loads%20Letter.pdf},
  note = {Issued Oct. 23, 2025}
}

@techreport{NERC_Incident_2025,
  author      = {{North American Electric Reliability Corporation (NERC)}},
  title       = {Incident Review: Considering Simultaneous Voltage-Sensitive Load Reductions},
  institution = {North American Electric Reliability Corporation},
  type        = {Event Analysis Report},
  month       = jan,
  year        = {2025},
  url         = {https://www.nerc.com/globalassets/our-work/reports/event-reports/incident_review_large_load_loss.pdf}
}

@techreport{NERC_SOR_2025,
  author      = {{North American Electric Reliability Corporation (NERC)}},
  title       = {2025 State of Reliability: Assessment Overview of 2024 Bulk Power System Performance},
  institution = {North American Electric Reliability Corporation},
  month       = Jun.,
  year        = {2025},
  url         = {https://www.nerc.com/globalassets/programs/rapa/pa/nerc_sor_2025_overview.pdf}
}

@techreport{NERC_GFM_2021,
  author      = {{North American Electric Reliability Corporation (NERC)}},
  title       = {Grid Forming Technology: Bulk Power System Reliability Considerations},
  institution = {North American Electric Reliability Corporation},
  type        = {White Paper},
  month       = Dec.,
  year        = {2021},
  url         = {https://www.nerc.com/globalassets/our-work/reports/white-papers/white_paper_grid_forming_technology.pdf}
}

@techreport{ESIG_GFM_Report_2022,
  author      = {{High Share of Inverter-Based Generation Task Force}},
  title       = {Grid-Forming Technology in Energy Systems Integration},
  institution = {Energy Systems Integration Group (ESIG)},
  address     = {Reston, VA, USA},
  month       = {Mar.},
  year        = {2022},
  url         = {https://www.esig.energy/wp-content/uploads/2024/10/ESIG-GFM-report-2022.pdf}
}

@misc{Siemens_SVCPLUS,
  author       = {{Siemens Energy}},
  title        = {{SVC PLUS} ({STATCOM})},
  howpublished = {Product Page},
  note         = {Accessed: Dec. 1, 2025},
  url          = {https://www.siemens-energy.com/global/en/home/products-services/product/svcplus.html}
}

@techreport{AESO_FFR_Pilot_2023,
  author      = {{Alberta Electric System Operator (AESO)}},
  title       = {Fast Frequency Response Pilot -- Lessons Learned},
  institution = {Alberta Electric System Operator},
  month       = Jun.,
  year        = {2023},
  url         = {https://www.aeso.ca/assets/FFR-Pilot-Lessons-Learned-2023_06_26.pdf}
}

@inproceedings{mughees2025short,
  title={Short-term load forecasting for AI-data center},
  author={Mughees, Mariam and Li, Yuzhuo and Chen, Yize and Li, Yunwei Ryan},
  booktitle={2025 IEEE Power \& Energy Society General Meeting (PESGM)},
  pages={1--5},
  year={2025},
  organization={IEEE}
}

@article{su2025grid,
  title={Grid-enhancing technologies for clean energy systems},
  author={Su, Tong and Zhao, Junbo and Gomez-Exposito, Antonio and Chen, Yousu and Terzija, Vladimir and Gentle, Jake P},
  journal={Nature Reviews Clean Technology},
  volume={1},
  number={1},
  pages={16--31},
  year={2025},
  publisher={Nature Publishing Group UK London}
}

@techreport{ESIG2025GETs,
  author      = {{Energy Systems Integration Group (ESIG)}},
  title       = {Utility Perspectives on Making Grid-Enhancing Technologies Work: Use Cases, Barriers, and Recommendations for Scalable Deployment},
  institution = {Energy Systems Integration Group},
  month       = Jul.,
  year        = {2025},
  url         = {https://www.esig.energy/reports-briefs/utility-perspectives-on-making-grid-enhancing-technologies-work/}
}

@misc{google_flexible_2025,
  author       = {M. Terrell},
  title        = {How we're making data centers more flexible to benefit power grids},
  howpublished = {Google Blog},
  month        = Aug.,
  year         = {2025},
  url          = {https://blog.google/innovation-and-ai/infrastructure-and-cloud/global-network/how-were-making-data-centers-more-flexible-to-benefit-power-grids/}
}

@misc{WRI2025_ForecastingTricky,
  author       = {I. Goldsmith and Z. Byrum},
  title        = {Powering the {US} Data Center Boom: Why Forecasting Can Be So Tricky},
  howpublished = {World Resources Institute},
  month        = Sep.,
  year         = {2025},
  url          = {https://www.wri.org/insights/us-data-centers-electricity-demand}
}

@misc{PJM2025_LTLoad,
  author       = {{PJM Interconnection}},
  title        = {2025 Long-Term Load Forecast Report Predicts Significant Increase in Electricity Demand},
  howpublished = {Inside Lines},
  month        = jan,
  year         = {2025},
  url          = {https://insidelines.pjm.com/2025-long-term-load-forecast-report-predicts-significant-increase-in-electricity-demand/}
}

@techreport{ge_dc_microgrid,
  author       = {{GE Vernova}},
  title        = {Greening the Future Data Center Infrastructure via the GE Aeroderivative Technology and Microgrid Controls: Powering Data Centers},
  institution  = {GE Vernova},
  type         = {White Paper},
  number       = {GEA35139A},
  year         = {2024},
  url          = {https://www.gevernova.com/electrification/white-papers-case-studies/greening-the-future-data-center-infrastructure-via-the-ge-aeroderivative-technology-and-microgrid-controls}
}

@misc{choukse_power_stabilization,
  author        = {E. Choukse and others},
  title         = {Power Stabilization for {AI} Training Datacenters},
  eprint        = {2508.14318},
  archiveprefix = {arXiv},
  primaryclass  = {cs.AR},
  month         = Aug.,
  year          = {2025},
  doi           = {10.48550/arXiv.2508.14318},
  url           = {https://arxiv.org/abs/2508.14318}
}

@article{LiHe2014_HarmonicComp,
  author = {Y. W. Li and J. He},
  title = {Distribution System Harmonic Compensation Methods: An Overview of {DG}-Interfacing Inverters},
  journal = {IEEE Industrial Electronics Magazine},
  volume = {8},
  number = {4},
  pages = {18--31},
  year = {2014}
}

@misc{LinMughees2025_ForecastMG,
  author = {Lin, Zhiheng and Mughees, Mariam and Li, Yuzhuo and Li, Yunwei Ryan},
  title = {Data-Driven Load-Forecast-Aided Microgrid for {AI} Data Center},
  howpublished = {TechRxiv preprint, version 1},
  month = {Nov.},
  year = {2025},
  doi = {10.36227/techrxiv.176300486.61825852/v1},
  url = {https://doi.org/10.36227/techrxiv.176300486.61825852/v1},
  note = {Posted Nov. 13, 2025}
}

@article{li2025ai,
  title={AI Load Dynamics--A Power Electronics Perspective},
  author={Li, Yuzhuo and Li, Yunwei},
  journal={arXiv preprint arXiv:2502.01647},
  year={2025}
}

@misc{IAEA2024_SMR,
  author       = {{International Atomic Energy Agency}},
  title        = {Small Modular Reactors: Advances in {SMR} Developments},
  howpublished = {IAEA, Vienna, Austria},
  year         = {2024},
  note         = {Accessed: Dec. 2, 2025},
  url          = {https://www.iaea.org/topics/small-modular-reactors}
}

@techreport{NERC2025FERCComments,
  author      = {{North American Electric Reliability Corporation (NERC)}},
  title       = {Comments of the {North American Electric Reliability
                 Corporation} on {FERC} Large Load {ANOPR}},
  institution = {North American Electric Reliability Corporation},
  type        = {Regulatory Filing},
  month       = jan,
  year        = {2026},
  url         = {https://www.nerc.com/globalassets/who-we-are/legal--regulatory/filings--orders/nerc-filings-to-ferc/2025/large_load_anopr_comments_final.pdf}
}

@techreport{FingridKJV2026,
  author      = {{Fingrid Oyj}},
  title       = {{KJV2026} -- Main Requirements for Demand Facility Connections},
  institution = {Fingrid Oyj},
  type        = {Draft stakeholder slide Dec.k/specification},
  month       = Jun.,
  year        = {2025}
}

@misc{ENTSOEDCC,
  author       = {{European Commission}},
  title        = {Commission Regulation ({EU}) 2016/1388 of 17 Aug. 2016 establishing a Network Code on Demand Connection},
  howpublished = {Official Journal of the European Union},
  year         = {2016},
  url          = {https://eur-lex.europa.eu/eli/reg/2016/1388/oj}
}

@article{takci2025,
  author  = {M. T. Takci and M. Qadrdan and J. Summers and J. Gustafsson},
  title   = {Data centres as a source of flexibility for power systems},
  journal = {Energy Reports},
  volume  = {13},
  pages   = {3661--3671},
  month   = Jun.,
  year    = {2025}
}

@misc{Eaton2025,
  author       = {K. Watson},
  title        = {Data Centers -- A Good Grid Citizen},
  howpublished = {Presentation for the NERC Reliability and Security Technical Committee (RSTC), Load Modeling Working Group (LMWG)},
  month        = Jul.,
  year         = {2025},
  url          = {https://www.nerc.com/globalassets/who-we-are/standing-committees/rstc/lmwg/20250715_lmwg_presentations.pdf}
}

@inproceedings{sakalkar2020data,
  author    = {V. Sakalkar and V. Kontorinis and D. Landhuis and S. Li and D. De Ronde and T. Blooming and A. Ramesh and J. Kennedy and C. Malone and J. Clidaras and others},
  title     = {Data center power oversubscription with a medium voltage power plane and priority-aware capping},
  booktitle = {Proc. 25th International Conference on Architectural Support for Programming Languages and Operating Systems (ASPLOS)},
  pages     = {497--511},
  year      = {2020}
}

@techreport{Xenergy_ERA_2025,
  author      = {{X-energy Canada Inc.}},
  title       = {Assessing Site and Distribution Infrastructure for Transitioning a Thermal Plant to an {Xe-100} Small Modular Reactor},
  institution = {Emissions Reduction Alberta (ERA)},
  type        = {Final Outcomes Report, Public Version},
  month       = jan,
  year        = {2025},
  url         = {https://www.eralberta.ca/wp-content/uploads/2024/03/ERA-Final-Report-Public-Version-XE-pre-FEED-Study-Rev-1.pdf}
}

@techreport{BloomLoadFollowing2024,
  author      = {{Bloom Energy}},
  title       = {Load Following Solid Oxide Fuel Cell},
  institution = {Bloom Energy},
  type        = {White Paper},
  month       = Feb.,
  year        = {2024},
  note        = {Accessed: Nov. 2025},
  url         = {https://www.bloomenergy.com/wp-content/uploads/load-following-solid-oxide-fuel-cell.pdf}
}

@inproceedings{stojkovic2025tapas,
  author    = {J. Stojkovic and C. Zhang and {\'I}. Goiri and E. Choukse and
               H. Qiu and R. Fonseca and J. Torrellas and R. Bianchini},
  title     = {{TAPAS}: Thermal- and Power-Aware Scheduling for {LLM}
               Inference in Cloud Platforms},
  booktitle = {Proc. 30th ACM International Conference on Architectural
               Support for Programming Languages and Operating Systems
               (ASPLOS '25), Vol. 2},
  pages     = {1266--1281},
  year      = {2025}
}

@inproceedings{liu2025greenllm,
  author    = {Liu, Qunyou and Huang, Darong and Zapater, Marina and Atienza, David},
  title     = {{GreenLLM}: {SLO}-Aware Dynamic Frequency Scaling for Energy-Efficient {LLM} Serving},
  booktitle = {Proceedings of the 63rd ACM/IEEE Design Automation Conference},
  series    = {DAC '26},
  year      = {2026},
  pages     = {1--7},
  publisher = {ACM},
  address   = {New York, NY, USA}
}

@article{basit2026dualscale,
  title={DualScale: Energy-Efficient Disaggregated LLM Serving via Phase-Aware Placement and DVFS},
  author={Basit, Omar and Liu, Yunzhao and Kong, Z Jonny and Hu, Y Charlie},
  journal={arXiv preprint arXiv:2602.18755},
  year={2026}
}

@article{hankendi2026pals,
  title={PALS: Power-Aware LLM Serving for Mixture-of-Experts Models},
  author={Hankendi, Can and Shahout, Rana and Yu, Minlan and Coskun, Ayse K},
  journal={arXiv preprint arXiv:2605.21427},
  year={2026}
}

@misc{fftrainer2025,
  author       = {B. Zhao and Y. Wang and C. Liu and J. Pan and G. Yang and R. Liu and T. Zhang and K. Luo and W. Xu},
  title        = {{FFTrainer}: Fast Failover in Large-Language Model Training with Almost-Free State Management},
  howpublished = {arXiv preprint arXiv:2512.03644},
  year         = {2025},
  doi          = {10.48550/arXiv.2512.03644}
}

@inproceedings{narayanan2020heterogeneity,
  title={$\{$Heterogeneity-Aware$\}$ cluster scheduling policies for deep learning workloads},
  author={Narayanan, Deepak and Santhanam, Keshav and Kazhamiaka, Fiodar and Phanishayee, Amar and Zaharia, Matei},
  booktitle={14th USENIX Symposium on Operating Systems Design and Implementation (OSDI 20)},
  pages={481--498},
  year={2020}
}

@inproceedings{jiang2024megascale,
  title={$\{$MegaScale$\}$: Scaling large language model training to more than 10,000 $\{$GPUs$\}$},
  author={Jiang, Ziheng and Lin, Haibin and Zhong, Yinmin and Huang, Qi and Chen, Yangrui and Zhang, Zhi and Peng, Yanghua and Li, Xiang and Xie, Cong and Nong, Shibiao and others},
  booktitle={21st USENIX Symposium on Networked Systems Design and Implementation (NSDI 24)},
  pages={745--760},
  year={2024}
}

@article{jaiswal2025sageserve,
  author  = {S. Jaiswal and K. Jain and Y. Simmhan and A. Parayil and
             A. Mallick and R. Wang and R. {St. Amant} and C. Bansal and
             V. R{\"u}hle and A. Kulkarni and S. Kofsky and S. Rajmohan},
  title   = {{SageServe}: Optimizing {LLM} Serving on Cloud Data Centers with
             Forecast Aware Auto-Scaling},
  journal = {Proc. ACM on Measurement and Analysis of Computing Systems
             (POMACS)},
  volume  = {9},
  number  = {3},
  year    = {2025}
}

@inproceedings{li2024sprout,
  author    = {B. Li and Y. Jiang and V. Gadepally and D. Tiwari},
  title     = {Sprout: Green Generative {AI} with Carbon-Efficient {LLM}
               Inference},
  booktitle = {Proc. 2024 Conference on Empirical Methods in Natural Language
               Processing (EMNLP)},
  pages     = {21799--21813},
  year      = {2024}
}

@article{sheshanarayana2026gar,
  title={GAR: Carbon-Aware Routing for LLM Inference via Constrained Optimization},
  author={Sheshanarayana, Disha and Pal, Rajat Subhra and Sinha, Manjira and Dasgupta, Tirthankar},
  journal={arXiv preprint arXiv:2605.11603},
  year={2026}
}

@inproceedings{moore2026marlin,
  author    = {Moore, Hayden and Qi, Sirui and Milojicic, Dejan and Bash, Cullen and Pasricha, Sudeep},
  title     = {{MARLIN}: Multi-Agent Game-Theoretic Reinforcement Learning for Sustainable {LLM} Inference in Cloud Datacenters},
  booktitle = {Proceedings of the 2026 International Green and Sustainable Computing Conference},
  series    = {IGSC '26},
  year      = {2026},
  publisher = {ACM},
  doi       = {10.1145/3797248.3815404},
  note      = {arXiv preprint arXiv:2605.13496},
  url       = {https://arxiv.org/abs/2605.13496}
}

@inproceedings{zhong2024distserve,
  title={$\{$DistServe$\}$: Disaggregating prefill and decoding for goodput-optimized large language model serving},
  author={Zhong, Yinmin and Liu, Shengyu and Chen, Junda and Hu, Jianbo and Zhu, Yibo and Liu, Xuanzhe and Jin, Xin and Zhang, Hao},
  booktitle={18th USENIX symposium on operating systems design and implementation (OSDI 24)},
  pages={193--210},
  year={2024}
}

@inproceedings{agrawal2024taming,
  title={Taming $\{$Throughput-Latency$\}$ tradeoff in $\{$LLM$\}$ inference with $\{$Sarathi-Serve$\}$},
  author={Agrawal, Amey and Kedia, Nitin and Panwar, Ashish and Mohan, Jayashree and Kwatra, Nipun and Gulavani, Bhargav and Tumanov, Alexey and Ramjee, Ramachandran},
  booktitle={18th USENIX symposium on operating systems design and implementation (OSDI 24)},
  pages={117--134},
  year={2024}
}

@article{zheng2026adaptive,
  title={Adaptive Multi-Objective Tiered Storage Configuration for KV Cache in LLM Service},
  author={Zheng, Xianzhe and Wang, Zhengheng and Ma, Ruiyan and Wang, Rui and Wang, Xiyu and Chen, Rui and Zhang, Peng and Pan, Sicheng and Huang, Zhangheng and Wu, Chenxin and others},
  journal={arXiv preprint arXiv:2603.08739},
  year={2026}
}

@inproceedings{leviathan2023fast,
  title={Fast inference from transformers via speculative decoding},
  author={Leviathan, Yaniv and Kalman, Matan and Matias, Yossi},
  booktitle={International conference on machine learning},
  pages={19274--19286},
  year={2023},
  organization={PMLR}
}

@inproceedings{dutta2026benchmarking,
  title={Benchmarking the Energy Savings with Speculative Decoding Strategies},
  author={Dutta, Rohit and Koley, Paramita and Poddar, Soham and Misra, Janardan and Podder, Sanjay and Balani, Naveen and Ghosh, Saptarshi and Ganguly, Niloy},
  booktitle={Findings of the Association for Computational Linguistics: EACL 2026},
  pages={4737--4748},
  year={2026}
}

@article{liu2026survey,
  title={A survey on inference optimization techniques for mixture of experts models},
  author={Liu, Jiacheng and Tang, Peng and Wang, Wenfeng and Ren, Yuhang and Hou, Xiaofeng and Heng, Pheng Ann and Guo, Minyi and Li, Chao},
  journal={ACM Computing Surveys},
  volume={58},
  number={10},
  pages={1--37},
  year={2026},
  publisher={ACM New York, NY}
}

@article{yang2026architectural,
  title={Architectural Implications of Agentic AI Workflows},
  author={Yang, Jirong and Liu, Peizhe and Zhang, Chaojie and Stojkovic, Jovan},
  journal={arXiv preprint arXiv:2608.04458},
  year={2026}
}

@inproceedings{pham2026agentstop,
  title={AgentStop: Terminating Local AI Agents Early to Save Energy in Consumer Devices},
  author={Pham, Dzung and Katevas, Kleomenis and Shahin Shamsabadi, Ali and Haddadi, Hamed},
  booktitle={Proceedings of the ACM Conference on AI and Agentic Systems},
  pages={1051--1069},
  year={2026}
}

@misc{Subzero2025,
  author       = {{Subzero Engineering}},
  title        = {Is liquid cooling becoming non-negotiable?},
  howpublished = {Blog},
  year         = {2025},
  url          = {https://www.subzeroeng.com/is-liquid-cooling-becoming-non-negotiable/}
}

@techreport{OCP_ORv3_2022,
  author      = {{Open Compute Project Foundation}},
  title       = {Open Rack Base Specification Version 3},
  institution = {Open Compute Project Foundation},
  note        = {Ver. 1.0},
  month       = Sep.,
  year        = {2022},
  url         = {https://www.opencompute.org/documents/open-rack-base-specification-version-3-pdf}
}

@misc{nvidia_800v_hvdc,
  author       = {M. Blake and M. Hsu and I. Goldwasser and H. Petty and J. Huntington},
  title        = {{NVIDIA} 800~{VDC} Architecture Will Power the Next Generation of {AI} Factories},
  howpublished = {NVIDIA Technical Blog},
  month        = May,
  year         = {2025},
  url          = {https://developer.nvidia.com/blog/nvidia-800-v-hvdc-architecture-will-power-the-next-generation-of-ai-factories/}
}

@article{She2013,
  author = {She, Xu and Huang, Alex Q. and Burgos, Rolando},
  title = {Review of solid-state transformer technologies and their application in power distribution systems},
  journal = {IEEE Journal of Emerging and Selected Topics in Power Electronics},
  volume = {1},
  number = {3},
  pages = {186--198},
  year = {2013},
  month = {Sep.}
}

@inproceedings{Lesnicar2003,
  author = {A. Lesnicar and R. Marquardt},
  title = {An innovative modular multilevel converter topology suitable for a wide power range},
  booktitle = {Proc. 2003 IEEE Bologna PowerTech Conference},
  address = {Bologna, Italy},
  pages = {272--277},
  year = {2003},
  volume = {3}
}

@misc{Zhang_SST_2026,
  author        = {J. Xu and X. Jiang and Y. Bao and Y. Zheng and X. Chen and
                   Q. Xu and S. Liao and D. Ke and X. Gao},
  title         = {Sequential Operating Simulation of Solid State
                   Transformer-Driven Next-Generation 800~{VDC} Data Center},
  eprint        = {2601.16502},
  archiveprefix = {arXiv},
  year          = {2026},
  doi           = {10.48550/arXiv.2601.16502},
  url           = {https://arxiv.org/abs/2601.16502}
}

@misc{Infineon800V,
  author = {{Infineon Technologies AG}},
  title = {Infineon advances leading-edge 800 {Volt} {AI} data center power architecture enabling better efficiency and serviceability},
  howpublished = {Press release},
  month = {Oct.},
  year = {2025},
  url = {https://www.infineon.com/press-release/2025/INFXX202510-003},
  note = {Published Oct. 13, 2025}
}

@article{MistriGaNReview,
  author = {S. Mistri and C. Langpoklakpam and S. Elangovan and H.-C. Kuo},
  title = {A comprehensive study on {GaN} power devices: Reliability, performance, and application perspectives},
  journal = {Electronics},
  volume = {14},
  number = {22},
  pages = {4430},
  month = {Nov.},
  year = {2025}
}

@misc{MouserGaN,
  author       = {{Mouser Electronics}},
  title        = {{GaN} power devices: Potential, benefits, and keys to successful use},
  howpublished = {Application note},
  note         = {Accessed: Dec. 2025}
}

@techreport{nvidia_h100_arch,
  author      = {{NVIDIA Corporation}},
  title       = {{NVIDIA} {H100} Tensor Core {GPU} Architecture},
  institution = {NVIDIA Corporation},
  type        = {White Paper},
  year        = {2022},
  url         = {https://resources.nvidia.com/en-us-hopper-architecture/nvidia-h100-tensor-c}
}

@inproceedings{GoogleTPU_Jouppi2017,
  author    = {N. P. Jouppi and C. Young and N. Patil and D. Patterson and others},
  title     = {In-Datacenter Performance Analysis of a Tensor Processing Unit},
  booktitle = {Proc. 44th ACM/IEEE International Symposium on Computer Architecture (ISCA)},
  pages     = {1--12},
  year      = {2017}
}

@misc{Narayanaswamy2025,
  author       = {S. Narayanaswamy and P. D. Patel and I. Karlin and A. Gupta and S. Saripalli and J. Guo},
  title        = {Datacenter Energy Optimized Power Profiles},
  howpublished = {arXiv preprint arXiv:2510.03872},
  year         = {2025},
  doi          = {10.48550/arXiv.2510.03872}
}

@inproceedings{Desai2025,
  author = {T. Desai and J. Shah and G. Alavani and S. Saha and S. Sarkar},
  title = {Adaptive {GPU} power capping: Balancing energy efficiency, thermal control and performance},
  booktitle = {Proc. 34th International Symposium on High-Performance Parallel and Distributed Computing (HPDC '25)},
  publisher = {ACM},
  address = {New York, NY, USA},
  year = {2025},
  pages = {1--3}
}

@misc{RedHatKepler2023,
  author       = {{Red Hat}},
  title        = {Introducing {Kepler}: Efficient power monitoring for {Kubernetes}},
  howpublished = {Red Hat Emerging Technologies Blog},
  month        = Aug.,
  year         = {2023},
  url          = {https://next.redhat.com/2023/08/22/introducing-kepler-efficient-power-monitoring-for-kubernetes/}
}

@misc{NERCProject2026_02,
  author       = {{North American Electric Reliability Corporation}},
  title        = {{Project 2026-02}: Computational Loads},
  year         = {2026},
  howpublished = {Reliability Standards Under Development},
  url          = {https://www.nerc.com/standards/reliability-standards-under-development/2026-02-computational-loads},
  note         = {Accessed Aug. 29, 2026}
}

@misc{ERCOTNOGRR282,
  author       = {{Electric Reliability Council of Texas}},
  title        = {{NOGRR282}: Board Priority---Large Computational Load
                  Ride-Through Requirements},
  year         = {2026},
  howpublished = {Nodal Operating Guide Revision Request},
  url          = {https://www.ercot.com/mktrules/issues/NOGRR282},
  note         = {Approved Jul. 9, 2026; effective Aug. 1, 2026;
                  accessed Aug. 29, 2026}
}

@misc{ERCOTNPRR1308,
  author       = {{Electric Reliability Council of Texas}},
  title        = {{NPRR1308}: Board Priority---Related to {NOGRR282},
                  Large Computational Load Ride-Through Requirements},
  year         = {2026},
  howpublished = {Nodal Protocol Revision Request},
  url          = {https://www.ercot.com/mktrules/issues/NPRR1308},
  note         = {Approved Jul. 9, 2026; effective Aug. 1, 2026;
                  accessed Aug. 29, 2026}
}

@misc{AESOLargeLoad2026,
  author       = {{Alberta Electric System Operator}},
  title        = {Large Load Projects},
  year         = {2026},
  howpublished = {Data-centre and large-load integration program webpage},
  url          = {https://www.aeso.ca/grid/connecting-to-the-grid/large-load-projects/},
  note         = {Accessed Aug. 29, 2026}
}

@misc{ferc2026largeload,
  author       = {{Federal Energy Regulatory Commission}},
  title        = {FERC Launches Aggressive Targeted Action to Speed Large Load Integration},
  month        = {June},
  year         = {2026},
  url          = {https://www.ferc.gov/news-events/news/ferc-launches-aggressive-targeted-action-speed-large-load-integration},
  note         = {Accessed: 2026-08-31}
}

@misc{NLRGenAIPower2026,
  author       = {Vercellino, Roberto and
                  Willard, Jared and
                  Campos, Gustavo and
                  Pereira, Weslley da Silva and
                  Hull, Olivia and
                  Selensky, Matt and
                  Mueller, Juliane},
  title        = {Dataset of Generative {AI} Workload Power Profiles},
  year         = {2026},
  howpublished = {{NLR} Data Catalog},
  publisher    = {National Laboratory of the Rockies},
  address      = {Golden, CO, USA},
  number       = {NLR/OT-2C00-99122},
  doi          = {10.7799/3025227},
  url          = {https://data.nlr.gov/submissions/312},
  note         = {Last updated Jul. 17, 2026;
                  accessed Aug. 29, 2026}
}

@inproceedings{Samsi2021MITSupercloud,
  author    = {Samsi, Siddharth and
               Weiss, Matthew L. and
               Bestor, David and
               Li, Baolin and
               Jones, Michael and
               Reuther, Albert and
               Edelman, Daniel and
               Arcand, William and
               Byun, Chansup and
               Holodnack, John and
               Hubbell, Matthew and
               Kepner, Jeremy and
               Klein, Anna and
               McDonald, Joseph and
               Michaleas, Adam and
               Michaleas, Peter and
               Milechin, Lauren and
               Mullen, Julia and
               Yee, Charles and
               Price, Benjamin and
               Prout, Andrew and
               Rosa, Antonio and
               Vanterpool, Allan and
               McEvoy, Lindsey and
               Cheng, Anson and
               Tiwari, Devesh and
               Gadepally, Vijay},
  title     = {The {MIT} Supercloud Dataset},
  booktitle = {2021 {IEEE} High Performance Extreme Computing
               Conference ({HPEC})},
  year      = {2021},
  pages     = {1--8},
  publisher = {IEEE}
}

@article{shehabi20242024,
  title={2024 united states data center energy usage report},
  author={Shehabi, Arman and Newkirk, Alex and Smith, Sarah J and Hubbard, Alex and Lei, Nuoa and Siddik, Md Abu Bakar and Holecek, Billie and Koomey, Jonathan and Masanet, Eric and Sartor, Dale},
  year={2024}
}

@inproceedings{Chung2025MLEnergy,
  author    = {Chung, Jae-Won and
               Ma, Jeff J. and
               Wu, Ruofan and
               Liu, Jiachen and
               Kweon, Oh Jun and
               Xia, Yuxuan and
               Wu, Zhiyu and
               Chowdhury, Mosharaf},
  title     = {The {ML.ENERGY} Benchmark: Toward Automated Inference
               Energy Measurement and Optimization},
  booktitle = {Advances in Neural Information Processing Systems 38:
               Datasets and Benchmarks Track},
  year      = {2025},
  volume    = {38},
  pages     = {121191--121219},
  publisher = {Neural Information Processing Systems Foundation, Inc.}
}

@inproceedings{Wang2025BurstGPT,
  author    = {Wang, Yuxin and
               Chen, Yuhan and
               Li, Zeyu and
               Kang, Xueze and
               Fang, Yuchu and
               Zhou, Yeju and
               Zheng, Yang and
               Tang, Zhenheng and
               He, Xin and
               Guo, Rui and
               Wang, Xin and
               Wang, Qiang and
               Zhou, Amelie Chi and
               Chu, Xiaowen},
  title     = {{BurstGPT}: A Real-World Workload Dataset to Optimize
               {LLM} Serving Systems},
  booktitle = {Proceedings of the 31st {ACM} {SIGKDD} Conference on
               Knowledge Discovery and Data Mining, Volume 2},
  year      = {2025},
  pages     = {5831--5841},
  address   = {Toronto, ON, Canada},
  publisher = {Association for Computing Machinery}
}

@article{Chung2026OpenG2G,
  author        = {Chung, Jae-Won and
                   Liang, Zhirui and
                   Mao, Yanyong and
                   Chen, Jiasi and
                   Chowdhury, Mosharaf and
                   Dvorkin, Vladimir},
  title         = {{OpenG2G}: A Simulation Platform for {AI}
                   Datacenter--Grid Runtime Coordination},
  journal       = {arXiv preprint arXiv:2605.05519},
  year          = {2026},
  month         = May,
  eprint        = {2605.05519},
  archiveprefix = {arXiv},
  primaryclass  = {cs.LG},
  doi           = {10.48550/arXiv.2605.05519},
  url           = {https://arxiv.org/abs/2605.05519}
}

@misc{MicrosoftGridInteractiveUPS2022,
  author       = {Roach, John},
  title        = {{Microsoft} Datacenter Batteries to Support Growth
                  of Renewables on the Power Grid},
  year         = {2022},
  month        = Jul.,
  howpublished = {Microsoft Source},
  organization = {Microsoft},
  url          = {https://news.microsoft.com/source/features/sustainability/ireland-wind-farm-datacenter-ups/},
  note         = {Published Jul. 7, 2022;
                  accessed Aug. 29, 2026}
}

@article{Colangelo2026GridInteractiveAI,
  author    = {Colangelo, Philip and
               Coskun, Ayse K. and
               Megrue, Jack and
               Roberts, Ciaran and
               Sengupta, Shayan and
               Sivaram, Varun and
               Tiao, Ethan and
               Vijaykar, Aroon and
               Williams, Chris and
               Wilson, Daniel C. and
               Records, Brandon and
               MacFarland, Zack and
               Dreiling, Daniel and
               Morey, Nathan and
               Ratnayake, Anuja and
               Vairamohan, Baskar},
  title     = {{AI} Data Centres as Grid-Interactive Assets},
  journal   = {Nature Energy},
  year      = {2026},
  month     = Feb.,
  volume    = {11},
  number    = {2},
  pages     = {254--261}
}

@misc{NationalGridEmerald2026,
  author       = {{National Grid}},
  title        = {{UK}-First Trial of {AI} Grid Technology Successfully
                  Demonstrates the Ability of Data Centres to Adjust
                  Power Needs},
  year         = {2026},
  month        = {Mar.},
  howpublished = {Press release},
  url          = {https://www.nationalgrid.com/uk-first-trial-ai-grid-technology-successfully-demonstrates-ability-data-centres-adjust-power-needs},
  note         = {Published Mar. 2, 2026;
                  accessed Aug. 29, 2026}
}

@misc{NLRAgora2026,
  author       = {{National Laboratory of the Rockies}},
  title        = {{NLR} Launches {Agora}, First-of-Its-Kind Large-Load
                  Grid Integration Test Bed},
  year         = {2026},
  month        = May,
  howpublished = {News release},
  url          = {https://www.nlr.gov/news/detail/press/2026/news-release-nlr-launches-agora-first-of-its-kind-large-load-grid-integration-test-bed},
  note         = {Published May 19, 2026;
                  accessed Aug. 29, 2026}
}

@techreport{IEEE_DCStandardsGap2026,
  author       = {{IEEE Standards Association}},
  title        = {Review of Industry Efforts and Standards of Grid Readiness for Data Center Deployment},
  institution  = {IEEE Standards Association},
  type         = {White Paper},
  number       = {IC25-004},
  month        = jan,
  year         = {2026},
  pages        = {1--62},
  url          = {https://standards.ieee.org/ieee/White_Paper/12415},
  note         = {Produced by the Data Centers: Standards Needs Analysis and Recommendations Industry Connections Activity}
}

@misc{AuroraAIFactory2026,
  author       = {{Electric Power Research Institute} and {NVIDIA} and {Emerald AI}},
  title        = {{Aurora AI Factory}, Manassas, Virginia},
  year         = {2026},
  howpublished = {\url{https://emeraldai.co}}
  
}

@misc{SVP_EmeraldPilot2026,
  author       = {{Silicon Valley Power} and {NVIDIA} and {Emerald AI}},
  title        = {Bidirectional Utility--Data Center Communication and Load Response Pilot},
  year         = {2026},
  howpublished = {\url{https://investor.nvidia.com/news/press-release-details/2026/NVIDIA-DSX-Gives-Infrastructure-Builders-the-Playbook-for-AI-Factories/default.aspx}},

}

@techreport{EPRI_FlexMOSAIC2026,
  author      = {{Electric Power Research Institute}},
  title       = {{Flex MOSAIC}: A Flexibility Classification Framework for Large Electric Loads},
  institution = {Electric Power Research Institute (EPRI)},
  type        = {Technical Brief},
  month       = {Mar.},
  year        = {2026},
  url         = {https://dcflex.epri.com/flex-mosaic},

}

@misc{CRU_LEU_Policy2025,
  author       = {{Commission for Regulation of Utilities}},
  title        = {Large Energy Users Connection Policy},
  howpublished = {CRU Decision Paper CRU/2025236},
  address      = {Dublin, Ireland},
  month        = Dec.,
  year         = {2025},
}

@misc{Ofgem_DemandConnections2026,
  author       = {{Ofgem}},
  title        = {Call for Input: Demand Connections Reform},
  howpublished = {Ofgem consultation},
  year         = {2026},
  month        = {Feb.},
  note         = {Responses were due 13 Mar. 2026. Outlined a connection queue surge from 41~GW to 125~GW against a 45~GW GB peak demand, heavily driven by ~50~GW of data center projects (~140 projects, 71 at FID)}
}

@misc{SPP_HILL2026,
  author       = {{Southwest Power Pool, Inc.}},
  title        = {High Impact Large Load Study Process},
  howpublished = {Tariff Revisions Approved by FERC, Docket No. ER26-247-000},
  year         = {2026},
  month        = jan,
  day          = {14},
  note         = {Effective January 15, 2026. Supplemental CHILLS framework approved June 5, 2026, under Docket No. ER26-1323.}
}

@misc{FERC_EL25_49,
  author = {{Federal Energy Regulatory Commission}},
  title = {Order on Show Cause Proceeding, Directing Compliance Filings, Establishing Paper Hearing, and Granting in Part and Denying in Part Complaint},
  howpublished = {193 FERC \P~61,217},
  month = {Dec.},
  year = {2025},
  url = {https://www.ferc.gov/sites/default/files/2025-12/E-1%20EL25-49-000.pdf},
  urldate = {2026-08-31},
  note = {Docket Nos. EL25-49-000, EL25-49-001, AD24-11-000 and EL25-20-000 (consolidated); issued Dec. 18, 2025}
}

@techreport{NERC_LLWG_GapPaper2026,
  author       = {{North American Electric Reliability Corporation}},
  title        = {Assessment of Gaps in Existing Practices, Requirements, and Reliability Standards for Emerging Large Loads},
  institution  = {NERC, Large Loads Working Group},
  type         = {White Paper},
  month        = {Mar.},
  year         = {2026},
  url          = {https://www.nerc.com/initiatives/large-loads-action-plan},
  note         = {Contains 11 core recommendations addressing bulk power system integration deficiencies for data centers and high-impact industrial loads.}
}

@misc{Niu2025TokenPowerBench,
  author       = {Niu, Chenxu and Zhang, Wei and Li, Jie and
                  Zhao, Yongjian and Wang, Tongyang and Wang, Xi and
                  Chen, Yong},
  title        = {{TokenPowerBench}: Benchmarking the Power Consumption
                  of {LLM} Inference},
  howpublished = {arXiv:2512.03024},
  year         = {2025},
  url          = {https://arxiv.org/abs/2512.03024},
  urldate      = {2026-08-31}
}

@misc{WattGPU2026,
  author       = {Fadel Argerich, Mauricio and F{\"u}rst, Jonathan and
                  Pati{\~n}o-Mart{\'i}nez, Marta},
  title        = {{WattGPU}: Predicting Inference Power and Latency on
                  Unseen {GPUs} and {LLMs}},
  howpublished = {arXiv:2607.02391},
  month        = Jul.,
  year         = {2026},
  url          = {https://arxiv.org/abs/2607.02391},
  urldate      = {2026-08-31},
  note         = {42 LLMs, 8 GPUs; median APE $\leq$3.4\,\% offline and
                  $\leq$13.5\,\% server on unseen GPUs}
}

@misc{WattCounts2026,
  author       = {Fadel Argerich, Mauricio and F{\"u}rst, Jonathan and
                  Pati{\~n}o-Mart{\'i}nez, Marta},
  title        = {{Watt Counts}: Energy-Aware Benchmark for Sustainable
                  {LLM} Inference on Heterogeneous {GPU} Architectures},
  howpublished = {arXiv:2604.09048},
  month        = Apr.,
  year         = {2026},
  url          = {https://arxiv.org/abs/2604.09048},
  urldate      = {2026-08-31},
  note         = {50 models across 10 GPU types}
}

@inproceedings{Hu2024AcmeTrace,
  author       = {Hu, Qinghao and Ye, Zhisheng and Wang, Zerui and
                  Wang, Guoteng and Zhang, Meng and Chen, Qiaoling and
                  Sun, Peng and Lin, Dahua and Wang, Xiaolin and
                  Luo, Yingwei and Wen, Yonggang and Zhang, Tianwei},
  title        = {Characterization of Large Language Model Development
                  in the Datacenter},
  booktitle    = {Proc.\ 21st USENIX Symp.\ Networked Systems Design and
                  Implementation (NSDI '24)},
  address      = {Santa Clara, CA, USA},
  pages        = {709--729},
  year         = {2024}
}

@misc{AzureLLMTrace,
  author       = {{Microsoft}},
  title        = {Azure {LLM} Inference Trace 2023},
  howpublished = {Azure Public Dataset},
  year         = {2023},
  url          = {https://github.com/Azure/AzurePublicDataset/blob/master/AzureLLMInferenceDataset2023.md},
  urldate      = {2026-08-31},
}

@misc{EPRI_DCFlex2026,
  author       = {{Electric Power Research Institute}},
  title        = {{EPRI}'s {DCFlex} Initiative Expands to Nine Demonstration Sites Across {U.S.}, {Europe}},
  howpublished = {Press release},
  month        = {Feb.},
  year         = {2026},
  url          = {https://www.epri.com/about/media-resources/press-release/rBbBPmK6zVt6eXN9rGQlWLMRwHRHQWIS},
  urldate      = {2026-08-31},
  note         = {Initiative launched Oct. 2024; programme page: https://dcflex.epri.com/},
}

@misc{Google1GWDR2026,
  author       = {{Google}},
  title        = {{Google} Signed 1~{GW} of Data Center Demand Response},
  howpublished = {Google Blog},
  month        = {Mar.},
  year         = {2026},
  url          = {https://blog.google},
  urldate      = {2026-08-31},
  note         = {Agreements with Indiana Michigan Power, TVA, Entergy Arkansas, Minnesota Power, and DTE Energy},
}

@misc{FERC_NERC_Directive2026,
  author = {{Federal Energy Regulatory Commission}},
  title = {Order Directing Reliability Standard(s) Pertaining to Computational Load Integration},
  howpublished = {FERC order, Docket No. RD26-7-000},
  month = {Jul.},
  year = {2026},
  note = {Issued Jul. 16, 2026; indexed by NERC in FERC Orders and Rules, 2026},
  url = {https://www.nerc.com/who-we-are/regulatory/ferc-orders-and-rules/2026}
}

@misc{chung2026mlenergy,
  author = {{ML.ENERGY}},
  title = {The {ML.ENERGY} Benchmark},
  howpublished = {Software repository},
  url = {https://github.com/ml-energy/benchmark},
  note = {Accessed: Sep. 15, 2026},
  urldate = {2026-09-15}
}

@inproceedings{patel2025n,
  author    = {Patel, Pratyush and Choukse, Esha and Zhang, Chaojie and
               Shah, Aashaka and Goiri, {\'I}{\~n}igo and Maleki, Saeed and
               Bianchini, Ricardo},
  title     = {Splitwise: Efficient Generative {LLM} Inference Using
               Phase Splitting},
  booktitle = {Proc. 51st Annual International Symposium on Computer
               Architecture (ISCA)},
  address   = {Buenos Aires, Argentina},
  pages     = {118--132},
  year      = {2024}
}

@article{cheng2022real,
  author  = {Cheng, Yunzhi and Fan, Lingling and Rose, Jonathan and
             Huang, Fred and Schmall, John and Wang, Xiaoyu and
             Xie, Xiaorong and Shair, Jan and Ramamurthy, Jayanth R. and
             Modi, Nilesh and others},
  title   = {Real-World Subsynchronous Oscillation Events in Power Grids
             With High Penetrations of Inverter-Based Resources},
  journal = {IEEE Transactions on Power Systems},
  volume  = {38},
  number  = {1},
  pages   = {316--330},
  month   = jan,
  year    = {2023}
}

@techreport{PNNL38817,
  author      = {Ross, B. A. and Follum, J. D.},
  title       = {Electromagnetic Transient Modeling of Large Data Centers
                 for Grid-Level Studies},
  institution = {Pacific Northwest National Laboratory},
  address     = {Richland, WA, USA},
  number      = {PNNL-38817},
  month       = jan,
  year        = {2026},
  url         = {https://www.pnnl.gov/publications/electromagnetic-transient-modeling-large-data-centers-grid-level-studies}
}

@techreport{psercDynamicLoads2024,
  author      = {{Power Systems Engineering Research Center (PSERC)}},
  title       = {Effects of Dynamic Power Electronic Load Models on Power
                 Systems Analysis Using {ZIP-E} Loads},
  institution = {Power Systems Engineering Research Center},
  type        = {Final Report},
  number      = {T-70G},
  month       = Dec.,
  year        = {2024},
  url         = {https://documents.pserc.wisc.edu/documents/publications/reports/2024_reports/T_70G_Final_Report.pdf}
}

@article{paananen2023grid,
  author  = {Paananen, Janne},
  title   = {Grid-Interactive Data Centers Enabling Energy Transition:
             Data Center\textquoteright s Hidden Potential to Provide
             Essential Grid Services of a Future Power System},
  journal = {IEEE Electrification Magazine},
  volume  = {11},
  number  = {3},
  pages   = {26--34},
  month   = Sep.,
  year    = {2023}
}

@article{jiang2021optimal,
  author = {Jiang, Chenwei and Tseng, Chung-Li and Wang, Yizheng and Lan, Zhou and Wen, Fushuan and Chen, Fei and Liang, Liang},
  title = {Optimal Pricing Strategy for Data Center Considering Demand Response and Renewable Energy Source Accommodation},
  journal = {Journal of Modern Power Systems and Clean Energy},
  volume = {11},
  number = {1},
  pages = {345--354},
  month = {jan},
  year = {2023}
}

@incollection{li2023smart,
  author = {Li, Yun Wei and Nejabatkhah, Farzam and Tian, Hao},
  title = {Smart Hybrid {AC/DC} Microgrids: Structures and Technical Challenges},
  booktitle = {Smart Hybrid {AC/DC} Microgrids: Power Management, Energy Management, and Power Quality Control},
  chapter = {1},
  pages = {1--20},
  publisher = {Wiley-IEEE Press},
  address = {Hoboken, NJ, USA},
  year = {2022}
}

@misc{FCHEADataCenters,
  author       = {Hanson, Tobias},
  title        = {Data Centers: More Power, More Fuel Cells},
  howpublished = {Fuel Cell \& Hydrogen Energy Association (FCHEA),
                  Catalyst Blog},
  month        = Nov.,
  day          = {14},
  year         = {2024},
  url          = {https://fchea.org/data-centers-more-power-more-fuel-cells/}
}

@techreport{siemens_onsite,
  author      = {{Siemens Energy}},
  title       = {On-site Power Generation to Leverage Maximum Uptime: How
                 Gas Turbines Make a Difference in Reliable Energy Supply
                 for Data Centers},
  institution = {Siemens Energy},
  type        = {White Paper},
  month       = Oct.,
  day         = {23},
  year        = {2019},
  note        = {Published via DatacenterDynamics},
  url         = {https://www.datacenterdynamics.com/en/whitepapers/-site-power-generation-leverage-maximum-uptime/}
}

@incollection{schlegel2024small,
  author    = {Schlegel, Joshua P. and Bhowmik, Palash K.},
  title     = {Small Modular Reactors},
  booktitle = {Nuclear Power Reactor Designs},
  editor    = {Wang, Jun and Talabi, Sola and {Bilbao y Le\'on}, Sama},
  pages     = {283--308},
  publisher = {Academic Press},
  month     = Dec.,
  year      = {2023}
}

@misc{wu2024lazarus,
  author        = {Wu, Yongji and Qu, Wenjie and Liu, Xueshen and
                   Tao, Tianyang and Qiao, Yifan and Wang, Zhuang and
                   Bai, Wei and Tian, Yuan and Zhang, Jiaheng and
                   Mao, Z. Morley and others},
  title         = {Lazarus: Resilient and Elastic Training of Mixture-of-Experts
                   Models},
  eprint        = {2407.04656},
  archiveprefix = {arXiv},
  primaryclass  = {cs.DC},
  month         = Jul.,
  year          = {2024},
  doi           = {10.48550/arXiv.2407.04656},
  url           = {https://arxiv.org/abs/2407.04656}
}

@misc{vicor_genai_2024,
  author       = {Wong, William G.},
  title        = {Tackling the Power Challenges of {GenAI} Data Centers},
  howpublished = {Electronic Design},
  month        = Jul.,
  year         = {2024},
  note         = {Sponsored by Vicor Corporation},
  url          = {https://www.electronicdesign.com/technologies/power/article/55126661/electronic-design-tackling-the-power-challenges-of-genai-datacenters-with-vicor-advanced-solutions}
}

@techreport{NavitasWhitepaper2025,
  author = {{Navitas Semiconductor}},
  title = {Redefining Data Center Power: {GaN} and {SiC} Technologies for Next-Gen 800~{VDC} Infrastructure},
  institution = {Navitas Semiconductor},
  type = {White Paper},
  month = {Oct.},
  year = {2025},
  url = {https://navitassemi.com/download/redefining-data-center-power-gan-and-sic-technologies-for-next-gen-800-vdc-infrastructure/}
}

@misc{nvidia_blackwell_nvcomp,
  author       = {{NVIDIA Corporation}},
  title        = {Speeding Up Data Decompression with {nvCOMP} and the
                  {NVIDIA} {Blackwell} Decompression Engine},
  howpublished = {NVIDIA Technical Blog},
  year         = {2025},
  url          = {https://developer.nvidia.com/blog/speeding-up-data-decompression-with-nvcomp-and-the-nvidia-blackwell-decompression-engine/}
}

@inproceedings{qin2024optimized,
   author    = {Qin, Zongyue and Hu, Ziniu and He, Zifan and
                Prakriya, Neha and Cong, Jason and Sun, Yizhou},
  title     = {Optimized Multi-Token Joint Decoding with Auxiliary Model
                for {LLM} Inference},
  booktitle = {Proc. 13th International Conference on Learning
               Representations (ICLR)},
   year      = {2025}
}

@inproceedings{zhao2024sustainable,
  author    = {Zhao, Dan and Samsi, Siddharth and McDonald, Joseph and
               Li, Baolin and Bestor, David and Jones, Michael and
               Tiwari, Devesh and Gadepally, Vijay},
  title     = {Sustainable Supercomputing for {AI}: {GPU} Power Capping
               at {HPC} Scale},
  booktitle = {Proc. 2023 ACM Symposium on Cloud Computing (SoCC '23)},
  publisher = {ACM},
  address   = {New York, NY, USA},
  year      = {2023}
}

@techreport{Ireland2026,
  author      = {Fearon, Se{\'a}n},
  title       = {{The Cost of Data Centres: Modelling the Impact of Data Centres on Electricity Bills for Irish Households}},
  institution = {Friends of the Earth Ireland and Beyond Fossil Fuels},
  type        = {Research report},
  year        = {2026},
  month       = May,
  url         = {https://www.friendsoftheearth.ie/publications/the-cost-of-data-centre-growth-in-ireland/},
  note        = {Published May 21, 2026; accessed Sep. 4, 2026}
}

@techreport{EUGrid2026,
  author      = {{European Commission, Directorate-General for Energy}},
  title       = {{Commission Staff Working Document Accompanying the Proposal for a Regulation of the European Parliament and of the Council Amending Regulation (EU) 2019/943, as Regards Future-Proofing Electricity Bills in the Union, through Reducing System Costs and Fostering Electrification and Digitalisation}},
  institution = {European Commission},
  number      = {SWD(2026) 600 final},
  address     = {Brussels, Belgium},
  year        = {2026},
  month       = Jul.,
  url         = {https://eur-lex.europa.eu/legal-content/EN/TXT/?uri=CELEX:52026SC0600},
  note        = {Published Jul. 17, 2026; accessed Sep. 4, 2026}
}

@misc{APSouthAfrica2026,
  author = {Magome, Mogomotsi and Gumede, Michelle},
  title = {{Civil Rights Groups Urge a Halt to South Africa Data Centers Boom amid Water and Power Fears}},
  howpublished = {AP News},
  year = {2026},
  month = {Sep.},
  url = {https://apnews.com/article/590b64141c2262bd9a1aacb6d313a343},
  note = {Published Sep. 4, 2026; accessed Sep. 4, 2026}
}

@misc{Kenya2026,
  author       = {Sheikh, Hafsa Abdiwahab},
  title        = {{Can Kenya's AI Ambitions Coexist with Naivasha's Water Needs?}},
  howpublished = {Al Jazeera},
  year         = {2026},
  month        = Aug.,
  url          = {https://www.aljazeera.com/economy/2026/8/24/can-kenyas-ai-ambitions-coexist-with-naivashas-water-needs},
  note         = {Published Aug. 24, 2026; accessed Sep. 4, 2026}
}

@misc{India2026,
  author       = {Vengattil, Munsif and Kalra, Aditya},
  title        = {{Google's \$15 Billion India Data Centre Project Battles Water, Wildlife Concerns}},
  howpublished = {Reuters},
  year         = {2026},
  month        = Aug.,
  url          = {https://www.reuters.com/world/asia-pacific/googles-15-billion-india-data-centre-project-battles-water-wildlife-concerns-2026-08-06/},
  note         = {Published Aug. 6, 2026; accessed Sep. 4, 2026}
}

@article{sreekumar2026multi,
  title={Multi-Scale Datacenter Power Modulation},
  author={Sreekumar, Akshay and Christianson, Nicolas and Kazhamiaka, Fiodar and Rajagopal, Ram},
  journal={arXiv preprint arXiv:2609.17809},
  year={2026}
}

@misc{ReutersAustralia2026,
  author = {Clark, Helen},
  title = {{Australia Outback Gas Players Eye Northern Territory to Skip Data Centre Power Queue}},
  howpublished = {Reuters},
  year = {2026},
  month = {Sep.},
  url = {https://www.reuters.com/business/energy/australia-outback-gas-players-eye-northern-territory-skip-data-centre-power-queue-2026-09-03/},
  note = {Published Sep. 3, 2026; accessed Sep. 4, 2026}
}

@misc{mclaughlin2026texasgrid,
  author       = {McLaughlin, Tim},
  title        = {{Texas} grid flags risks as data centers, crypto sites fail voltage tests},
  howpublished = {Reuters},
  year         = {2026},
  month        = jun,
  day          = {5},
  url          = {https://www.reuters.com/business/energy/texas-grid-flags-risks-data-centers-crypto-sites-fail-voltage-tests-2026-06-05/}
}

@misc{NZ2026,
  author       = {Corlett, Eva},
  title        = {{'A Lot of Red Flags': Plans for New Zealand's First AI Datacentre Spark Concern as Locals Demand Greater Transparency}},
  howpublished = {The Guardian},
  year         = {2026},
  month        = Jul.,
  url          = {https://www.theguardian.com/world/2026/Jul./10/new-zealand-first-datacentre-concern-locals-makarewa-invercargill-datagrid},
  note         = {Published Jul. 10, 2026 (New Zealand date); accessed Sep. 4, 2026}
}

@misc{Brazil2026,
  author       = {{Minist{\'e}rio P{\'u}blico Federal}},
  title        = {{MPF e DPU Pedem Adequa{\c{c}}{\~o}es em Licenciamento Ambiental de Data Center Antes do In{\'i}cio da Opera{\c{c}}{\~a}o no Cear{\'a}}},
  howpublished = {Procuradoria da Rep{\'u}blica no Cear{\'a}},
  year         = {2026},
  month        = May,
  url          = {https://www.mpf.mp.br/o-mpf/unidades/pr-ce/noticias/mpf-e-dpu-pedem-adequacoes-em-licenciamento-ambiental-de-data-center-no-ceara-antes-do-inicio-da-operacao},
  note         = {Published May 20, 2026; accessed Sep. 4, 2026}
}

@misc{Japan2026,
  author       = {Otake, Tomoko},
  title        = {{Japan's AI Goldrush Faces Backlash as Data Centers Sprout Up in Urban Areas}},
  howpublished = {The Japan Times},
  year         = {2026},
  month        = Jun.,
  url          = {https://www.japantimes.co.jp/environment/2026/06/21/data-center-pushback/},
  note         = {Published Jun. 21, 2026; accessed Sep. 4, 2026}
}

@misc{Uruguay2026,
  author       = {{El Observador}},
  title        = {{Vecinos del Data Center de Google en Canelones Piden Frenar Avances: Reclaman Dejar sin Efecto Autorizaci{\'o}n Ambiental y Denuncian Molestias}},
  howpublished = {El Observador},
  year         = {2026},
  month        = Aug.,
  url          = {https://www.elobservador.com.uy/nacional/vecinos-del-data-center-google-en-canelones-piden-frenar-avances-reclaman-dejar-efecto-autorizacion-ambiental-y-denuncian-molestias-n6053924},
  note         = {Published in Aug. 2026}
}

@techreport{pjm_lar_impl,
  author      = {{PJM Interconnection}},
  title       = {Load Adjustment Request Implementation},
  institution = {PJM Resource Adequacy Planning Department},
  address     = {Audubon, PA, USA},
  note        = {Manual 19, Attachment B},
  year        = {2025},
  url         = {https://www.pjm.com/committees-and-groups/subcommittees/las},
  urldate     = {2026-09-15}
}

@techreport{pjm_lf2026,
  author      = {{PJM Interconnection}},
  title       = {2026 {PJM} Load Forecast Report},
  institution = {PJM Interconnection},
  address     = {Audubon, PA, USA},
  month       = jan,
  year        = {2026},
  url         = {https://www.pjm.com/-/media/DotCom/library/reports-notices/load-forecast/2026-load-report.pdf},
  urldate     = {2026-09-15}
}

@misc{li2024unseen,
    author = {Yuzhuo Li and Mariam Mughees and Yize Chen
              and Yunwei Ryan Li},
    title = {The Unseen {AI} Disruptions for Power Grids:
             {LLM}-Induced Transients},
    year = {2024},
    eprint = {2409.11416},
    archivePrefix = {arXiv},
    primaryClass = {cs.AR},
    url = {https://arxiv.org/abs/2409.11416}
  }

@article{pope2023efficiently,
  title={Efficiently scaling transformer inference},
  author={Pope, Reiner and Douglas, Sholto and Chowdhery, Aakanksha and Devlin, Jacob and Bradbury, James and Heek, Jonathan and Xiao, Kefan and Agrawal, Shivani and Dean, Jeff},
  journal={Proceedings of machine learning and systems},
  volume={5},
  pages={606--624},
  year={2023}
}

@inproceedings{haghbayan2020thermal,
  title={Thermal-cycling-aware dynamic reliability management in many-core system-on-chip},
  author={Haghbayan, M-H and Miele, Antonio and Zouv, Z and Tenhunen, Hannu and Plosila, Juha and others},
  booktitle={Proceedings of the 2020 Design, Automation and Test in Europe Conference and Exhibition, DATE 2020},
  pages={1229--1234},
  year={2020},
  organization={Institute of Electrical and Electronics Engineers Inc.}
}

@techreport{CEC2026DataCenterMethodology,
    author      = {{California Energy Commission}},
    title       = {Supporting Document for the 2025 {IEPR} Forecast},
    institution = {California Energy Commission},
    type        = {Technical methodology document},
    year        = {2026},
    month       = Apr.,
    note        = {Published April 15, 2026},
    url         = {https://www.energy.ca.gov/sites/default/files/2026-04/Data_Center_Methodology_Memo_ada.pdf}
  }

@misc{NOVEC2025DataCenterForecast,
    author       = {{Northern Virginia Electric Cooperative}},
    title        = {{NOVEC} Data Center Load Forecast},
    howpublished = {Presentation to the PJM Load Analysis Subcommittee},
    year         = {2025},
    month        = Sep.,
    note         = {September 16, 2025},
    url          = {https://www.pjm.com/-/media/DotCom/committees-groups/subcommittees/las/2025/20250916/20250916-item-04b---novec-large-load-request.pdf}
  }

@misc{Roumpani2025GPCIRPTestimony,
    author       = {Maria Roumpani},
    title        = {Direct Testimony of {Maria Roumpani, PhD}
                    on Behalf of {Georgia Conservation Voters}},
    howpublished = {Georgia Public Service Commission,
                    Docket Nos. 56002 and 56003},
    year         = {2025},
    month        = May,
    note         = {May 2, 2025},
    url          = {https://services.psc.ga.gov/api/v1/External/Public/Get/Document/DownloadFile/222501/103611}
  }

@techreport{NERC2025LTRA2026,
    author      = {{North American Electric Reliability Corporation}},
    title       = {{2025 Long-Term Reliability Assessment}},
    institution = {NERC},
    year        = {2026},
    month       = jan,
    url         = {https://www.nerc.com/globalassets/our-work/assessments/nerc_ltra_2025.pdf},
    note        = {Table 2, p. 14}
  }

\end{document}